\documentclass[11pt,a4paper,twoside,titlepage,openright]{report}

\usepackage{graphicx}
\usepackage{amsmath,amsfonts,amssymb}
\usepackage{bbm}
\usepackage{bm}
\usepackage{fancyhdr}

\newcommand{\INT}[3]{{\ensuremath{\int_{#1}^{#2}\,{\sf{d}}{#3}}}}
\newcommand{\E}{{\ensuremath{\sf{e}}}}
\newcommand{\I}{{\ensuremath{\sf{i}}}}
\newcommand{\D}{{\ensuremath{\sf{d}}}}

\newcommand{\RR}{{\ensuremath{\mathbbm{R}}}}
\newcommand{\NN}{{\ensuremath{\mathbbm{N}}}}
\newcommand{\ZZ}{{\ensuremath{\mathbbm{Z}}}}
\newcommand{\bond}[2]{{\ensuremath{ {(#1,#2)}}}}
\newcommand{\dbond}[2]{{\ensuremath{ {[#1,#2]}}}}
\newcommand{\define}[1]{{\bfseries \textit{#1}}}
\newcommand{\evol}{{\ensuremath{\mathcal{U}_B}}}
\newcommand{\zetab}{{\ensuremath{\zeta_B}}}
\newcommand{\zetah}{{\ensuremath{\zeta_h}}}
\newcommand{\bondprop}{{T}}
\newcommand{\scattering}{{S}}
\newcommand{\sigX}{\ensuremath{\sigma_1}}
\newcommand{\sigY}{\ensuremath{\sigma_2}}
\newcommand{\sigZ}{\ensuremath{\sigma_3}}

\newcommand{\DIR}{\ensuremath{\mathbf{dir}}}

\newcommand{\SPIN}{\ensuremath{\mathbf{spin}}}

\newcommand{\ONE}{\ensuremath{\mathbbm{1}}}
\newcommand{\EEE}{\ensuremath{\mathbbm{E}}}
\renewcommand{\vec}[1]{\ensuremath{{\bm{#1}}}}
\renewcommand{\hbar}{\ensuremath{\hslash}}

\begin{document}

\title{\bfseries Quantum Graphs:\\ Applications to Quantum Chaos\\ and Universal Spectral Statistics}

\author{
	Sven Gnutzmann$^{1,2,*}$ and Uzy Smilansky$^{2,3}$\\[1cm]
	$^1$ Freie Universit\"at Berlin,\\ Germany\\[0.3cm]
	$^2$ The Weizmann Institute of Science,\\ Rehovot, Israel\\[0.3cm]
	$^3$ School of Mathematics, Bristol University,\\  Bristol, United Kingdom\\[1cm]
	\small $^*$ new address:\\ \small
 School of Mathematical Sciences, University of Nottingham,\\ \small
Nottingham,  United Kingdom
}


\maketitle

\newpage
\thispagestyle{empty}
\pagenumbering{roman}
\vspace*{1cm}

\newpage

\begin{abstract}
During the last years quantum graphs have become a paradigm 
of quantum chaos with applications from spectral statistics 
to chaotic scattering
and wave function statistics. In the first part of
this review
we give a detailed introduction to the
spectral theory of quantum graphs and
discuss exact trace formulae for the spectrum and the quantum-to-classical
correspondence. The second part of this review is
devoted to the spectral statistics of
quantum graphs as an application to quantum chaos.
Especially, we summarise recent developments on the 
spectral statistics
of generic large quantum graphs based on two approaches: the periodic-orbit
approach and the supersymmetry approach. The latter provides a
condition and a proof for universal spectral statistics
as predicted by random-matrix theory.
\end{abstract}

\newpage\thispagestyle{empty}
\tableofcontents

\newpage
\pagenumbering{arabic}

\chapter{Introduction}
\label{chapter:Introduction}

The general mathematical concept of a graph (network)
as a set of elements which
are connected by some relation has found applications
in many branches of science, engineering, 
and also social
science. 
A street network of a traffic engineer, the network of neurons studied by a
neuroscientist, the structure of databases in
computer science can all be described by graphs.

Recently the Laplacian on a metric graph has gained a lot of attention
in physics and mathematics in terms of the
diffusion equation or Schr\"odinger equation. They have 
now become known as \define{quantum graphs} but 
different aspects are studied under various names 
such as quantum networks or quantum wires. They have a long history
in mathematics and physics.
  
In physics, the first application has probably been in the context of free
electron models for organic molecules about seventy years ago 
by Pauling \cite{pauling}, an approach which has been further
developed in subsequent years \cite{kuhn,kuhn2,platt,rudenberg2,coulson,montroll,richardson}. Quantum graphs have also been applied successfully to
superconductivity in granular and artificial materials \cite{alexander}, 
acoustic and electromagnetic waveguide networks \cite{flesia,mitra}, 
the Anderson transition in a disordered wire \cite{anderson,shapiro},
quantum Hall systems \cite{chalker,klesse,klesse:diss}, 
fracton excitations in fractal
structures \cite{avishai,nakayama}, and mesoscopic quantum systems
\cite{Book:Imry,kowal,texier2,texier3,texier4}. 
Quantum graphs have also been simulated experimentally
\cite{hul:2004}.
 
The construction of self-adjoint operators, or wave equations with
appropriate boundary conditions on graphs has first
been addressed by Ruedenberg and Scherr \cite{rudenberg2} 
(see also \cite{richardson}). They considered graphs as an
idealisation of networks of wires or wave guides of finite cross-section
in the limit where the diameter of the wire is much smaller than any
other length scale. Similar approaches to graphs 
as networks of thin wires with a finite diameter, or \define{fat quantum graphs} as they are now called, have been
a topic in mathematical physics recently \cite{saito,saito2,rubinstein,kuchment2,exner:2003a,post}.

Another interesting aproach which has been discussed mainly by Exner and his
coworkers is based on \define{leaky graphs}
\cite{exner:2001,exner13,exner14,exner:2003,exner:2004,exner16,exner15,exner:2005,exner:2005e}. 
Here, a finite attractive potential
in the Schr\"odinger
equation is centered on a metric graph. A leaky graph is a generalisation of the Schr\"odinger equation with $\delta$-function
potentials. A quantum graph can be realised in a limit of infinitely strong
attracting potentials of this type.

Probably the first mathematical approaches to the
Laplacian on a metric graph were by Roth who derived a trace formula
for the spectrum of the Laplacian \cite{roth,roth:trace} and by
von~Below \cite{below:1985,below:1988,below:1993}. The spectral theory
of quantum graphs was mainly developed on the basis of the von Neumann
theory of self-adjoint extensions for formal differential operators
\cite{avron_leshouches,albeverio,exnerseba,novikov,exner,exner2,exner3,exner12,avronexner,carlson,carlson2,carlson3,kostrykin:1999,kostrykin2,kostrykin4,cheon,exner:2005b,fulling}.
Other recent topics in mathematics and mathematical physics
include the spectral theory of infinite periodic graphs
\cite{carlson:2004,bentosela,bruning,cattaneo,exner2,exner,friedlander,kostrykin:2004,pankrashkin1,pankrashkin2}, 
tree graphs \cite{aizenman,aizenman2,aizenman3,aizenman:2005,solomyak:2004a,carlson4,naimark} 
and Sierpinski graphs
\cite{teplyaev,meyers,shirai}, 
scattering and bound states in open graphs
\cite{barra:2002,exner4,exner10,exner11,gerasimenko,kostrykin:2001,texier,texier5}, 
diffusion and localisation
\cite{below:1989,comtet:2002,desbois:2000,desbois:2002,texier4,dittrich:1999},
random walks \cite{severini:2004,kostrykin},
approximations of quantum systems by quantum graphs \cite{exner17},
some inverse problems  \cite{below:1985,below:2001,mehmeti:1986,cvetcovic:1980,roth:trace,nicaise:1987,gutkin:2001,kostrykin3,kostrykin2,kurasov,kurasov2,carlson,pivorarchik,harmer,harmer2,gerisimenko2}, extremal spectral
properties \cite{friedlander2},
and models of dissipative graphs
\cite{smilansky:2004,solomyak:2004,evans:2005,evans:2005a}.

Recent review papers on various
mathematical approaches to quantum graphs
can be found in  
\cite{kuchment,kuchment:2004,kuchment:2005}.
Special issues of research journals 
\cite{special_issue_wrm,special_issue_jpa} and  
recent conference proceedings \cite{proceedings} have been
devoted to quantum graphs. 

The relevance of quantum graphs to the study of quantum
chaos was brought to light by the work of
Kottos and Smilansky \cite{kottos:1997,kottos:1998}.
They analysed the spectral statistics for simple graphs,
and showed that their spectral statistics follow very closely
the predictions of random-matrix theory.
They proposed an alternative derivation of the
trace formula and pointed out its similarity
to the famous Gutzwiller trace formula \cite{gutzwiller:trace,Book:Gutzwiller} 
for chaotic Hamiltonian systems.
While
quantum graphs do not have a deterministic classical limit they still
share a lot of important properties with classically chaotic Hamiltonian
systems, e.g.~ periodic-orbit theory in the semiclassical regime
is completely analogous to periodic-orbit theory on a quantum graph.
While semiclassical approaches are approximations the analogous 
approaches for quantum graphs are exact. More importantly,
quantum graphs are not as resistant to analytical approaches.
Following these pioneering beginnings of Kottos and Smilansky 
quantum graphs have become a new paradigm of quantum chaos
and have been
applied to various problems, including also disorder and diffusion
\cite{winn:diss,comtet:2005}.
All aspects of quantum chaos have been covered: spectral statistics
in finite \cite{barra:2000,berkolaiko:1999,berkolaiko:2003a,berkolaiko:2004,berkolaiko:diss,berkolaiko:2001,berkolaiko:2002,bolte,bolte:2003,kottos:2001,pakonski:2001,pakonski,schanz:comb,schanz:comb2,severini:2004,tanner:2000a,tanner:2001,tanner:2002,gnutzmann:2004a,gnutzmann:2004b,gnutzmann:2004c,gnutzmann:2005} and infinite periodic structures \cite{dittrich:1999},
localisation and wavefunction statistics 
\cite{bercioux:2004,kaplan:2001,schanz:2000,gnutzmann:2004,stargraphs,schanz:2003a,berkolaiko:2003,berkolaiko,winn:diss}, chaotic scattering \cite{kottos:2000,kottos:2003}, transport through
chaotic devices \cite{schanz:2003,cohen:2005},
resonances 
and decay in open chaotic systems \cite{puhlmann:2005,kottos:2004,kottos:2005}.

The main purpose of this paper is to review
recent approaches to spectral statistics in quantum
graphs and discuss the relation to quantum chaos in general.
To make the paper self-contained special effort was made
to introduce the necessary concepts in a clear and consistent way.

Recently important progress
has been achieved by M{\"u}ller \textit{et al} 
\cite{muller:2004,muller:2005} in 
the periodic-orbit theory for spectral statistics
in chaotic Hamiltonian systems 
building on the pioneering work of Sieber and Richter
\cite{richter,sieber,sieber:2002}.
A similar approach for quantum graphs has been developed
by Berkolaiko, Schanz and Whitney \cite{berkolaiko:2002,berkolaiko:2003a,berkolaiko:2004}.
None of these works captured the low energy sector of spectral
statistics.
An alternative approach which so far was successfully
applied to quantum graphs \cite{gnutzmann:2004c,gnutzmann:2005}
was able to close the above mentioned gap,
and it will be discussed in detail in the second
part of this review.

Several other applications to quantum chaos will not be 
addressed in this review. Functional integration approaches
to statistical properties of wave functions and localisation
are thoroughly reviewed by Comtet, Desbois and Texier
\cite{comtet:2005} (see also \cite{akkermans:2000}). 
Chaotic scattering and transport in quantum graphs 
is discussed in \cite{puhlmann:2005,kottos:2005,kottos:2004,schanz:2003,cohen:2005}.

This review is arranged in the following way.
The first chapters provide a detailed introduction
to quantum graphs and their spectral theory. Though the
choice of material is biased by the application to quantum chaos the
first chapters are rather general.
In \ref{sec:definitions}
we define graphs and their topological description.
In Chapter \ref{chapter:Schrodinger}
we quantise graphs in a straight forward way.
The reformulation in Chapter \ref{chapter:QUE} in terms
of a unitary quantum evolution map will
allow us to generalise the quantisation procedure.
This point of view will be the foundation for the remaining
chapters.
Chapter \ref{chapter:classical} gives account of the classical dynamics that
corresponds to a quantum graph and Chapter \ref{chapter:spectral_theory} is devoted
to spectral theory with the introduction of the trace formula and its
discussion.

The remaining chapters are devoted to spectral statistics in quantum
graphs. After a general introduction of spectral correlation
functions (and other statistical measures of the spectrum) in
Chapter \ref{chapter:spectral_statistics} we will give some background from
quantum chaos and random-matrix theory on universality in the
spectral fluctuations of complex quantum systems in Chapter \ref{chapter:universality}.
In the last two chapters we present two analytical approaches to universal
spectral statistics in large quantum graphs.
The periodic-orbit approach is presented
in Chapter \ref{chapter:periodic_orbit}
which also includes sections on quantum graphs with spin
and  quantum graphs which are coupled to
superconductors (Andreev graphs).
Finally, we summarise
the supersymmetry approach with a proof
of universality in large quantum graphs
in Chapter \ref{chapter:susy}.
In Appendix \ref{app:symmetry_classes} we add some background
on the symmetry classification of quantum systems and in
Appendix \ref{app:random_matrix_theory} we summarise some
relevant results from random-matrix theory.

\section{Graphs and their topology}
\label{sec:definitions}

\begin{figure}[ht]
  \begin{center}
    \includegraphics[width=0.6\textwidth]{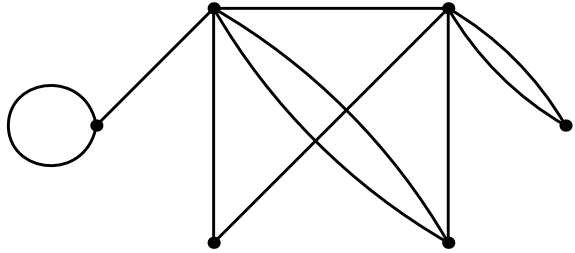}
  \end{center}
  \caption{A graph with $V=6$ vertices and $B=10$ bonds.}
  \label{fig:v6graph}
\end{figure}

A graph  $\mathcal{G}(V,B)$ consists of $V$
\define{vertices} connected by $B$ \define{bonds}
(or \define{edges}). A graph
with six vertices and ten bonds ($V=6,\ B=10$)
is shown in figure \ref{fig:v6graph}.

The graphs are not necessarily embedded in the plane, and the
fact that in the figure bonds cross each other at points which are
not vertices is completely immaterial. A physical realisation of a
graph is a network of coaxial cables (bonds) connected by
junctions (vertices). The topology of the graph, that is, the way
the vertices and bonds are connected is given in terms of the
$V \times V$
\define{connectivity matrix} $C_{i,j}$ (sometimes referred to as the
\define{adjacency matrix}) which is defined as
\begin{equation}
  C_{i,j}=C_{j,i}=
  \begin{cases}
    m & \text{if $i\neq j$ where $i$ and $j$ are connected by $ m$ bonds,} \\
    2m & \text{if $i=j$ and there are $m$ loops at vertex $i$}\\
    0 & \text{if $i$ and $j$ are  not connected.}
  \end{cases}
  \label{cmat}
\end{equation}
This definition allows for vertices to be connected by several
bonds, and also for a vertex to be connected to itself by one or
several loops (in which case the corresponding diagonal element of
the connectivity matrix equals twice the number of loops).

The \define{valency} $v_{{i}}$ of a vertex $i$ is the number of
vertices $j$ connected to $i$, each weighted by the number 
of parallel bonds (loops). Thus, $m$
parallel bonds ($m$ parallel loops) contribute $m$ ($2m$) to the
valency. In terms of the connectivity matrix,
\begin{equation}
  v_{{i}} = \sum_{j=1}^V C_{i,j}\ .
  \label{eq:valency}
\end{equation}

The \define{neighbourhood} $\Gamma_i$ of the vertex $i$ consists of
the vertices $j$ connected to $i$. 

The \define{boundary} of a subgraph $\hat{\mathcal{G}} \subset
\mathcal{G}$, $\Gamma(\hat{\mathcal{G}})$, consists of the vertices
which are not in $\hat{\mathcal{G}}$ but which are in the union of the
neighbourhoods of the vertices of $\hat{\mathcal{G}}$.

The number of bonds is expressed by
\begin{equation}
  B = \frac{1}{2}\sum_{i=1}^V \sum_{j= i }^V  C_{i,j}\ .
  \label{eq:Bonds}
\end{equation}

Unless otherwise specified, we shall always consider
\define{connected} graphs, for which the vertices cannot be divided
into two non-empty subsets such that there is no bond connecting
the two subsets. That is, for a connected
graph the connectivity matrix cannot be brought
into a block-diagonal form by permuting the vertices.

\begin{figure}[ht]
  \begin{center}
    \includegraphics[scale=0.5]{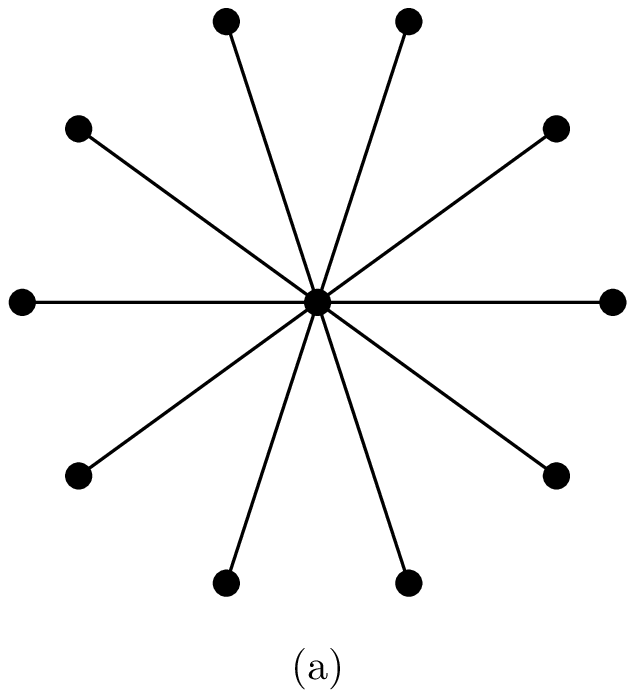}\;\;\;
    \includegraphics[scale=0.5]{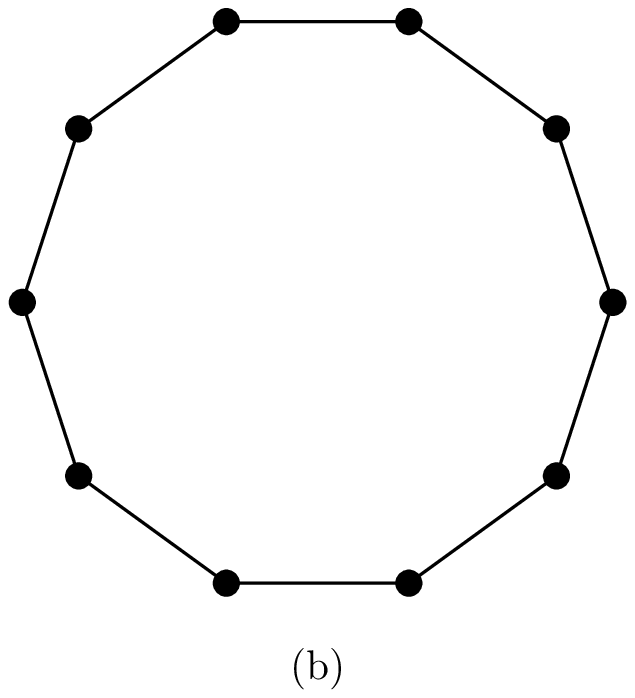}\;\;\;
    \includegraphics[scale=0.5]{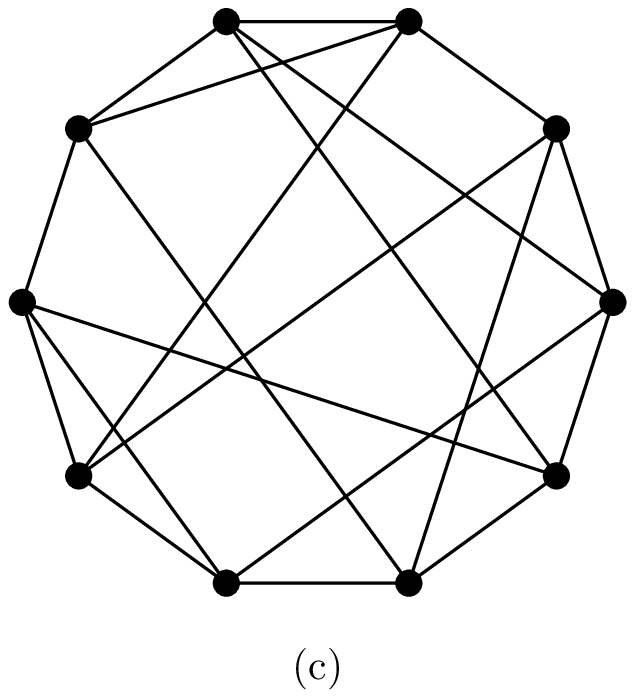}\\
    \includegraphics[scale=0.5]{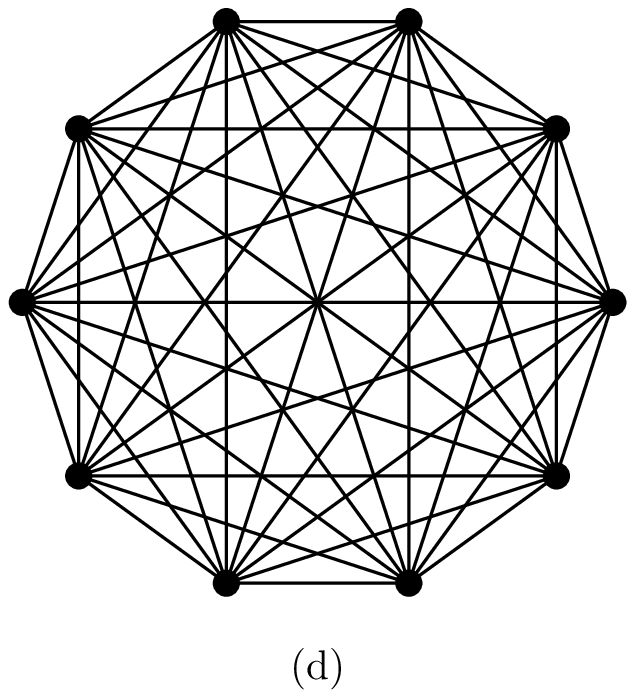}\;\;
    \includegraphics[scale=0.5]{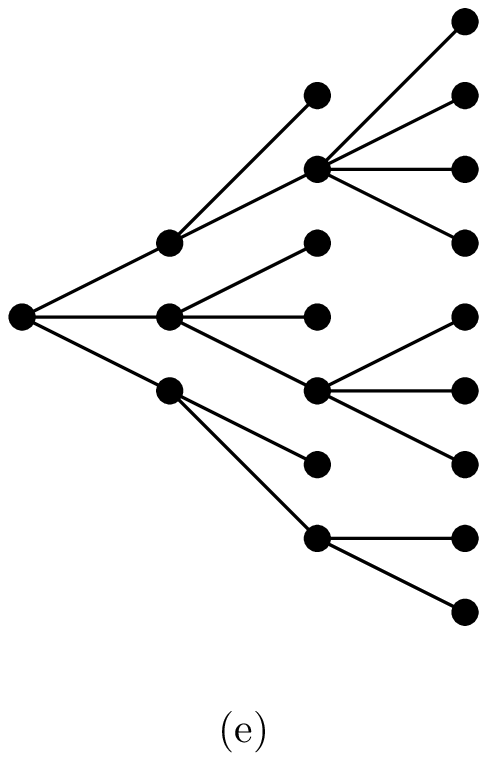}
  \end{center}
  \caption{Some examples of graphs: (a) star graph ($B=10$, $V=11$), (b)
    ring graph ($B=10$, $V=10$), (c) $v$-regular graph with $v=4$ ($B=20$,
    $V=10$),
    (d) complete graph ($B=45$, $V=10$), (e) tree graph ($B=19$, $V=20$).}
  \label{fig:examples}
\end{figure}

There are a few classes of graphs which often appear in the
literature. They are characterised by their connectivity (see
figure \ref{fig:examples} for some examples):

\begin{itemize}
\item \define{Simple graphs} are graphs which have no loops and
  no parallel bonds connecting their vertices (no multiply connected
  vertices). In this case, for all
  $i$ and $j$, $C_{i,j} \in \{0,1\}$, and in particular all the
  diagonal elements vanish $C_{i,i}=0$. 
  For simple graphs, the cardinality of $\Gamma_i$ is the valency $v_i$
  for each vertex.
  When we define quantum graphs
  in chapter \ref{chapter:Schrodinger} we will show that every
  connected quantum graph can be turned into a graph of simple
  topology by adding some vertices
  without changing the spectrum or the wave functions.
  This will allow us to significantly simplify the notation in
  the remaining chapters where a simple topology will be assumed
  without loss of generality. 
  \item \define{$v$-regular graphs} are simple graphs whose
  vertices have the same valency $v$. The simplest $v$ regular
  graphs are the \define{rings} for which $v=2$ and $V=B$. A non-trivial
  ring has at least two vertices.
  A \define{$v$-regular}
  graph is \define{complete} when $v=V-1$.
\item \define{Simply connected graphs} do not contain any nontrivial ring
  as a subgraph.
\item \define{Tree graphs} are simple, connected and simply connected.
\item \define{Star graphs} are trees that consist of a
  main (central) vertex with
  valency $v$, connected to $v$ peripheral vertices of valency one.
\end{itemize}

In many applications it is convenient to refer to bonds directly,
and  we shall use lowercase letters to denote the bonds of the
graph. If the graph is simple, we can use the end points of the
bond as its label : $b =\bond{i}{j} =\bond{j}{i}$.
If a graph is not simple $\bond{i}{j}$ will denote the set of all
bonds that connect the vertices $i$ and $j$.
The round brackets will always
be used to denote such a set of undirected bonds unless otherwise stated.

All the bonds
which emerge from a vertex $i$ form a \define{star},
\begin{equation}
S^{(i)}=\bigcup_{j\in \Gamma(i)}\bond{i}{j}.
\end{equation}
If the
graph is simple, the bonds in a star $S^{(i)}$ are
$\{ b = \bond{i}{j}:
j\in \Gamma(i)\}$.

\define{Directed bonds} (also referred to as \define{arcs} in the literature)
are bonds on which a direction is specified.
For a simple graph we
denote them by the ordered pair of vertex indices enclosed in
square brackets $\dbond{j}{i}$, and the direction is from the right to
the left index.
Again, for non-simple graphs $\dbond{j}{i}$ is the set of all
directed bonds starting at $i$ and ending at $j$.
Lowercase Greek letters will be used to
distinguish them from the undirected bonds.
The reverse direction
will be denoted by a hat so that e.g., if $\alpha\in\dbond{j}{i}$ then
$\hat\alpha \in \dbond{i}{j}$.

Alternatively, we will denote a directed bond $\alpha\in \dbond{j}{i}$
as a pair $\alpha=(b,\omega)$ of a bond $b \in \bond{j}{i}$ and
a direction index $\omega=\pm 1$ where $\omega=+1$ if $j>i$ and
$\omega=-1$ if $j<i$. If the bond is
a loop ($i=j$)
the direction $\omega$ has to be assigned to avoid ambiguities. We will
use the notation $\omega_\alpha$  ($b_\alpha$) to refer to the 
direction index (bond)
of the directed bond $\alpha$.
If $\alpha=(b,\omega)$ is a directed bond the reverse direction
is $\hat\alpha=(b,-\omega)$.
All directed bonds that start at a vertex $i$ form an \define{outgoing
star},
\begin{equation}
  S^{(i)}_+=\bigcup_{j\in \Gamma(i)}\dbond{j}{i}.
\end{equation}
The \define{incoming star} $S^{(i)}_-$ is defined analogously as
the set of directed bonds that end at $i$.

The set of directed bonds  $\dbond{k}{l}$ \define{ follows} the
set of directed bonds $\dbond{r}{s}$ if $r=l$. With $\alpha \in
\dbond{k}{r}$ and $\beta \in \dbond{r}{s}$ we then write $\alpha
\in \mathcal{F}_{r}(\beta)$ which means that the directed bond
$\alpha$ follows $\beta$ at the vertex $r$. A \define{trajectory}
$ t=(\alpha_1,\dots,\alpha_{n})$ from vertex $i$ to $j$ is a
sequence of directed bonds $\alpha_l=(b_l,\omega_l)$ such that
$\alpha_{l+1}$ follows $\alpha_l$ where $\alpha_1$ starts at
vertex $i$ and $\alpha_{n}$ ends at vertex $j$. The
\define{topological length} of the trajectory is the number
$n\in \NN$ of directed bonds in the path. A
\define{closed trajectory} starts and ends at the same vertex $i=j$ and
a \define{periodic orbit} $p\equiv
\overline{\alpha_1,\dots,\alpha_{n}}$ of \define{period} $n$ is
the equivalence class of closed trajectories that are equal up to
cyclic permutation. The \define{code} of the periodic orbit is the
equivalence class $\overline{\alpha_1,\dots,\alpha_{n}}$ of
visited directed bonds. A \define{primitive periodic orbit} has a
code which cannot be written as a repetition of a shorter code.
Each trajectory $t$ defines a subgraph $\mathcal{G}_{t}$ which
consists of all bonds and vertices visited by the path.
The number of different points on periodic
periodic orbits with period $n$ is exactly equal 
$\mathrm{tr}\, C^n$. As a consequence, if $n$ is prime
the number of periodic
orbits $\#(n)$ of period $n$ is exactly
\begin{equation}
  \#(n) =
  \frac{1}{n}\mathrm{tr}\, C^n= \frac{1}{n}\sum_{j=1}^V \gamma_j^n
  \label{eq:no_po}
\end{equation}
where $\gamma_j$ are the eigenvalues of the connectivity matrix.
If $n$ is not prime \eqref{eq:no_po} is a good approximation.
In the limit of large $n$, $\#(n)$ is dominated by the maximum eigenvalue
which shows that this number grows exponentially with $n$.
In analogy to the theory of dynamical systems we define
the \define{topological entropy} $\lambda_T$ as the logarithm
of the largest eigenvalue. One can easily show that for fully
connected graphs $\lambda_T=\log V$ and for star graphs
$\lambda_T=\frac{1}{2}\log B$.
\newpage
\thispagestyle{empty}

\chapter{The Schr\"odinger operator on graphs}
\label{chapter:Schrodinger}

In chapter \ref{chapter:Introduction}
we defined and discussed the graphs from a topological point of
view, where concepts like connectivity and neighbourhood played the
main r\^oles. At this point we would like to endow the graphs with
a \define{metric} which will enable us to define the Schr\"odinger
operator on the graph.

We assign the natural metric to the bonds. The position $x$ of a
point on the graph is determined by specifying on which bond $b$
it is, and its distance $x_b$ from the vertex with the
\emph{smaller} index such that $x_b$ increases in direction $\omega=+1$
and decreases in direction $\omega=-1$. If the bond $b$ is a loop this
defines the starting point for $x_b$.
The length of a bond is denoted by $L_b$ and, $0\le x_b\le L_b$.
The length of a path $t=(\alpha_1,\dots,\alpha_{n})$
is the sum over all
bond lengths $L_{t}=\sum_{l=1}^{n} L_{b_l}$ (where $\alpha_l=(b_l,\omega_l)$)
along the path. For a bond $b=(i,j)$ we will also use the notation
$x_b^i$ and $x_b^j$ for the values of $x_b$ at the vertices. That is, if $i<j$
one has $x_b^i=0$ and $x_b^j=L_b$. For later use, we
define \define{incommensurable} (or \define{rationally independent})
bond lengths on the graph. For these the equation
\begin{equation}
  \sum_b m_b L_b=0
\end{equation}
where $m_b\in \ZZ$ only has the trivial solution $m_b=0$ for
all $b$.

The Schr\"odinger operator on ${\mathcal{G}}$ consists of the one
dimensional operators associated with each bond:
\begin{equation}
  H_b = \left[\left( \frac{1}{\I}\frac{\D\ }{\D x_b}
      +A_b \right)^2 + w_b(x_b)\right]
 \label{eq:bondschroedinger}
\end{equation}
Here, $w_b(x_b)$ is a potential function assumed to be non
negative  and smooth on the interval $[0,L_b]$.  $A_b$ are real,
positive constants.
If the graph contains a nontrivial ring as a subgraph
(that is, it is not simply connected),
and $A_b$ do not vanish on this subgraph, time reversal
invariance is broken, as can be seen from the eigenfunctions of
(\ref{eq:bondschroedinger}) with $w_b=0$ on the line,
\begin{equation}
  \left(\frac{1}{\I}\frac{\ \D \ }{\D x } +A \right)^2 \psi (x) =
k^2\psi \qquad  \Rightarrow \qquad \psi(x) = \E^{-\I A
x}(c_1\E^{\I k x}+c_2\E^{-\I k x})\ .
 \label{eq:free}
\end{equation}
($c_1$ and $c_2$ are arbitrary constants).  The complex conjugate
of $\psi(x)$ above is not a solution of the same equation - a
hallmark of a system which violates time reversal invariance.  We
shall refer to the constants $A_b$ as \define{magnetic fluxes}
because they play the same r\^ole as the vector potentials in the
Schr\"odinger operator in higher dimensions.  The presence of $A$
implies also that the wave functions are intrinsically not
symmetric under the reflection $x\rightleftharpoons-x$.

In the physics literature, and in particular, in the quantum
chaos connection the bond potentials $w_b(x_b)$ are commonly set
to zero. We included them in the general framework for
the sake of generality, and also because they appear in some of
the mathematical literature on the topic. Graphs with non vanishing
potentials are sometimes referred to as \define{dressed graphs}
\cite{dabaghian:2003,blumel:2002,blumel:2002a,schmidt}.

Next, we have to identify the space of wave functions and the
boundary conditions which renders the operator self-adjoint. In
physical terms, this implies that the evolution induced by the
operator conserves probability and the vertices cannot be either
sinks or sources. In other words, the boundary conditions at the
vertices should be such that the total probability current
vanishes when summed over all the bonds which emerge from any of
the vertices. This is similar to the well known Kirchoff rule in
the theory of electric networks, and we shall now derive it for
quantum graphs.

\section {Vertex boundary conditions -- self-adjoint extension}
\label{sec:boundary}

We consider the set of functions ${\mathcal{D}} $ which have the
following properties: $\Psi(x)\in {\mathcal{D}} $  are continuous
and complex valued functions of $x\in {\mathcal{G}}$, with
$\Psi(x)= \psi_b(x_b)$ for $x \in b$, and $0 \le x_b \le L_b$.
Continuity is implied here also at the vertices. This means that
at each vertex $i$ the limit $\lim_{x_{b}\rightarrow x_b^i}
\psi_{b}(x_{b})=\phi_i$ does not depend on $b\in S^{(i)}$. The
functions $\psi_b(x_b)$ are complex valued, bounded with piecewise
continuous and square integrable first derivatives. The set
${\mathcal{D}} $ is the domain of the \emph{positive definite}
quadratic form
\begin{equation}
  \label{eq:quadform}
  \begin{split}
    Q_{\Lambda}[\Psi] =&
    \INT{\mathcal{G}}{}{x} \left(
      \left|
        \left(\frac {1}{{\I}}\nabla+A
        \right)
        \Psi(x)
      \right|^2 +
      \Psi(x)\cdot W \Psi(x)\right)+\\
    &+
    \vec{\phi}
      \cdot \Lambda \vec{\phi}
    \\
    \equiv&
    \sum_{b=1}^B   \INT{0}{L_b}{x_b}
    \left(
      \left|
        \frac {1}{{\I}}
        \frac{\D \psi_b}{\D x_b}+ A_b\psi_b(x_b)
      \right|^2
      +w_b(x_b)|\psi_b(x_b) |^2
    \right)+\\
    &+
    \sum_{j=1}^V\lambda_j
    \left|\phi_j\right|^2 \ .
  \end{split}
\end{equation}
Here,  $\vec \phi$ denotes the $V$ dimensional array of
$\phi_j$. $\Lambda$ is a positive diagonal matrix with elements
$\lambda_j \ge 0$. The r\^ole of these parameters will become
clear in the sequel and their physical significance will be
discussed below.

The self-adjoint extension of the Schr\"odinger operator, $H$, is
determined by the Rayleigh-Ritz extremum principle. For this
purpose we compute the variation of the quadratic form
$Q_{\Lambda}[\Psi]$ with respect to $\Psi$ under the condition
that $\|\Psi\|^2 \equiv\int_{\mathcal{G}}{\rm d}x|\Psi(x)|^2 = 1$.
This constrain is introduced by considering the variation of the
modified quadratic form $\tilde{Q}[\Psi]=
Q_{\Lambda}[\Psi]-k^2\|\Psi\|$, where $k^2$ is a positive Lagrange
multiplier. The variations of the modified quadratic form with
respect to both $\Psi$ and its complex conjugate have to vanish
identically. Writing the variation with respect to $\Psi^{*}$
explicitly, and performing partial integration where necessary, we
get
\begin{equation}
  \label{eq:varquadform}
  \begin{split}
    \delta \tilde{Q}_{\Lambda}[\Psi]  =&
    \sum_{b=1}^B   \INT{0}{L_b}{x_b}
    \delta\psi_b^{*}(x_b)
    \left(
      \left(
        \frac {1}{{\I}}\frac{\D\ }{\D x_b}+A_b
      \right)^2
      +w_b(x_b) -k^2
    \right)
    \psi_b(x_b)
    \\
    &+
    \left[
    \delta\psi_b^{*}(x_b)
      \left(
        \frac{\D \ }{\D x_b  }+ {\I} A_b
      \right)
      \psi_b(x_b)
    \right]_0^{L_b} +   \sum_{j=1}^V\delta
    \phi^{*}_j\lambda_j \phi_j
    \ .
  \end{split}
\end{equation}
A similar expression can be obtained for the variation of
$\tilde{Q}$ with respect to $\Psi(x)$. Requiring that both
variations vanish for every $\delta \Psi(x)$ and
$\delta\Psi^{*}(x)$, we find that the domain ${\mathcal{D}}_H$ of
the Schr\"odinger operator consists of functions in
${\mathcal{D}}$, with twice differentiable $\psi_b(x_b)$, which
satisfy the boundary conditions
\begin{equation}
  \forall \ 1 \le   i \le  V  :\ \sum_{ (b,\, \omega)\, \in\,  S^{(i)}_+}
  \left.
    \omega
    \left(
      \frac{\D \  }{\D x_b} + {\I}
      A_b
    \right)
    \psi_b \
  \right|_i =  \lambda_i\phi_i \ .
  \label{eq:boundary}
\end{equation}
The derivatives in \eqref{eq:boundary} above are computed at the
common vertex $i$. That is at $x_b=x_b^i=0$ if $\omega=+1$ and $x_b=x_b^i=L_b$ if $\omega=-1$ (if $b$ is a loop, both directions appear in the sum).
These conditions are obtained by reorganising the
second line in \eqref{eq:varquadform} according to the vertices
and their associated (outgoing) stars.

The eigenfunction are solutions of the bond Schr\"odinger
equations
\begin{equation}
  \forall \ 1 \le  b  \le  B  :\ H_b\psi_b = k^2 \psi_b\ ,
\end{equation}
which satisfy the boundary conditions \eqref{eq:boundary}. The
spectrum $\{ k_n^2\}_{n=1}^{\infty}$ is discrete, non-negative and
unbound. It consists of the values of the Lagrange multiplier
$k^2$ for which a non-trivial solution is found. The sequence of
eigenvalues is conveniently arranged by increasing order, so that
$k_n\le k_m$ if $n < m$.

The eigenfunctions (ordered by increasing eigenvalues)
have the following  property. Let
${\mathcal{D}}_n$ denote the subspace of functions in
${\mathcal{D}}$ which are orthogonal to the first $n-1$
eigenfunctions of $H$. Then, for any non-zero $\Phi\in
{\mathcal{D}}_n$
\begin{equation}
  Q [\Phi] \ge k_n^2  \|\Phi\|^2\ .
  \label{eq:RR}
\end{equation}
Equality holds if and only if $\Phi$ is the $n$'th eigenfunction
of $H$.

At this point it will be good to familiarise
oneself with 
the general formalism
above by considering a few examples.

The boundary conditions \eqref{eq:boundary} take a very simple
form for vertices with valency $v=2$, when the magnetic fluxes
on the two bonds are the same and the potentials on the two bonds
take the same value at the vertex. We can now think of the two
bonds as two adjacent intervals on the line with a common boundary
which we choose as the point $x=0$. Writing \eqref{eq:boundary}
explicitly we get
\begin{equation}
  \lim_{\epsilon \rightarrow 0^+}
  \left\{
    \left.\frac{\D \psi}{\D x}
    \right|_{0+\epsilon}
    -
    \left. \frac{\D \psi}{\D x}
    \right|_{0-\epsilon}
  \right\}
  = \lambda \psi(0)\ .
  \label{eq:kroning}
\end{equation}
This is the well known boundary condition one obtains from a
``$\delta$-potential'' of strength $\lambda$. In physics
textbooks this boundary condition is derived by integrating the
Schr\"odinger equation along an interval of size $2\epsilon$
centred at $x=0$, and the self-adjoint extension of the operator
is obtained automatically. This is not the case when we are
dealing with a more complex vertex, and the road we chose to
derive the boundary conditions must be taken.

Even though the parameters $\lambda_i$ can take arbitrary values,
the limiting values $\lambda_i=0$ (\define{Neumann} boundary
conditions) or $1/ \lambda_i=0$ (\define{Dirichlet} boundary
conditions) are of special interest.

The importance of Neumann boundary conditions (also known as \define{Kirchhoff boundary conditions} in the mathematical literature) comes from the fact
(to be proven shortly) that the spectra of systems with finite
(but not vanishing) $\lambda_i$ approach to the Neumann spectrum
as one looks higher up in the spectrum. A similar situation is
well known for the spectra of Schr\"odinger operators on domains
with boundaries, where the boundary conditions intermediate
between Dirichlet and Neumann are studied \cite{mixedbc}. Note that the case
studied in the example above of a vertex with $v=2$ is trivial
under the Neumann condition. Indeed, \eqref{eq:kroning} implies
that the wave function and its first derivative are continuous so
that the point $x=0$ becomes an ordinary point on the interval.
Thus, a Neumann vertex with $v=2$ can simply be erased from the
graph, without any effect on the spectrum or wave functions. This
property can be used in the reverse direction as well. In
particular, every non simple quantum graph can be turned into an
equivalent (with respect to spectra and wave functions) quantum
graph  of simple topology (no loops, no multiple connections) by
adding two Neumann vertices on each loop and one Neumann vertex on
each bond which is responsible for multiple connectivity. Because
of this reason, and unless stated otherwise, we shall henceforth
\emph{consider only graphs with simple topology}:
\begin{equation}
\label{eq:simple}
 C_{i,j} \in \{0,1\}\qquad \text{and}\qquad C_{i,i}=0\ .
\end{equation}
This significantly simplifies the notation while
the transition back to
the original connectivity is straight forward.

The Dirichlet boundary conditions imply
that the value of the wave function vanishes at all the vertices.
This isolates the various bonds and the spectrum of the
Schr\"odinger of the graph reduces to the union of the bond
spectra, with Dirichlet boundary conditions on each of the ends of
the bonds. The wave function corresponding to a nondegenerate eigenvalue
$k_n^2$ is
identically zero on all bonds but one and the spectrum is given
by the union of the independent spectra on the bonds
\begin{equation}
  \sigma_{\mathrm{Dirichlet}}= \bigcup_{b=1}^B \bigcup_{n=1}^\infty
    \left\{
    \left(\frac{\pi n}{L_b}\right)^2 
 \right\} \ .
\end{equation}

\section {The secular equation}
\label{sec:secular}

In the previous section the Schr\"odinger operator and the
accompanying boundary conditions were defined and discussed. Here,
we shall derive the secular function - the real valued function
whose zeros (the values of the argument where the function
vanishes) stand in one to one correspondence with the spectrum of
the graph. The spectrum of Dirichlet
graphs was computed previously and we shall exclude this case 
in the following.

We shall first assume that the bond potentials vanish, $w_b(x_b)=0$.
We will comment on non-vanishing bond potentials at the end
of this section.

The starting point is the potential-free solution
(\ref{eq:free}). On a bond $b = \bond{i}{j}$ with  $i\le j$, the
general solution of the bond Schr\"odinger equation is a
superposition of the two solutions \eqref{eq:free}, which can be
written as
\begin{equation}
  \psi_b(x_b;k) = \frac{\E^{- \I A_b x_b}}{\sin k
    L_b}
  \left(
    \phi_j \E^{\I A_b L_b} \sin kx_b  +
    \phi_i\sin k(L_b-x_b)
  \right) \ .
  \label{eq:wavefunct}
\end{equation}
At the two endpoints $x_b=0$ and $x_b=L_b$, the wave function $\psi_b(x_b;k)$ assumes the values
$\phi_i$ and $\phi_j$, respectively. These same values appear in the
wave functions on bonds which connect $i$ or $j$ to other
vertices, and therefore, the resulting graph wave function is
continuous. Thus, the most general form of $\Psi \in
\mathcal{D}_H$ is completely determined (up to a scalar factor) by
$V$ complex parameters $\vec \phi = (\phi_1,\cdots,\phi_V)$. The
appearance in the denominator of $\sin k L_b$ for all $b$ will
eventually lead to poles of the secular function at the Dirichlet
spectrum.

The eigenfunctions and eigenvalues of the Schr\"odinger operator
are found by demanding that the $B$ functions \eqref {eq:wavefunct} 
satisfy the
boundary conditions \eqref{eq:boundary}. Upon substitution, we
obtain a set of $V$ homogeneous and linear equations for the
vertex wave functions $\vec\phi$,
\begin{equation}
  \sum _{j=1}^V \phi_j\, h_{j,i}(k)=0 \qquad \forall\ i=1,
  \dots, V\ ,
  \label{eq:secmat1}
\end{equation}
where,
\begin{equation}
  \label{eq:secmat}
  \begin{split}
    h_{i,i}(k)=&
    \frac{\lambda_i}{k}\ +\sum_{(b,\, \omega)\, \in
    S^{(i)}_+} \cot
    kL_b
    \\
    h_{j,i}(k)=& -
    C_{j,i}
    \frac{\E^{\I\, \omega A_b L_b }}
    {\sin kL_b}\ ,\ \ \ \ (b,\omega)=[j,i] \    .
  \end{split}
\end{equation}
Note that the matrix $h$ is hermitian because the indicators
$\omega$ change their sign when $i$ and $j$ are
interchanged. Equations \eqref{eq:secmat1} have a non-trivial
solution if and only if
\begin{equation}
\zeta_h(k)\equiv \det\ h(k) = 0.
 \label{eq:secularh}
\end{equation}
The function $\zeta_h(k)$ is the \define{secular function}. Because
of the symmetry $\psi_b(x;k)=-\psi_b(x,-k)$, the zeros of the
secular function appear symmetrically on the negative and positive
$k$ half-lines at $\pm k_n$. The Schr\"odinger spectrum is
$k_n^2$. In Chapter \ref{chapter:QUE}, an alternative secular
function will be derived. The suffix $h$ in $\zeta_h(k)$ is added
to distinguish between this and the other secular function.

In the previous section we commented (without proof) that in the
limit of large eigenvalues, the spectra of the Schr\"odinger
operators with $\lambda_b \ne 0$ converge to the Neumann spectrum.
This is immediately apparent from \eqref{eq:secmat}, since the
parameters $\lambda_b$ appear in the combination
$\frac{\lambda_b}{k}$ which vanishes in the limit of large $k$.

Finally, it will be instructive to write explicitly the secular
function for a simple example. We do it for a star graph which was
defined in Section \ref{sec:definitions}. We denote the central
vertex by the index $0$, and label the emanating bonds as well as
their exterior vertices by $i,\ i=1,\cdots,B$. The only non
vanishing matrix elements of $h$ are
\begin{equation}
  \label{eq:starh}
  \begin{split}
    h_{0,0}=& \frac{\lambda_0}{k} +
    \sum_{i=1}^B\cot k L_i
    \\
    h_{j,j}=&\frac{\lambda_j}{k} + \cot kL_j
    \\
    h_{0,j}=&-\frac{\E^{-\I A_j L_j}}{\sin k L_j}
  \end{split}
\end{equation}
The secular function can be computed directly,
\begin{equation}
  \zeta_h(k) = \left(  h_{0,0} - \sum_{i=1}^B  
  \frac{ |h_{0,i}|^2 }{ h_{i,i} }\right) 
  \left(\prod_{j=1}^B h_{j,j}\right) \ .
\label{eq:starzeta}
\end{equation}
It is worth while noticing that the magnetic fluxes $A_j$ do not
appear in the secular function. The reason for this is that the star graph
is simply connected. Moreover, $\zeta_h$ can only vanish if the first factor 
on the right hand side of equation
\eqref{eq:starzeta} vanishes.
The secular equation reduces to
\begin{equation}
 \zeta_{h_\bigstar}=\frac{\lambda_0}{k}+\sum_{i=1}^B \cot k L_i 
-\sum_{i=1}^B \frac{1}{\sin^2 kL_i  \left(\frac{\lambda_i}{k} + \cot k
L_i    \right)}
 =0 \ .
 \label{eq:zetastar}
 \end{equation}

To end this section we shall indicate how the computation of the
secular equation can be generalised to include non-vanishing
potentials $w_b(x_b)$ on the bonds. In the general case one cannot
write an explicit solution of the bond wave equation as in
\eqref{eq:wavefunct}. However, since it is of the Sturm-Liouville type,
it is possible to find two independent solutions, which satisfy
independent \emph{initial conditions}. One function vanishes at the
vertex $i$ and its derivative is set to $1$, and the other
vanishes at the vertex $j$ and its derivative is $1$.  The most
general solution which takes the values $\phi_i$ and $\phi_j$ at
both ends of the bond can be written as a linear combination of
the two solutions, and from here on one can proceed as in the
potential free case. Use should be made of the Wronskian relation,
which in the presence of the magnetic flux takes the form
\begin{equation}
W(f,g)\equiv f\left(\frac{{\rm d}g^*}{{\rm d}x_b}-{\rm
i}A_bg^*\right)- g^*\left(\frac{{\rm d}f}{{\rm d}x_b}+{\rm i}A_b
f\right) =Const.
 \label{eq:wronsk}
\end{equation}
\newpage
\thispagestyle{empty}

\chapter{The Quantum Evolution Map}
\label{chapter:QUE}

It is quite common in theoretical physics
that a given subject can be formulated and studied from several
points of view which use different concepts and tools. Consider
for instance, the Lagrangian and Hamiltonian approaches to classical
mechanics. The phase-space description reveals the symplectic
structure of classical dynamics in a natural way, which is only
implicit in the configuration-space description. Following this
example, we shall present in this chapter another formulation of
the spectral theory of graphs which uses other concepts and
structures then those used in the previous chapter.  The main
observation and intuition which underlays the new formulation is
that a wave function on the graph can be written as a
superposition of waves travelling in opposite directions on the
bonds. The waves which propagate towards a given vertex are
scattered from the vertex and emerge as outgoing waves, which
scatter again and again. A wave function is an eigenfunction if it
is \emph{stationary} under the multiple scattering scenario
described above, and this, in turn can happen only if the wave
number $k$ of the propagating wave is correctly selected.
Thus, a new form of the secular equation can be obtained, based on
this multiple scattering approach \cite{kottos:1998}. 
The advantages gained by
developing the alternative theory are both conceptual and
technical, and they will be revealed as the theory is
systematically unfolded in the following sections. These
advantages are gained at a price:  the number of parameters
necessary to represent a wave function increases from  $V$  to
$2B$, which is the number of amplitudes of waves which propagate
back and forth on the bonds.

{\em Note}: In the present chapter we consider the
``bare'' graphs so that on all the bonds  $w_b=0$.
We also want to remind that we assume a simple topology
\emph{without loss of generality} (see Section \ref{sec:boundary}).

\section{Vertices as scattering centres}
\label{sec:v-scat}

Given a graph $\mathcal{G}$, the bond lengths $L_b$,
magnetic fluxes $A_b$ and the boundary condition parameters
$\lambda_b$, one can write a general solution of the Schr\"odinger
equation \eqref{eq:wavefunct} (for  $i<j$)
\begin{equation}
  \begin{split}
  \label {eq:wavefunct1}
    \psi_b(x_b;k) =& \frac{\E^{- \I A_b x_b}}{\sin k L_b}
    \left(
      \phi_j \E^{\I A_b L_b} \sin kx_b  +
      \phi_i\sin k(L_b-x_b)
    \right) \\
    =& \E^{- \I A_b x_b+\I  k x_b } a_{[j,i]}^{\mathrm{out}}
    +\E^{- \I A_b x_b-\I  k x_b } a_{[i,j]}^{\mathrm{in}}
   \end{split}
\end{equation}
where, $a_{[i,j]}^{\mathrm{in}}\equiv a^{\mathrm{in}}_{\beta}$ and $a_{[j,i]}^{\mathrm{out}}\equiv a^{\mathrm{out}}_{\hat{\beta}}$
are the (complex valued) amplitudes
of the counter propagating waves (incoming and outgoing at vertex $i$)
along the bond $b$,
and they can
be readily written in terms of the parameters $\phi_i$ and
$\phi_j$.
In Section \ref{sec:definitions} we defined the outgoing (incoming)
stars $S_+^{(i)}$ ($S_-^{(i)}$) as the set of directed
bonds which point away (point at) the vertex $i$ which are
convenient for the following discussion.
Consider now a single vertex $i$ and the bonds $b \in S^{(i)}$.
For the present discussion we also use the convention that the
vertex $i$ corresponds to the point $x_b=0$ for all the bonds
in the star $S^{(i)}$.
From \eqref{eq:wavefunct1} we see that $\phi_i$ can be expressed
as
\begin{equation}
  \phi_i = a^{\mathrm{out}}_{b,\omega}+ a^{\mathrm{in}}_{b,-\omega},
  \label{eq:wavefunct2}
\end{equation}
on all the bonds $b\in S^{(i)}$.
These $v_i$ relations provide $v_i-1$
homogeneous equations
\begin{equation}
  a^{\mathrm{out}}_{b,\omega}+a^{\mathrm{in}}_{b,-\omega} 
  = a^{\mathrm{out}}_{b',\omega}+a^{\mathrm{in}}_{b',-\omega}
  \qquad
  \text{for }\qquad b,b' \in S^{(i)} \ .
\end{equation}
Another homogeneous expression is derived by substitute both \eqref{eq:wavefunct1}
 and \eqref{eq:wavefunct2} in the boundary
condition at vertex $i$ \eqref{eq:boundary},
\begin{equation}
  \sum_{\beta \in S_{-}^{(i)} } \left(\ 1-\frac {\lambda _i}{\I v_i
  k}\ \right )\ a_{\beta }^{\mathrm{in}} = 
  \sum_{\alpha \in S_{+}^{(i)}}\ \left
  (1\ +\ \frac{\lambda_i}{\I v_i k}\ \right )\ a_{\alpha }^{\mathrm{out}}
\end{equation}
These $v_i$ relations enable us to write the \emph{outgoing}
amplitudes $a_{\alpha}^{\mathrm{out}},\ \alpha \in S_{+}^{(i)}$ 
in terms of the
\emph{incoming} amplitudes $a_{\beta}^{\mathrm{in}},\ 
\beta \in S_{-}^{(i)}$ in
the form
\begin{equation}
  \forall\ \alpha \in S_{+}^{(i)} \ \ : \ \
  a_{\alpha}^{\mathrm{out}} =\
  \sum_{\beta \in S_{-}^{(i)}}\ \sigma^{(i)}_{\alpha,\beta}\
  a_{\beta}^{\mathrm{in}}\ ,
  \label{eq:inout}
\end{equation}
where,
\begin{equation}
  \sigma^{(i)}_{\alpha,\beta} =
  \frac{1+\E^{i\rho_i (k)}}{v_i}-\delta_{\hat{\alpha}, \beta}
  \label{eq:vertexsig}
\end{equation}
and,
\begin{equation}
  \E^{i\rho_i (k)}=\frac{ 1-\I \frac { \lambda _i\ }{ v_i  k}}{
  1+\I \frac {\lambda _i}{v_i k}} \ .
\end{equation}
This can be written shortly as
$\sigma^{(i)}=\frac{2 k }{v_i k +\I \lambda_i}\EEE-\ONE$
where $\EEE$ is a full matrix with unit entries.

It is not difficult to prove that the $v_i\times v_i$ symmetric
matrix $\sigma^{(i)}$ is \emph{unitary}, which implies that the total
outgoing probability current equals the total incoming probability
current, as stated (without proof) in a previous chapter. The
matrices $\sigma^{(i)}$ will be referred to as the \define{vertex
scattering matrices}. The simple topology allows us to use
the previous and the next
vertex as indices of the vertex scattering matrix
instead of the corresponding directed bonds,
\begin{equation}
  \sigma_{[j,i],[i,j']}^{(i)}\equiv\sigma^{(i)}_{jj'}
\end{equation}
for all $j,j'$ in the neighbourhood of $i$. We shall use
both notations at convenience.

Consider as an example a Neumann vertex with $v =2$ and $S^{+}_i=\{\alpha,\beta\}$. It follows
from \eqref{eq:vertexsig} that
$\sigma_{\alpha,\hat
\alpha}=\sigma_{\beta,\hat \beta}= 0$,
but
$\sigma_{\alpha,\hat
\beta}=\sigma_{\beta,\hat \alpha}= 1$.
In other words, a $v =2$
Neumann vertex transmits waves without any reflection. This is
consistent with a previous remark that such vertices have no
effect on the quantum dynamics on graphs, and can be added or
removed at will.

The vertex scattering matrices are the building blocks of much of
the subsequent theory, and they deserve some further discussion.
The vertex scattering matrices presented above are but one of many
possible examples, and they can be constructed to model various
physical systems. For example, an engineer considers the graph as
a model of a network of wave-guides connected by junctions. In the
ideal situation, there is no dissipation at the junctions, and
therefore the transmission and reflection at the junctions are
characterised by unitary scattering matrices. Their details (such
as their dependence on the frequency), reflect the particular way
by which the junction is designed and constructed.
Also, the quantisation procedure
of Chapter \ref{chapter:Schrodinger} can be generalised
in several ways.
We shall
dedicate the special Section  \ref{sec:vertexscat} to present
and discuss the most commonly used classes of vertex matrices
beyond the present quantisation scheme. We
place this section at the end of the chapter not to interrupt the
flow of the exposition.

\section {The quantum evolution map}
\label{sec:qmap}

The graph is a network of connected vertices, and the waves are
scattered between them along the connecting bonds. When the graph
is assembled from all its vertices and bonds, it is important to
remember that given two connected vertices $i$ and $j$, a directed
bond $\alpha = [i,j]$ is \emph{outgoing} from $j$, $\alpha \in
S_+^{(j)}$ but it is \emph{incoming} to $i$, $\alpha \in
S_-^{(i)}$. That is, when $C_{i,j} =1$
\begin{equation}
  [i,j] =  S_{+}^{(j)} \bigcap  S_{-}^{(i)} \qquad
  \text{and} \qquad [j,i] =
   S_{-}^{(j)} \bigcap  S_{+}^{(i)} \ .
  \label{eq:directed}
\end{equation}
The wave function \eqref{eq:wavefunct1} on the bond $b=(i,j)$
was written down by adopting the convention that $x_b=0$ at the
vertex $i$ - this choice emphasises the vertex $i$ as the
scattering centre. However, we can write the same wave function in
terms of the coordinate $x_{\hat b} = L_b-x_{b}$, so that $x_{\hat
b} = 0$ marks the vertex $j$,
\begin{equation}
  \label {eq:wavefunct3}
  \psi_{\hat b}(x_{\hat b};k) = \E^{+ \I A_b x_{\hat b}+\I k x_{\hat
  b} } a_{[i,j]}^{\mathrm{out}}
  + \E^{+ \I A_b x_{\hat b}-\I k x_{\hat b} }\hat{a}_{[j,i]}^{\mathrm{in}}\ .
\end{equation}
Again, the wave function is expressed in terms of counter
propagating waves with
incoming and outgoing amplitudes with respect to vertex $j$. 
However, the magnetic flux appears now with a
different sign, as befits its r\^ole in the theory as the element
which breaks time reversal invariance. It is therefore only
natural to consider the magnetic flux as a quantity which is
associated with a directed bond, with $A_{\alpha} = -A_{\hat
\alpha}$, in contrast with the bond length which does not depend
on orientation.
Comparing the two expressions \eqref{eq:wavefunct1}
and \eqref{eq:wavefunct3} for the wavefunction on the bond $b$
one can read off that the outgoing amplitude at the
starting vertex of a directed bond $\alpha=(b,\omega)$ and the
incoming amplitude at the next vertex differ by a
phase factor
\begin{equation}
  a^{\mathrm{in}}_{b,\omega}=\E^{\I kL_b +\I \omega A_b L_b}
  a^{\mathrm{out}}_{b,\omega}
\end{equation}
which the wave aquires from one end to the other of the bond.

Equipped with all the above notations and remarks, we can now
demand that the bond wave functions \eqref{eq:wavefunct1} or
\eqref{eq:wavefunct3} satisfy the continuity and the boundary
conditions \eqref{eq:boundary} at all the vertices. This results
in $2B$ homogeneous linear equations for the $2B$ coefficients
$a_{\alpha}$ which can be written
explicitly as
\begin {equation}
\label{eq:consistency}
 \forall\ \alpha \ :\
 a_{\alpha}^{\mathrm{in}} = 
 \sum_{\beta}\evol(k)_{\alpha,\beta}a_{\beta}^{\mathrm{in}}\
 ,
\end{equation}
where the sum over $\beta$ is over all $2B$ directed bonds
and the matrix  $\evol(k)$ is written as
\begin{equation}
  \evol(k)= \bondprop(k) \scattering(k) \ ,
\end{equation}
with $\bondprop (k)$, the \define{bond propagation matrix}, 
which is the diagonal $2B \times 2B$ matrix
\begin{equation}
  \bondprop (k)_{(b,\omega),(b',\omega')}
  =\delta_{bb'}\delta_{\omega \omega'}
  \E^{\I (k+\omega A_b) L_b}\ .
\end{equation}

The $2B \times 2B$-matrix $\scattering (k)$ contains the vertex
scattering coefficients
\begin{equation}
  \scattering_{\alpha',\alpha }(k)=
  \begin{cases}
    \sigma^{(i)}_{\alpha', \alpha }&
    \text{if $\alpha'$ follows $\alpha $
    at vertex $i$, $\alpha'\in \mathcal{F}_i\alpha$, }\\
    0 & \text{else.}
  \end{cases}
\end{equation}
We will refer to $\scattering$ as the \define{graph scattering
matrix}. A non trivial solution of (\ref{eq:consistency}) exists
for the wave number $k>0$ for which
\begin{equation}
\label{eq:secularb}
 \zetab(k)   \equiv \det (\ONE-\evol(k)) = 0 \ .
\end{equation}
$\zetab(k)$ is another secular function, whose zeros define the
spectrum of the graph. In contrast with $\zetah(k)$, it involves a
determinant of a matrix of larger dimension, yet it has many
advantages, the most outstanding is that $\zetab(k)$ has no poles
on the real $k$ axis, in contrast to $\zetah(k)$ which has poles
at the Dirichlet spectrum of the graph. 
The $\zeta_B$-function can be thought of as the characteristic polynomial
$\mathrm{det}\left( \lambda\ONE -\evol(k)\right)=\sum_{j=0}^{2B} a_j \lambda^j$.
A consequence of the unitarity of $\evol(k)$ is
\begin{equation}
  \mathrm{det}\left( \lambda \ONE -\evol(k)\right)=
  \mathrm{det}\left( \frac{1}{\lambda}\ONE -\evol^\dagger(k)\right) 
  \mathrm{det}\left(-\evol(k)\right) \lambda^{2B}
\end{equation}
which implies
\begin{equation}
  a_j=a_{2B-j}^* \mathrm{det}\left(-\evol(k)\right)\ .
\end{equation}
This identity is the analogue of the Riemann-Siegel `look-alike'
symmetry which holds only approximately for spectral
$\zeta$-functions of chaotic quantum systems.
Other properties of
$\zetab(k)$ will be discussed as the theory is developed.

The matrix $\evol(k)$ will be referred to as \define{the quantum
evolution map} because of the following reasons. Being an
involution of two unitary matrices, it is unitary - which is a
basic requirement for a quantum map. The action of the map is a
composition of two successive operations: scattering followed by a
propagation along the bonds: The scattering map operates on \emph{
incoming} amplitudes at all the vertices and produces the
corresponding \emph{outgoing} amplitudes. For finite $\lambda_i>0$
the graph scattering matrix depends on the wavenumber -- for
$\lambda_i=0$ and for some of the generalisations to discuss later
it becomes independent. The matrix $\bondprop(k)$ propagates the
outgoing waves along the bonds and provides the correct phase for
the next scattering event. Starting with an arbitrary distribution
of amplitudes $a_{\alpha}^{\mathrm{in}}$ of waves with a wave number $k$, one
can examine the evolution of the wave pattern as a function of the
number $n$ of scattering events by applying $\evol^n(k)$ to the
initial distribution. The consistency condition
\eqref{eq:consistency} can be interpreted as a requirement that at
an eigenvalue of the graph, the wave function is \emph{stationary}
with respect to the quantum evolution map - a very natural
requirement. More care has to be taken for vanishing wave number
$k=0$ where $\E^{\pm \I k L_b}\rightarrow 1$ is not propagating.
That is, only the strictly positive part of the spectrum $k_n>0$
is equivalent to positive zeros of the secular function while this
one-to-one correspondence may fail at $k=0$.
The simplest
example for such a failure is the line of length $L$ with Dirichlet boundary
conditions, for which $\zetab(k)=1-\E^{\I 2 k L}$. Obviously
$\zetab(0)=0$ in this case, but $k=0$ is  not in the spectrum.
Note, that for Neumann boundary conditions $\zetab(k)$ is not
changed but now $k=0$ is in the spectrum. 
Such ``false zeros''
of the secular function at $k=0$ appear quite often. For instance,
quantum graphs with Neumann boundary conditions have a
non-degenerate ground state at $k=0$ which is just the constant
function on the graph and the secular function vanishes
$\zetab(k=0)=0$. In most topologies this is however a highly
degenerate zero of the secular function (for instance in all cases
where all valencies are larger than two). In spite of
this possible failure of the secular function at $k=0$ the graph
scattering matrix $\scattering$ completely determines the boundary
conditions at all vertices of the graph and the complete evolution
operator $\evol(k)$ determines the complete spectrum if the limit
$k\rightarrow 0$ is considered correctly\footnote{This limit has
to take into account, that for $k=0$ the general solution of the
one-dimensional Schr\"odinger equation (without magnetic field) is
a linear function $a+bx$. The rest is straight forward. In the
following we will always imply a full knowledge of the quantum map
when referring to the condition \eqref{eq:secularb} such that the
oddities at $k=0$ can always be removed by the correct limit. We
will only come back to this point if it has some non-trivial
consequences.}.

So far we have shown that the condition
\eqref{eq:secularb} is
equivalent to the boundary conditions
\eqref{eq:boundary}.
However, with the above interpretation, we can obtain the spectrum
of a wider class of graphs, whose vertex scattering matrices are
unitary matrices derived or postulated in other ways, as discussed
in the next section.

\section{Examples of generalised vertex scattering matrices}
\label {sec:vertexscat}

In the quantisation procedure of Chapter \ref{chapter:Schrodinger}
we have considered only boundary
conditions for which the wave function is continuous through
each vertex and we have given the most general account of them.
The continuity condition may however be relaxed, and more general
boundary conditions corresponding to a self-adjoint operator
on a metric graph have been derived for this case 
by Kostrykin and Schrader \cite{kostrykin:1999}.
Generalised boundary conditions at the vertex $i$ can be written in
the form
\begin{equation}
  \sum_{j} C^{(i)}_{j'j}\phi_{(i,j)}(x_{(i,j)}^i)+
  D^{(i)}_{j'j}\frac{\D}{\D x_{(i,j)}}
  \phi_{(i,j)}(x^i_{(i,j)})=0
  \label{eq:gen_bc}
\end{equation}
where the sum runs over all vertices $j$ in the neighbourhood of $i$.
The ($k$ independent) $v_i\times v_i$ matrices $C^{(i)}$ and $D^{(i)}$
can be chosen
arbitrarily for each vertex, restricted only by the
conditions that \textit{(i)}
the $v_i \times 2 v_i$ matrix $(C^{(i)},D^{(i)})$
has maximal rank and
\textit{(ii)}  $C^{(i)} {D^{(i)}}^\dagger$ is hermitian.
It then follows that the Schr\"odinger operator on the graph
with these boundary conditions is self-adjoint.
A small exercise
gives the corresponding vertex scattering
matrices.
\begin{equation}
  \sigma^{(i)}(k)=(\I k D^{(i)}+C^{(i)})^{-1}(\I k D^{(i)} - C^{(i)})\ .
  \label{eq:gen_vertex_scattering}
\end{equation}
The vertex scattering matrices defined in this way are
unitary and generally $k$-dependent. With the given restrictions
on $C^{(i)}$ and $D^{(i)}$ the matrices $\I k D^{(i)}\pm
C^{(i)}$ are always invertible \cite{kostrykin:1999}.
However the matrices $D^{(i)}$ and $C^{(i)}$ are in general not
invertible and the limits $k\rightarrow 0 $ and $k\rightarrow \infty$
in \eqref{eq:gen_vertex_scattering} are non-trivial.

As an example, let us show that the boundary conditions from
Chapter \ref{chapter:Schrodinger} are a subset
of this generalised approach. Choosing the $v_i \times v_i$ matrices
\begin{equation}
  C^{(i)}=-\lambda_i \ONE \qquad D^{(i)}=\EEE
\end{equation}
where $\EEE_{jj'}=1$ for each matrix element
is equivalent to the boundary conditions \eqref{eq:boundary} and
satisfies the conditions on $C^{(i)}$ and $D^{(i)}$.
The vertex scattering matrix
$\sigma^{(i)}=-\ONE+ \frac{2 k}{v_i k +\I \lambda_i}\EEE$
given by \eqref{eq:vertexsig} is easily seen to
satisfy $(\I k D^{(i)} +C^{(i)})\sigma^{(i)}=\I k D^{(i)}-C^{(i)}$
using the property $\EEE^2 = v_i \EEE$.
In the limit $\lambda_i\rightarrow 0$ our choice of $C^{(i)}$
and $D^{(i)}$ does not satisfy the required condition that
$(C^{(i)},D^{(i)})$ has maximal rank. However, the choice
of $C^{(i)}$ and $D^{(i)}$ is not unique and it is easy to find
a choice that
works for $\lambda_i=0$. To be explicit, for $v_i=3$ choose
\begin{equation}
  C^{(i)}=
  \begin{pmatrix}
    1 & -1 & 0\\
    0 & 1 & -1\\
    0 & 0 & 0
  \end{pmatrix}
  \qquad
  D^{(i)}=
  \begin{pmatrix}
    0 & 0 & 0\\
    0 & 0 & 0\\
    1 & 1 & 1
  \end{pmatrix}\ .
\end{equation} Here $C^{(i)} {D^{(i)}}^\dagger=0$ which is
self-adjoint and it is obvious how to extend it to higher valencies
$v_i>3$.

Instead of defining a self-adjoint Schr\"odinger operator one may
as well quantise a metric graph by requiring that a unitary
quantum evolution operator of the form $\evol(k)= \bondprop(k)
\scattering(k) $ exists. This is equivalent to assigning an
arbitrary unitary vertex scattering matrix $\sigma^{(i)}(k)$ to
each vertex $i$. Such an arbitrary choice is in general
not equivalent to a self-adjoint Schr\"odinger operator on a
metric graph with boundary conditions specified by
\eqref{eq:gen_bc}. With little abuse of language
physicists nonetheless speak of boundary conditions which are
defined by such arbitrary vertex scattering matrices -- and so
will we in the sequel. While mathematically less satisfactory
this is a physically well motivated and well defined
generalisation. Physically, a vertex is some scattering centre
which may have some internal structure. While the details of this
internal structure are not relevant one just describes it through
its unitary scattering matrix which ensures probability
conservation (unitarity of the quantum evolution map). The
scattering at the vertex may or may not depend on the wave-number
$k$ depending on the internal structure of the vertex. In most
physical applications they are just chosen to be constant. In the
remaining chapters we will stick to an arbitrary but
$k$-independent choice just for simplicity. 
Note, that a $k$-independent choice is also mathematically 
satisfying. It has been shown by Carlson \cite{carlson}
that such a choice
defines a self-adjoint Schr\"odinger operator on a (directed)
metric graph.
An example of a widely
used vertex scattering matrix is the Discrete Fourier Transform
(DFT) matrix for which
\begin{equation}
  \sigma^{(i),\mathrm{DFT}}_{jj'}=\frac{1}{\sqrt{v_i}}\E^{2\pi \I
  \frac{n(j)n(j')}{v_i}}
\end{equation}
where $n(j)$ is a one-to-one mapping of the $v_i$ neighbouring
vertices $j$ to the numbers $0,1,\dots,v_i-1$. In contrast to a
vertex with Neumann boundary conditions where backscattering is
favoured for large valency $v>4$ the scattering amplitudes of a
vertex with a DFT vertex scattering matrix has equal absolute
value for all incoming and outgoing bonds.

In other applications one may have a specific scattering system
in mind (e.g. the bonds that are connected to a vertex are really
channels that couple to a quantum billiard). Quite general
vertex scattering matrices may also arise starting from a
quantum graph with Neumann boundary conditions. In that case
one may define a \define{composite vertex} which combines some subset
of the vertices of the graph and combines them to a single one.
This may simplify the topology (or calculations) considerably.
For the new composite vertex one may derive a vertex scattering
matrix by eliminating all bonds that now belong to the internal
structure.
As a simple example let us look at a composite vertex
defined by replacing
\begin{equation*}
  \begin{array}{c}
  \includegraphics[width=4cm]{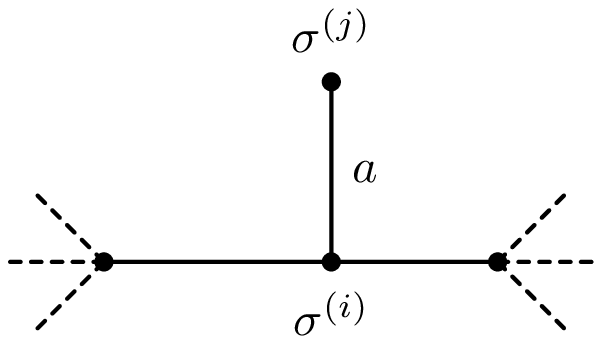}
  \end{array}
  \; \Longrightarrow \;
  \begin{array}{c}
  \includegraphics[width=4cm]{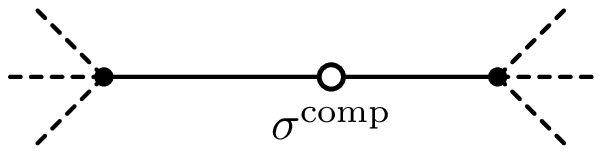}
  \end{array}\,
\end{equation*}
where one bond of length $a$ is eliminated. For Neumann boundary
conditions $\sigma^{(j)}=1$ and $\sigma^{(i)}=\frac{2}{3}\EEE_3
-\ONE_3$ one gets the $k$-dependent composite vertex scattering
matrix
\begin{equation}
  \begin{split}
  \sigma^{\mathrm{comp}}=&
  \frac{1}{3+\E^{\I 2 a k}}
  \begin{pmatrix}
    \E^{\I 2ak}-1 & 2 (\E^{\I 2ak} +1)\\
    2 (\E^{\I 2ak} +1) & \E^{\I 2ak}-1
  \end{pmatrix}\\
  =& -\ONE_2 + \EEE_2\left( \frac{2}{3} + \frac{2}{3}
  \frac{\E^{\I 2ak}}{1+\frac{1}{3}\E^{\I 2ak}}
  \frac{2}{3}\right)\ .
  \end{split}
  \label{eq:comp_vertex}
\end{equation}
Here, we have written the last line in such a way that an
interpretation in terms of trajectories is apparent. The first two
terms $-\ONE_2+\frac{2}{3}\EEE_2$ describe direct processes for
which the particle does not visit the eliminated bond. In the
third term the two factors $2/3$ are the amplitudes for entering
and exiting the eliminated bond and the factor
\begin{equation}
  \frac{\E^{  \I 2ak}}{1+\frac{1}{3}\E^{\I 2ak}}
  =\E^{\I 2ak}\sum_{n=0}^\infty \left(- \frac{1}{3} \E^{\I 2 ak}\right)^n
\end{equation}
is a geometric sum over trajectories for which the particle is
scattered back to the eliminated bond $n$ times.

Eliminating bonds by defining composite vertices is very useful
in numerical calculations (it reduces the matrix
dimension), and also in analytical approaches.
For instance, if there
is a length scale separation between small bond lengths inside
the composite vertex and large bond lengths in the rest of the graph
like in
\begin{equation*}
  \begin{array}{c}
  \includegraphics[width=4cm]{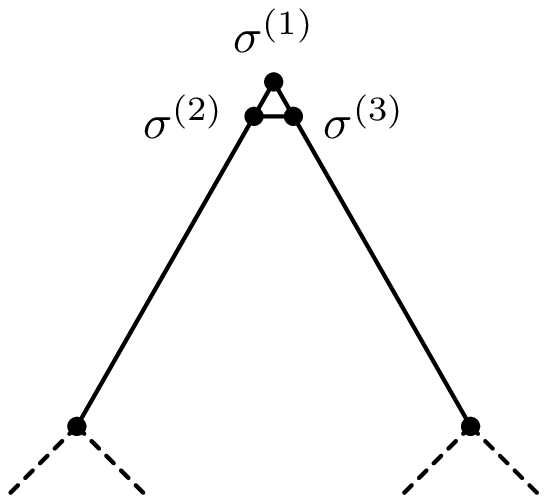}
  \end{array}
  \; \Longrightarrow \;
  \begin{array}{c}
  \includegraphics[width=4cm]{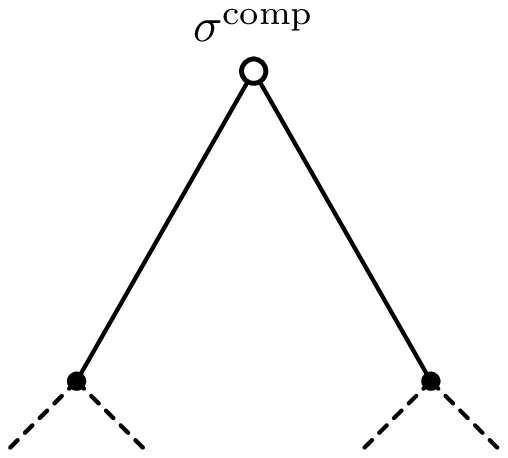}
  \end{array}
\end{equation*}
the vertex scattering matrix of the composite vertex will
weakly depend on the wavenumber $k$.
\newpage
\thispagestyle{empty}
\chapter{Classical evolution on graphs}
\label{chapter:classical}

So far we discussed  graph dynamics from a quantum mechanical point of
view.  At the present stage, we would like to study graphs from a
different point of view, which  provides the classical counterpart
of the quantum theory \cite{kottos:1998,barra:2001,pakonski:2001}. 
Usually the connection between the quantum
and the classical description is provided by \emph{quantising}  a
classical system. Here we take the process in the reverse
direction, for reasons which will be explained below.

\section{Classical phase-space and transition probabilities}
\label{sec:transition_prop}

Given a graph, we consider a classical particle which moves freely
as long as it is on any of the bonds. The vertices are singular
points, and it is not possible to write down the analogue of the
Newton equations at the vertices. To circumvent this intrinsic
difficulty we employ a Liouvillian approach, based on the study of
the evolution of phase-space densities. The phase space evolution
operator assigns transition probabilities between phase space
points, and it is the classical analogue of the quantum evolution
operator. We shall employ this analogy to construct the classical
dynamics. For this purpose we should first establish what is the
graph classical phase-space, and second, construct the classical
analogue of the quantum evolution operator.

 The phase-space description will be constructed on a
Poincar\'{e} section which is defined in the following way.
Crossing of the section is registered as the particle encounters a
vertex, thus the ``coordinate'' on the section is the vertex
label. The corresponding ``momentum'' is the direction at which
the particle emerges from the vertex. This is completely specified
by the label of the next vertex to be encountered. In other words,
\begin{equation}
  \begin{Bmatrix}
    \text{ position} \\
    \text{momentum}
  \end{Bmatrix}
  \qquad \Longleftrightarrow \qquad
  \begin{Bmatrix}
  \text{vertex index} \\
  \text{next index}
  \end{Bmatrix}
  \equiv  (b,\omega) .
\end{equation}

The set of all possible vertices and directions is equivalent to
the set of $2B$ directed bonds. Thus, the classical phase space
densities are defined on the same space as the corresponding
quantum evolution map.

The evolution on the Poincar\'{e} section is described in discrete time
steps (topological time) and is well defined once we postulate the
transition probabilities $P_{\alpha^{\prime }\leftarrow
\alpha}^{(i)}\ge 0$ between the directed bonds $\alpha$ and
$\alpha^{\prime }$ where $\alpha^{\prime}$ follows $\alpha$
at the vertex $i$,
$\alpha^\prime \in \mathcal{F}_i\alpha$. Probability conservation
then requires
\begin{equation}
  \sum_{\alpha^\prime: \alpha^\prime \in \mathcal{F}_i \alpha}
  P^{(i)}_{\alpha^\prime\leftarrow \alpha}=1
  \label{eq:prop_cons1}
\end{equation}
for all $\alpha$.

For a general classical dynamics (Markov process) on a graph one
may postulate any transition probabilities $P^{(i)}_{\alpha'
\leftarrow \alpha}\ge 0$ that satisfies \eqref{eq:prop_cons1}. To
construct the classical analogue of the quantum graphs, we choose
the classical transition probabilities to be equal to the quantum
transition probabilities, expressed as the absolute squares of the
$\evol$ matrix elements
\begin{equation}
  P_{\alpha^{\prime }\leftarrow \alpha}^{(i)}  =\left|
  \sigma_{\alpha^{\prime },\alpha}^{(i)}(k)\right| ^2\ .
  \label{eq:classical_propability}
\end{equation}
In addition to the general condition \eqref{eq:prop_cons1} the
transition probabilities on a quantum graph also satisfy
\begin{equation}
  \sum_{\alpha: \hat\alpha \in \mathcal{F}_i \hat{\alpha}^\prime}
  P^{(i)}_{\alpha^\prime\leftarrow \alpha}=1
  \label{eq:prop_cons2}
\end{equation}
for all $\alpha^\prime$ because the $\sigma^{(i)}$ are unitary.
 
As examples, we quote
explicitly the transition probability which correspond to a DFT,
Neumann and Dirichlet vertex scattering matrices.
\begin{equation}
  P_{\alpha^{\prime }\leftarrow \alpha}^{(i)} =
  \begin{cases}
    \frac{1}{v_i} & \text{for a DFT vertex scattering matrix,}\\
    \\
    \left(1-\frac{4}{v_i}\right)
    \delta_{\hat\alpha,\alpha^{\prime }}+
    \frac{4}{v_i^2}  & \text{for Neumann boundary conditions,}
    \\ \\
    \delta_{\hat\alpha,\alpha^{\prime }} &\text{
    for Dirichlet boundary conditions.}
  \end{cases}
  \label{eq:classical_Neumann_Dirichlet}
\end{equation}

In the DFT case a particle on a directed bond $\alpha\in
S^{(i)}_-$ is scattered with equal probability to all outgoing
bond $\alpha^\prime \in S^{(i)}_+$. In contrast, for Neumann
boundary conditions at a vertex of high valency $v_i>4$
backscattering $\alpha'=\hat\alpha$ is strongly favoured. One may
thus expect that the classical decay of correlations is much
faster in a graph with DFT vertex scattering matrices compared to
a Neumann graph. The transition probability for the Dirichlet
case, admits the following physical interpretation. The particle
is confined to the bond where it started and thus the phase space
is partitioned  to non-overlapping components (``tori'' - since
the dynamics on each bond is periodic). Thus, the Dirichlet
boundary conditions correspond to an integrable classical
analogue.

In general, a graph is \define{dynamically connected} if the
directed bonds cannot be split into two nonempty subsets such that
all transition probabilities to go from one set to the other
vanish. Dynamical connectivity for quantum graphs may equivalently
be defined via the scattering amplitudes
$\sigma^{(i)}_{\alpha,\alpha^\prime}$.

\section{The classical evolution operator}
\label{sec:classical_evolution_operator}

The transition probabilities $P^{(i)}_{\alpha^\prime \leftarrow
\alpha}$ can be combined into a $2B \times 2B$ matrix in the same
way that the vertex scattering matrices have been used to
construct the graph scattering matrix. This defines the
\define{classical evolution  operator} (also called
\define{Frobenius-Perron operator})
\begin{equation}
  \mathcal{M}_{\alpha^\prime,\alpha }=\left|\scattering_{\alpha^\prime,\alpha}
  \right|^2 = \left| \evol_{\alpha^\prime,\alpha}(k)\right|^2=
  \begin{cases}
    | \sigma^{(i)}_{\alpha^\prime, \alpha }|^2 &
    \text{if $\alpha' \in \mathcal{F}_i \alpha$,}\\
    0 & \text{else.}
  \end{cases}
 \label{eq:classical_evolution_operator}
\end{equation}
$\mathcal{M} $ does not involve any metric information on the
graph, and for Dirichlet or Neumann boundary conditions it is
independent of $k$. The evolution is a discrete Markov process
with transition probabilities $\mathcal{M}_{\alpha',\alpha }$. The
unitarity of the quantum evolution map  $\evol(k)$  guarantees
that $\mathcal{M}$ is a \define{bistochastic matrix} (also known
as \define{doubly stochastic matrix}) which means
\begin{equation}
  \sum_{\alpha} \mathcal{M}_{\alpha^{\prime },\alpha}=\sum_{\alpha^\prime}
  \mathcal{M}_{\alpha^{\prime },\alpha}=1
  \qquad \text{and} \qquad
  0\leq \mathcal{M}_{\alpha^{\prime },\alpha}\leq 1\ .
  \label{eq:bistochastic}
\end{equation}
Bistochastic matrices that are defined by a unitary matrix in
the form \eqref{eq:classical_evolution_operator} are also
called \define{unistochistic} matrices\footnote{In general unistochastic
matrices are a subset of bistochastic matrices. A unistochastic matrix
defines a Markov process which can be quantised (though not uniquely)
\cite{bengtson:unistochastic,zyczkowski:2003,berkolaiko:2001a}.}.
These properties of the classical evolution map are equivalent to
the condition \eqref{eq:prop_cons1} and \eqref{eq:prop_cons2}
which ensure probability conservation.

If $\rho _{\alpha}(n)\ge 0$ denotes the probability to occupy the
directed bond $\alpha$ at the (topological) time $n$, then we can
write down a Markovian Master equation for the classical density:
\begin{equation}
  \vec{\rho}(n+1)=
  \mathcal{M} \vec{\rho}(n)\ ,
  \label{master}
\end{equation}
with the $2B$-dimensional vector $\vec{\rho}=(\rho
_{1},\cdots,\rho _{2B})$. The total probability to be anywhere on
the graph is unity thus
\begin{equation}
  ||\vec{\rho}||\equiv \sum_{\alpha} \rho_\alpha =1\ .
\end{equation}

From the bistochastic property of $\mathcal{M}$ follows that the
uniformly distributed density on the directed bonds
\begin{equation}
  \rho^{\mathrm{inv}}_{\alpha}=\frac{1}{2B}
\end{equation}
is invariant under the classical evolution
\begin{equation}
  \vec{\rho}^{\mathrm{inv}}=\mathcal{M} \vec{\rho}^{\mathrm{inv}}\ .
\end{equation}
In most cases there exists only one invariant probability
distribution. It then follows that the Markov process is
\define{ergodic} which means that the time average of any
\define{classical observable} $\vec{f}\in \RR^{2B}$ (that is,
$f_\alpha$ is the value of the observable on the directed bond
$\alpha$) is equal to an average over the invariant uniform
probability distribution
\begin{equation}
  \lim_{N\rightarrow \infty}\frac{1}{N+1}\sum_{n=0}^N \vec{f} \cdot
  \mathcal{M}^n
  \vec{\rho}(0)= \vec{f} \cdot \vec{\rho}^{\mathrm{inv}}\equiv\sum_\alpha
  \frac{f_\alpha}{2B}
  \label{eq:ergodic1}
\end{equation}
for every initial distribution $\vec{\rho}(0)$. This is equivalent
to the statement that the time averaged occupation probability on
a directed bond $\alpha$ is uniformly distributed over $\alpha$
\begin{equation}
  \lim_{N\rightarrow \infty}\frac{1}{N+1}\sum_{n=0}^N \vec{\rho}_\alpha(n)
  =\frac{1}{2B}\ .
  \label{eq:ergodic2}
\end{equation}
This is not a very strong statement on the classical
dynamics of a graph.
It is known that
the classical dynamics for
every dynamically connected graph is ergodic
\cite{barra:2001}.
A Markov process may have the stronger dynamical property of being
\define{mixing} which is defined by
\begin{equation}
  \lim_{n\rightarrow \infty} \mathcal{M}^n \vec{\rho}(0)=
  \vec{\rho}^{\mathrm{inv}}
  \label{eq:mixing}
\end{equation}
for any initial probability distribution $\vec{\rho}(0)$.

The properties of the graph which determine whether it is ergodic
or mixing are encoded in the classical spectrum $\nu_\ell\ ,\
\ell=1,\cdots,2B$ which is defined by
\begin{equation}
  \mathcal{M} \vec{\chi}_\ell= \nu_\ell \vec{\chi}_\ell\ .
\end{equation}
The classical spectrum is restricted to the interior of the unit
circle $|\nu_\ell|\le 1$ and $\nu_1 = 1$ corresponds to the
uniform distribution $\vec{\chi}_1=\vec{\rho}^{\textrm{inv}}$. Any
probability distribution can be written in the form
\begin{equation}
  \vec{\rho}=\sum_{\ell=1}^{2B} a_\ell \chi_\ell
\end{equation}
and $a_1=1$. For $\nu_\ell\neq 1$ one has
$||\vec{\chi}_\ell||=\sum_\alpha \chi_{\alpha,\ell}=0$ due to the
probability conservation of the dynamics.

Ergodicity implies that $\nu_1=1$ is the only unit eigenvector.
Thus there is a \define{spectral gap}
\begin{equation}
  \Delta_g=\min_{\ell=2,\dots,2B}\, \left|1 - \nu_\ell\right|\equiv
  \frac{1}{n^{\mathrm{erg}}}
  \label{eq:spectral_gap}
\end{equation}
which determines the time scale
$n^\mathrm{erg}$ on which the left hand side of
\eqref{eq:ergodic2} decays to the uniform distribution. Thus,
ergodicity allows for eigenvalues on the unit circle
$\nu_\ell=\E^{\I \theta}$ which are damped by the time average. In
practice, such eigenvalues are non-generic -- however there are
some example, such as star graphs which have an eigenvalue
$\nu=-1$\footnote{Star graphs are an example of
\define{bipartite}  graphs for which the directed bonds can be
split into two nonempty sets and the transition probabilities
within each of the two sets vanish. For star graphs due to its
topology there is no transition from any outgoing (incoming) bond
to another outgoing (incoming) bond. One may easily construct
other examples which rely on the dynamical connectivity as well.
Dynamically connected star graphs are generally mixing
if the classical dynamics is defined on bonds instead of directed bonds as
in Chapters \ref{chapter:periodic_orbit} and \ref{chapter:susy}.}.

If there is no eigenvalue on the unit circle apart from $\nu_1=1$
all non-uniform modes of the probability distribution decay
\begin{equation}
  \mathcal{M}^n \vec{\rho}(0)=
  \vec{\rho}^\mathrm{inv}+\sum_{\ell=2}^{2B} \nu_\ell^n a_\ell \vec{\chi}_\ell
  \xrightarrow{ {n\rightarrow \infty }} \vec{\rho}^\mathrm{inv}
\end{equation}
and the dynamics is mixing.
The rate at which equilibrium is
approached is determined by the gap
\begin{equation}
  \tilde{\Delta}_g=\min_{\ell=2,\dots,2B} 1-|\nu_\ell|\equiv 
  \frac{1}{n^{\mathrm{mix}}}
\end{equation}
between the next largest
eigenvalue and $1$. In general ergodicity
sets in faster than mixing $n^{\mathrm{erg}}< n^{\mathrm{mix}}$.

We shall end this section by showing how the trajectories  which
we defined formally in Section \ref{sec:definitions} acquire a
dynamical significance, and emerge naturally in the classical
framework developed above.

The classical probability to make a transition from a directed
bond $\alpha$ to a directed bond $\beta$ after $n$ steps equals
\begin{equation}
(\mathcal{M}^n)_{\beta,\alpha} = \sum_{t\in
\mathcal{T}_n(\beta,\alpha)} V_{\beta,\alpha}(t)
\end{equation}
where the sum extends over the set $\mathcal{T}_n(\beta,\alpha)$
of trajectories  which start at $\alpha$ and get to $\beta$ in $n$
steps, and
\begin{equation}
 V_{\beta,\alpha}(t)
= \prod_{s=1}^{n-1 } \mathcal{M}_{\alpha_{s+1},\alpha_{s}}
\end{equation}
are the classical \define {weights} or probabilities contributed by
the trajectory $t$ to the transition probability. This form of the
transition probability emphasises the similarity between the
classical dynamics on a graph and the evolution under Hamiltonian
maps.

One of the characteristic features of chaotic (mixing) classical
dynamics is the exponential proliferation of the number of
classical trajectories which connect the same initial and final
points as the transition time increases. This property is also
shared by the trajectories on the graphs. This follows immediately
from the fact that we have a natural code -- the
sequence of directed bonds -- which associates a
string of numbers 
with a trajectory $t=(\alpha_1,\dots,\alpha_n)$
(by numbering the directed bonds from $1$ to $2B$), 
and a connectivity matrix which established a
Markovian grammar on the codes. 
In the language used in the theory
of dynamical systems this is a Bernoulli code, which in Hamiltonian
systems guarantees chaotic dynamics (see \cite{Book:gaspard,webbook}
for more details).
On a simple graph, the sequence of vertices is often used to define
an alternative (equivalent) Bernoulli code.

\section{The return probability and a classical sum rule}
\label{sec:return}

Of prime importance in the discussion of the relation between the
classical and the quantum dynamics are the traces
\begin{equation}
  u_n=\mathrm{tr}\,\mathcal{M}^n =\sum_{\alpha=1}^{2B}
  (\mathcal{M}^n)_{\alpha,\alpha}\ .
\end{equation}
This is the \define{return probability} - the classical
probability to perform $n$-periodic motion. The mean probability
to return  is $u_n/2B$. For mixing dynamics where only one
eigenvalue $\nu$ is on the unit circle, one has
\begin{equation}
  u_n=\sum_{\ell=1}^{2B} \nu_\ell^n
  \xrightarrow{ {n\rightarrow \infty }} 1\ .
  \label {eq:classerg}
\end{equation}

Following the discussion at the end of the previous section, $u_n$
can be written also as a classical sum over closed trajectories of
length $n$. The classical weight  of a closed trajectory $t_c$
with a code $(\alpha_1,\dots,\alpha_{n},\alpha_1)$ is the product
\begin{equation}
  V_{t_c}= \prod_{s=1}^{n} \mathcal{M}_{\alpha_{s+1},\alpha_{s}}
  \label{eq:classical_amplitude}
\end{equation}
where $\alpha_{n+1}\equiv\alpha_1$. The weight can be interpreted
as the probability to remain on the trajectory. Any cyclic
permutation of the bond indices in the code of a closed trajectory
is also a closed trajectory with identical classical weight.
Hence, instead of summing over closed trajectories, we may express
the return probability as a sum over \emph{periodic orbits }. A
code  of an $n$-periodic orbit  can be written as a repetition  of
primitive codes whose lengths $n_p$ divides $n$. The set of all
the primitive orbits which build $n$ periodic orbits will be
denoted by $\mathcal{P}(n)$. Denote the weight of a primitive
orbit $p$ by $W_p$. Then, each primitive orbit contributes to the
return probability a term which consists of its weight taken to
its $r$ power, multiplied by $n_p=n/r$,
\begin{equation}
u_n = \sum_{p\in \mathcal{P}(n)} n_p (W_{p})^r
 \label{eq:classreprb}
\end{equation}
This is a basic formula in all our subsequent discussion and it
deserves a few comments.

The weights $W_p$ can also be defined for codes which
violate the dynamical connectivity, by assigning them the value
$W_p=0$. Sums over periodic orbits that involve the weight as a
factor can immediately be replaced by sums over all arbitrary
cyclic codes for this reason.

The code gives rise to a variety of classifications of
periodic orbits. For instance, a periodic orbit is either
\define{reducible} or \define{irreducible}. An {irreducible periodic orbits}
visits each directed bond at most once. All directed bonds appearing in its
code are different $\alpha_i\neq \alpha_j$. The maximal period of
an irreducible periodic orbit is $n_{max}=2 B$. All other periodic
orbits are reducible, that is at least one directed bond is
visited more than once. All periodic orbits of period larger $2B$
are reducible\footnote{The definition of reducible and irreducible
periodic orbits depends on the code which is used to define a
periodic orbit unambiguously. For a simple graph one may use the
the sequence of vertices $p=\overline{i_1i_2\dots i_n}$ as a code
(symbolic dynamics) which uniquely defines a periodic orbit. In
the directed bond code the same orbit is given by
$p=\overline{[i_2,i_1],[i_3,i_2],\dots,[i_1,i_n]}$. Irreducible
periodic orbits with respect to the vertex code
(\define{vertex-irreducible periodic orbits}) do not intersect at
any vertex such that all $i_j$ are different. Our definition which
is based on the directed bond code is \emph{not equivalent} to the
one based on the vertex code. In general there are
bond-irreducible orbits of period $n>V$ and the latter cannot be
vertex-irreducible. Both codes have their advantages and
disadvantages. The vertex code has the advantage of a smaller
number of symbols and a grammar which is readily encoded in the
connectivity matrix. In the present context the grammar is less
relevant since a code which does not correspond to a periodic
orbit has zero weight. The advantage of using the bond code to
define reducible orbits, is that weights of reducible orbits are
products of weights of non-irreducible orbits.}. Reducible orbits
have the remarkable property that their weights can be written as
products of weight of shorter orbits $W_p=W_{p_1}W_{p_2}$.

We are now equipped with the background to obtain a classical
sum-rule for mixing dynamics by substituting a periodic-orbit expansion of
$u_n$. 
Combining \eqref{eq:classerg} and \eqref{eq:classreprb} we get\footnote{
In the non-mixing case a dynamically connected graph
is ergodic. With an
additional time average over an interval $\Delta n \ll n$ one then gets
the sum rule
\begin{equation}
  \frac{1}{\Delta n}\sum_{n^{\prime}=n}^{n+\Delta n-1} u_{n^{\prime}}
  =\frac{1}{\Delta n}\sum_{p: n\le r_p n_p < n+\Delta n} n_p W_p^{r_p}
  \xrightarrow{{n, \Delta n \rightarrow \infty }}\ 1\ .
  \label{eq:classicalsum_ergodic}
\end{equation}}
\begin{equation}
  u_n=\sum_{p\in \mathcal{P}(n)} n_p (W_{p})^r
  \xrightarrow{{n\rightarrow \infty }}\ 1.
  \label{eq:classicalsum}
\end{equation}

This sum rule is analogous to the \textbf{Sinai-Bowen-Ruelle} 
\cite{sinai,bowen,ruelle} sum
rule for classically chaotic dynamics
(in the physics literature also known as Hannay-Ozorio de Almeida sum rule
\cite{ozorio:sumrule}). The weights $W_p$ are the
counterparts of the stability weights $|\det (I-M_p)|^{-1}$ for
hyperbolic periodic orbits in Hamiltonian systems, where $M_p$ is
the monodromy matrix. Graphs, however, are one dimensional and the
motion on the bonds is simple and stable. Ergodic or mixing
dynamics is generated because at each vertex a (Markovian) choice
of one out of $v$ directions is made. Thus, chaos on graphs
originates from the multiple connectivity of the (otherwise
linear) system.

Using the expression \eqref{eq:classicalsum} for $u_n$ one can easily
write down the complete thermodynamic formalism for the graph.
Here, we shall only quote the periodic orbit expression for the
Ruelle $\zeta$-function 
\begin{equation}
  \begin{split}
  \zeta_R(z) \equiv \det (\ONE-z \mathcal{M})^{-1}
  =& \E^{ -\mathrm{tr}\,\mathrm{ln}\,  (\ONE-z \mathcal{M}) }
   =\E^{ \sum_n\frac{z^n}{n} u_n}\\
  =& \prod_{p }
    \frac{1}{\left(1-z^{n_{ p }}\exp (-n_{ p } \gamma_{ p })
    \right)}
  \end{split}
  \label{eq:Ruelle}
\end{equation}
where the product extends over all primitive periodic orbits $ p $
and we have set $W_{ p }=\E^{- n_{ p }\gamma_{ p }}$.
The Ruelle $\zeta$-function is an important tool to analyse dynamical
systems  \cite{Book:gaspard,webbook}.\newpage
\thispagestyle{empty}  
\chapter{Spectral theory for quantum graphs}
\label{chapter:spectral_theory}

Trace formulae are an important tool in spectral theory. They
express spectral sums in terms of sums over periodic orbits of the
underlying classical dynamics. In this chapter our main goal is to
introduce the trace formula for the density of states of a quantum
graph. The first trace formula for a quantum graph was derived by
Roth \cite{roth:trace,roth} who showed that a spectral
determinant (which can be interpreted as a variant of the secular
function $\zetab(k)$) can be written as a sum over periodic
orbits. The trace formula for the density of states goes back to
Kottos and Smilansky \cite{kottos:1997,kottos:1998}
and is formulated in analogy to the semi-classical Gutzwiller
trace formula \cite{gutzwiller:trace,Book:Gutzwiller} for chaotic
(hyperbolic) Hamiltonian quantum systems. In contrast to the
latter, the trace formula for quantum graphs is exact. Apart from
the density of states we will also discuss how other spectral
functions are related to sums over periodic orbits.

The starting point for the derivation is the secular function
$\zetab(k)$ which is expressed in terms of the quantum evolution
map $\evol (k)$ (see Chapter \ref{chapter:QUE}). Thus, the
boundary conditions at the vertices are defined by a given set
of vertex scattering matrices and the wave functions are not
necessarily continuous across a vertex. Without loss of
generality we will restrict the discussion to dynamically
connected quantum graphs of simple topology. If the graph is not
dynamically connected (e.g.~ a graph with Dirichlet boundary
conditions) the graph can be divided into subgraphs which have
independent spectra and wave functions -- that is, the complete
spectrum is the superposition of the independent spectra. For
simplicity, we will also assume throughout this chapter that the
vertex scattering matrices do not depend on the wave number $k$
(the generalisation to $k$ dependent scattering matrices follows
in a straight forward way).

\section{The density of states and the counting function}
\label{sec:DOS}

If $\{ k_n\}$ ($n\in  \mathrm{N}$) is the spectrum of a quantum
graph (degenerate eigenvalues appear according to their
multiplicity) we define the \define{density of states} as
\begin{equation}
  d(k)=\sum_{n=1}^\infty \delta(k-k_n)\ .
  \label{eq:define_dos}
\end{equation}
The \define{spectral counting function} which provides the number
of eigenvalues $k_n$ which are smaller than 
$k$ is given by the
integral
\begin{equation}
  N(k)= \theta(k) N_0 +\lim_{\delta \rightarrow 0^+} 
  \INT{\delta}{k}{k'} d(k')=
  \sum_{n=1}^\infty \theta(k-k_n)\ ,
  \label{eq:define_counting_function}
\end{equation}
where $N_0$ is the number of eigenvalues at $k=0$ and 
\begin{equation}
  \theta(k)=
  \begin{cases}
    0 & \text{for $k<0$}\\
    1/2 & \text{for $k=0$}\\
    1 & \text{for $k>0$}
  \end{cases}
\end{equation}
is the Heaviside step function.
The limit
$\delta \rightarrow 0^+$ is needed only when $N(k)$ is computed at
$k\rightarrow 0$, and it ensures that $\lim_{k\rightarrow 0}
N(k)=N_0$. When the secular function $\zetab(k)$ does not
have a zero at $k=0$ and $N_0=0$ one can set $\delta=0$ from the
start in the following discussion.

We will now express the counting functions $N(k)$ in terms of the
secular function $\zetab(k)=\mathrm{det}\, (\ONE-\evol(k))$.
To this end, we introduce the eigenvalues $\{ \E^{\I
\phi_\ell(k)}\}$ ($\ell=1,2,\dots,2B$) of the quantum evolution
map $\evol(k)$ at fixed wave number $k$ and first discuss some of
their properties. The phases $\phi_\ell(k)$ are important in the
present context since the quantisation condition $\zetab(k)=0$ is
equivalent to the requirement
\begin{equation}
  \phi_\ell(k)=2\pi z \qquad \text{with} \qquad z=0,1,2,\dots
  \label{eq:quant_condition_phi}
\end{equation}
for one of the $\phi_\ell(k)$. Each of the quantisation conditions
\eqref{eq:quant_condition_phi} yields a discrete subset
$\{k^{(\ell)}_n\}\subset \{k_n\}$ of the complete spectrum and
this subset is free of degeneracies since $\frac{\D \phi_{\ell}}{\D
k}>0$. Indeed, a direct computation reveals that
\begin{equation}
\frac{\D \phi_{\ell}}{\D k}= \sum_{b=1}^B L_b
\left(|a^{(\ell)}_{(b,\omega)}(k)|^2+
|a^{(\ell)}_{(b,-\omega)}(k)\right|^2) \ > \ 0 \ ,
\end{equation}
where $a^{(\ell)}_{(b,\omega)}$ are the components of the $\ell$
eigenvector. This proof of the monotonic increase of the
eigenphases depends crucially on the assumption that the vertex
scattering matrices are independent of $k$. 
It follows immediately
that the maximal degeneracy in a quantum graph is $2B$.
Direct inspection shows $L_{\mathrm{min}}\le \frac{\D \phi_{\ell}}{\D k}
\le L_{\mathrm{max}}$ where $L_{\mathrm{min}}$ ($L_{\mathrm{max}}$)
is the smallest (largest) bond length, and $\sum_{\ell=1}^{2B }
\frac{\D \phi_{\ell}}{\D k}= 2 B \overline{L}$ where
$\overline{L}=\frac{1}{B}\sum_{b=1}^B L_b$ is the mean bond length.
For Neumann graphs with rationally independent bond
lengths a much stronger statement has been proven: their spectra
are generically devoid of degeneracies \cite{interlacing}.

We also note that $\evol(k+\I \epsilon)$ is
\define{subunitarity} for any finite $\epsilon>0$.
That is, $|\E^{\I \phi_\ell(k+\I \epsilon)}|<1$. This is an
immediate consequence of the obvious subunitarity of the bond
propagator $\bondprop (k+\I \epsilon)$.

A counting function for the subspectrum $\{k^{(\ell)}_{n}\}$ is
obtained through the correspondence
to the zeros of the real function  $\sin
\frac{\phi_\ell(k)}{2}$. A sum over step functions at the (strictly
positive part of the) subspectrum can thus be written as
\begin{equation}
  \sum_n \theta(k -k^{(\ell)}_{n })= -\frac{1}{\pi}
  \,\mathrm{Im}\,\left(\mathrm{ln}\,\sin
  \frac{\phi_\ell(k+\I\epsilon)}{2} - \mathrm{ln}\,\sin
  \frac{\phi_\ell(\delta)}{2}
  \right)
  \label{eq:subcount}
\end{equation}
where the limit $\epsilon\rightarrow 0^+$  is always implied. The
$\epsilon$-shift ensures that the complex number $\sin
\frac{\phi_n(k+\I\epsilon)}{2}$ rotates counterclockwise around
the origin of the complex plane. The term $\mathrm{ln}\,\sin
\frac{\phi_n(\delta)}{2}$ ensures that one starts to count at zero
and $\epsilon$ has been set to zero to indicate that the limit
$\epsilon\rightarrow 0^+$ has to be performed before the limit
$\delta\rightarrow 0^+$.

The full counting function is obtained by summing over all $2B$
subspectra. Using  $\sin \frac{\phi_\ell(k)}{2}=\frac{\I}{2}\E^{
\frac{-\I \phi_\ell(k)}{2}}(1-\E^{ \frac{\I \phi_\ell(k)}{2}}) $
we get,
\begin{equation}
  \begin{split}
    N(k)=& N_0+
      \frac{1}{2\pi}\mathrm{Im}\,\mathrm{ln}\,\mathrm{det}\frac{\evol(k+\I
      \epsilon)}{\evol(\delta)} -\frac{1}{\pi }
      \mathrm{Im}\,\mathrm{ln}
      \frac{\zetab(k+\I\epsilon)}{\zetab(\delta)}\\
      =& N_0+\frac{B \overline{L}}{ \pi}
      k+\frac{1}{\pi}\mathrm{Im}\,\mathrm{ln}\,
      \zetab(\delta)-\frac{1}{\pi}\mathrm{Im}\,\mathrm{ln}\,
      \zetab(k+\I \epsilon) \ .
  \end{split}
  \label{eq:counting_zeta}
\end{equation}
We have used the assumption that the vertex scattering matrices do
not depend on $k$ in the second line.

The counting function can be decomposed into a
smooth term (the so-called \define{Weyl term})
and an oscillatory term
\begin{equation}
  N(k)=N^\mathrm{Weyl}(k)+N^\mathrm{osc}(k)\ .
  \label{eq:counting_decompose}
\end{equation}

The smooth term is given by
\begin{equation}
  N^\mathrm{Weyl}(k)=\frac{B\overline{L}}{\pi } k+N^\mathrm{Weyl}(0)
  \label{eq:counting_Weyl}
\end{equation}
where the first term $\frac{B\overline{L}}{\pi}k$
describes the linear increase of the counting function and
\begin{equation}
  N^\mathrm{Weyl}(0)=N_0+\frac{1}{\pi}\mathrm{Im}\,\mathrm{ln}\,
  \zetab(\delta).
  \label{eq:nzero}
\end{equation}
is a constant that depends on the boundary conditions at the
vertices. This constant ensures that the complete counting reduces
to $N_0$ at $k=\delta$. When one wants to calculate
$N^{\mathrm{Weyl}}(0)$ from \eqref{eq:nzero} one should be aware
that the logarithm is a multivalued function
and it is not always obvious which is the correct sheet. The
correct sheet can be identified by the condition
$\frac{1}{K}\INT{0}{K}{k}
\left(N^\mathrm{Weyl}(k)-N(k)\right)\rightarrow 0 $ which just
means that the difference is oscillating around zero with a
bounded amplitude. For general $\evol(0)$ the shift
$N^\mathrm{Weyl}(0)$ can take real values. In many cases (e.g.~
for Neumann boundary conditions) $\evol(0)=\scattering$ is a real
unitary matrix. Then, $N^{\mathrm{Weyl}}(0)$ is some half-integer
number. This can be calculated directly by observing that all
eigenvalues of $\scattering$ with non-vanishing imaginary part
appear in complex conjugated pairs such that their contributions
cancel each other. Real eigenvalues take the values $\pm 1$.
$-1$ does not contribute since the imaginary part of
$\mathrm{ln}\left(1-(-\E^{\I\delta})\right)$ vanishes for $\delta\rightarrow
0$. Only eigenvalues $+1$ contribute $\frac{1}{\pi}
\mathrm{Im}\,\mathrm{ln }\left( 1-\E^{\I\delta} \right)\sim \frac{1}{\pi}
\mathrm{Im}\, \mathrm{ln}\ \E^{-\I \pi/2} \delta=-\frac{1}{2}$ and
$N^{\mathrm{Weyl}}(0)=N_0 - \frac{z}{2}$ where $z$ is the number
of unit eigenvalues.

The linear dependence of $N^{\mathrm{Weyl}}(k)$ on $k$ is a
consequence of the fact that the graph is a one dimensional
object. The fact that it is not simply connected appears only in
the oscillatory part of the counting function - the second term in
\eqref{eq:counting_decompose} -
\begin{equation}
  N^\mathrm{osc}(k)=-\frac{1}{\pi}\mathrm{Im}\,\mathrm{ln}\,
  \zetab(k+\I \epsilon)\ .
  \label{eq:counting_osc}
\end{equation}
The oscillations are bounded by the number of bonds
$|N^\mathrm{osc}(k+\I \epsilon)| \le B$. Equation
\eqref{eq:counting_osc} will be the starting point for our
derivation of the trace formula in Section
\ref{sec:periodic_orbit_theory}.

It is now straight forward to derive the
density of states by differentiation with respect to $k$ as a sum
\begin{equation}
  d(k)=d^\mathrm{Weyl}(k)+d^\mathrm{osc}(k)
  \label{eq:DOS1}
\end{equation}
of a smooth Weyl term
\begin{equation}
  d^\mathrm{Weyl}(k)=\frac{B\overline{L}}{\pi}
  \label{eq:weyl_dos}
\end{equation}
and an oscillatory part
\begin{equation}
  d^\mathrm{osc}(k)=-\frac{1}{\pi}
  \frac{\D}{\D k} \mathrm{Im}\,\mathrm{ln}\,
  \zetab(k+\I \epsilon)\ .
  \label{eq:weyl_osc}
\end{equation}

\subsection{The spectrum of the quantum evolution map}

The $2B$ eigenvalues $\E^{\I \phi_\ell(k)}$ of the quantum
evolution map $\evol(k)$ at fixed $k$ can also be used to define a
density and a corresponding counting function which will turn out
to be interesting objects in their own right. We define the
\define{density of eigenphases} as
\begin{equation}
  \tilde{d}(\phi; k)=\sum_{\ell=1}^{2B}
  \delta_{2\pi}(\phi-\phi_\ell(k))
  \label{eq:define_doe}
\end{equation}
and the \define{eigenphase counting function} as
\begin{equation}
  \tilde{N}(\phi; k)=\tilde{N}_0\ \theta(\phi)+
  \INT{\delta}{\phi}{\phi'}
  \tilde{d}(\phi'; k) =
  \sum_{n=1}^N \sum_{\nu=0}^\infty
  \,\theta(\phi-\phi_n(k)+2\pi\nu)\ .
\end{equation}
In \eqref{eq:define_doe}
$\delta_{2\pi}(\phi)=\sum_{\nu=-\infty}^\infty\, \delta(\phi+2\pi \nu)$
denotes the $2\pi$-periodic $\delta$-function.

In complete analogy to the derivation for the spectral
counting function above one may write
\begin{equation}
  \tilde{N}(\phi;k)=\tilde{N}^\mathrm{Weyl}(\phi;k)
  +\tilde{N}^\mathrm{osc}(\phi;k)
\end{equation}
and express the smooth and oscillatory
parts in terms of the $\phi$-dependent
secular function
\begin{equation}
  \zetab(\phi;k)=\mathrm{det}\,(\ONE - \E^{-\I \phi}\,\evol(k))\ .
\end{equation}
As a result one obtains
\begin{equation}
  \tilde{N}^\mathrm{Weyl}(\phi;k)=\frac{B}{\pi}\phi+\tilde{N}_0
  -\frac{1}{\pi}
  \mathrm{Im}\, \mathrm{ln}\, \zetab(\delta,k)
  \label{eq:counting_phi_weyl}
\end{equation}
for the smooth part, and
\begin{equation}
  \tilde{N}^\mathrm{osc}(\phi;k)=
  \frac{1}{\pi }
  \mathrm{Im}\, \mathrm{ln}\, \zetab(\phi-\I\epsilon,k)
  \label{eq:counting_phi_osc}
\end{equation}
for the oscillating part. These expressions for the eigenphase
counting function can alternatively be derived using Poisson's
formula. Note that the second term of the Weyl part
$-\frac{1}{\pi} \mathrm{Im}\, \mathrm{ln}\, \zetab(\delta,k)$ is
independent of $\phi$, however,  as a function of $k$ it can
oscillate with a maximal amplitude of $B$.

The smooth Weyl part of the density of eigenphases
\begin{equation}
  \tilde{d}^\mathrm{Weyl}(\phi;k)= \frac{B}{\pi}
  \label{eq:dos_phi_weyl}
\end{equation}
is consistent with having $2B$ eigenvalues
distributed in an interval of length $2\pi$.
Finally, the oscillating part of the density of
eigenphases is given by
\begin{equation}
  \tilde{d}^\mathrm{osc}(\phi;k)=\frac{1}{\pi} \frac{\D}{\D \phi}
  \mathrm{Im}\, \mathrm{ln}\, \zetab(\phi-\I \epsilon,k)\ .
  \label{eq:dos_phi_osc}
\end{equation}

\section{Periodic orbits and the trace formula}
\label{sec:periodic_orbit_theory}

We will now consider the oscillating parts of the spectral
functions $N(k)$ and $d(k)$ from a periodic-orbit perspective. One
of the reasons for the interest in quantum graphs in the quantum
chaos community is the analogy of an exact trace formula for
quantum graphs to Gutzwiller's semiclassical trace formula for the
density of states of chaotic Hamiltonian systems. While there is
no underlying \emph{deterministic} classical dynamics for graphs
they still display the generic behaviour of chaotic Hamiltonian
systems. At the same time they are less resistant to either
rigorous or numerical approaches.

The logarithm of the secular equation appearing in
\eqref{eq:counting_osc} can directly be written in terms of traces
of powers of the quantum evolution map
\begin{equation}
\mathrm{ln}\,\mathrm{det}(\ONE-\evol (k))=\mathrm{tr}\,
\mathrm{ln}(\ONE-\evol (k))=-\sum_{n=1}^\infty \frac{1}{n}\mathrm{tr}\,
\evol^n(k)\ .
\end{equation}
We can now follow the  path we used in Section \ref{sec:return}
to compute the classical return probability $u_n=\mathrm{tr}\,
\mathcal{M}^n $ in terms of periodic orbits.  The result is again
a sum over primitive periodic orbits \cite{kottos:1998}
\begin{equation}
  \begin{split}
    N^{\mathrm{osc}}(k)=&
    \frac{1}{\pi}\mathrm{Im}\,
    \sum_{n=1}^\infty
    \frac{1}{n}\mathrm{tr}\,
    \evol^n (k+\I \epsilon)\\
    =&
    \mathrm{Im}\,
    \sum_{ {p}} \;
    \sum_{r=1}^\infty \frac{1}{\pi\,r}
    \mathcal{A}_{p}^r \E^{\I r {L}_{ {p}}
    (k+\I\epsilon)+ \I r \Phi_{ {p}}}\\
    =&
    \sum_{ {p}} \;
    \sum_{r=1}^\infty \frac{1}{\pi \, r}
    |\mathcal{A}_{ {p}}|^r \sin r({L}_{ {p}} k+
    \Phi_{ {p}} + \mu_{ {p}})
    \E^{-L_p \epsilon}.
  \end{split}
  \label{eq:tf_counting}
\end{equation}
In the second line we have introduced the sum over all primitive
periodic orbits $ {p}$ of the graph and their repetitions $r$.
Here, ${L}_{ {p}}\equiv\sum_{l=1}^{n_{ {p}}} L_{b_l}$ is the
length of the primitive periodic orbit
$p=\overline{\alpha_1,\dots,\alpha_{n_{ {p}}}}$,
\begin{equation}
  \mathcal{A}_p=|\mathcal{A}_p| \E^{\I \sigma_p}\equiv
  S_{ \alpha_1 \alpha_{n_p}}
  S_{\alpha_{n_p} \alpha_{n_p-1}}
  \dots S_{\alpha_2 \alpha_1}
  \label{eq:ppo_amplitude}
\end{equation}
is the quantum amplitude of the primitive orbit defined as the
product of all scattering amplitudes along the orbit,
and
\begin{equation}
  \Phi_{\hat{p}}=\sum_{l=1}^{n_{\hat{p}}} A_{\alpha_l}
  =\sum_{l=1}^{n_{\hat{p}}} \omega_l A_{b_l}
\end{equation}
is the overall magnetic flux through the periodic orbit. The step
from the first line of \eqref{eq:tf_counting} to the second is
performed by collecting all contributions to the $n$-th trace that
stem from the same primitive periodic orbit such that $n= r n_{
{p}}$. After a resummation over the repetitions $r$ in the second line
of \eqref{eq:tf_counting} the 
oscillatory part of the counting function can be expressed as
\begin{equation}
 N^{\mathrm{osc}}(k)=
 \mathrm{Im}\,\frac{1}{\pi}
 \sum_{ {p}}\,\mathrm{ln}\left(
 1-
 \mathcal{A}_{ {p}} \E^{\I  {L}_{ {p}}
 (k+\I\epsilon)+ \I  \Phi_{ {p}}}\right)\ .
\end{equation}
It should be noted, that these period-orbit sums are not
absolutely convergent due to the exponential proliferation of
periodic orbits. That is, the number of periodic orbits of period
$n$ grows exponentially in $n$. Absolute convergence is only
obtained for complex $k$ beyond an entropy barrier
$\mathrm{Im}\,k> \epsilon_\mathrm{crit}$. For real $k$ the limit
$\epsilon\rightarrow 0$ implies that the infinite sum over
periodic orbits is ordered with respect to the length $r L_p$ of
the periodic orbit.

The derivative of \eqref{eq:tf_counting} with respect
to $k$ leads to the, equally exact, trace formula
\begin{equation}
  \begin{split}
    d^{\mathrm{osc}}(k)=&
    \mathrm{Re}\,
    \sum_{ {p}} \;
    \sum_{r=1}^\infty
    \frac{L_{ {p}}}{\pi}
    \mathcal{A}_{ {p}}^r
    \E^{\I r {L}_{ {p}} (k+\I\epsilon) + \I r \Phi_{ {p}}}
    \\
    =&
    \sum_{ {p}}
    \sum_{r=1}^\infty
    \frac{L_{ {p}}}{\pi}
    |\mathcal{A}_{ {p}}|^r
    \cos r(L_{ {p}} k +\Phi_{ {p}}+
    \sigma_{ {p}})
    \E^{- L_{{p}} \epsilon}
    \\
    =&
    \mathrm{Re}\,
    \sum_{ {p}} \,
    \frac{L_{{p}} }{\pi}
    \frac{
      \mathcal{A}_{ {p}}
      \E^{\I  {L}_{ {p}}(k+\I\epsilon) +\I
      \Phi_{ {p}}}
    }{
      1-\mathcal{A}_{ {p}}
      \E^{
    \I  {L}_{ {p}}(k+\I\epsilon) +\I
    \Phi_{ {p}}
      }
    }
  \end{split}
  \label{eq:tf_DOS}
\end{equation}
for the oscillatory part of the density of states. Instead of the
sum over primitive periodic orbits and its repetitions, one may
combine the two sums in a single sum over \emph{all} periodic
orbits $d^\mathrm{osc}(k)=\mathrm{Re}\, \sum_{p} \frac{L_p
\mathcal{A}_p}{\pi  r_p} \E^{\I L_p k}$. In that case
$\mathcal{A}_p$ is the amplitude of the full orbit, and $L_p$ its
full length while the primitive length is $L_p/r_p$ where $r_p$ is
the repetition number of the periodic orbit. Note, that the
classical probability to stay on a periodic orbit
\eqref{eq:classical_amplitude} is just the absolute square of the
quantum amplitude
\begin{equation}
  W_p=|\mathcal{A}_p|^2\ .
\end{equation}
Comparing the trace formula \eqref{eq:tf_DOS} to the
Gutzwiller trace formula reveals a complete analogy.

\subsection{Trace formulae for the eigenphase spectrum}

The oscillatory parts of the spectral functions
$\tilde{N}(\phi;k)$ and $\tilde{d}(\phi;k)$
for the eigenphase spectrum of quantum evolution map at fixed
wavenumber $k$ can be written in terms of primitive periodic
orbits in an analogous way.
This leads to the trace formulae
\begin{equation}
  \tilde{N}^\mathrm{osc}(\phi;k)=-
    \mathrm{Im}\,
    \sum_{{p}} \,
    \sum_{r=1}^\infty
    \frac{1}{\pi \, r} \mathcal{A}_{{p}}^r
    \E^{\I r
    \left(
      {L}_{{p}} k + \Phi_{{p}}
      - r n_{{p}}
      (\phi-\I \epsilon)
    \right)}
  \label{eq:counting_phases}
\end{equation}
and
\begin{equation}
  \tilde{d}^\mathrm{osc}(\phi;k)=
    \mathrm{Re}\,
    \sum_{{p}} \,
    \sum_{r=1}^\infty
    \frac{n_{{p}}}{\pi } \mathcal{A}_{{p}}^r
    \E^{\I r \left(
      {L}_{{p}} k + \Phi_{{p}}- n_{{p}}
    (\phi-\I \epsilon)\right)}\ .
\end{equation}

\section[The length spectrum and the quan\-tum-to-classical duality]{The length spectrum and\\ the quan\-tum-to-classical duality}
\label{sec:length_spectrum}

The trace formula \eqref{eq:tf_DOS} for the density of states may be written
in the form
\begin{equation}
  d(k)=  \mathrm{Re}\,\INT{-\delta}{\infty}{L}\ \mathcal{A}(L)
  \E^{\I k
  L}
\end{equation}
where $\delta\rightarrow 0$ will always be implied, and
\begin{equation}
  \mathcal{A}(L)=\frac{B\overline{L}}{\pi}
  \delta(L)+\frac{1}{\pi}\sum_{p}\sum_{r=1}^\infty L_p 
  \mathcal{A}^r_p \E^{\I r \Phi_p}
  \delta(L-r L_p)
\end{equation}
is a sum over primitive periodic orbits and their repetitions
plus a zero length contribution
(related to the Weyl term)
that is proportional to the length $B\overline{L}$ of the graph.
The complex function
$\mathcal{A}(L)$
is a weighted sum over $\delta$-functions
located at the length spectrum $\{L_p\}$ (and their repetitions).
This reveals a quantum-to-classical
duality between the length spectrum $\{L_p\}$
and the spectrum $\{ k_n \}$ of the quantum graph.
The relation may also be inverted
\begin{equation}
  \mathcal{A}(L)=\frac{1}{\pi}\INT{-\delta}{\infty}{k}\, d(k)
  \E^{-\I k L} = \frac{1}{\pi}\sum_{k_n} \E^{-\I k_n L}\ .
\end{equation}

The length spectrum is highly degenerate. This follows from the
fact that the length of any periodic orbit of period $n$
is an integer combination of
the bond lenths 
\begin{equation}
  L_p=\sum_{b=1}^B q_b L_b \qquad q_b=0,1,2,\dots ,
\end{equation}
with $\sum_{b=1}^B q_b=B$.
Periodic orbits with the same length differ only
in the order by which the bonds are traversed.
Note, that not every combination of integers $q_b$
is consistent with the connectivity of the graph. 
The degeneracy class will be denoted by the
vector $\{\vec{q}\}$ where $ \vec{q}=(q_1,\dots,q_B)$. For star graphs
the degeneracies are maximal, only even $q_b$ are
allowed by the connectivity without any further restriction.

The presence of the exact `quantum-to-classical' duality
imprinted by the relation between the \define{length spectrum}
$\{L_p\}$
and the quantum spectrum $\{k_n\}$ has been used efficiently
in the context of isospectrality. In a variant of
Kac' famous question `Can one hear the shape of a drum?'
\cite{kac:drums}
one may ask the question if there are topolgically
or metrically different quantum graphs
which have the same spectrum $\{k_n\}$ \cite{below:2001}. 
Such graphs are called
\define{isospectral}. Kac' question can be reformulated as
`Can one hear the shape of a graph?'. Exploiting the
quantum-to-classical duality an affirmative
answer has been proven in \cite{gutkin:2001} 
if the class of graphs is restricted to simple
topologies and rationally independent bond lengths.
In contrast, if the restriction of rationally independent bond lengths
is dropped, a lot of examples of isospectral
graphs are known today. Still, isospectrality is not understood
in sufficient detail. Recently, it has been asked \cite{talia}
how one may resolve
isopectrality for quantum graphs by adding non-trivial
additional information such
as nodal counting \cite{gnutzmann:2004}.

\section{The secular function is a $\zeta$-function}
\label{sec:Riemann_Siegel}

Let us set $t_{p}=\mathcal{A}_{p} \E^{\I k L_{p}}$
for a primitive orbit $p$.
For a fixed primitive orbit $p$ the sum over
repetitions $r$ in the trace formula for the
counting function
\eqref{eq:ppo_amplitude} can easily be performed
$\sum_{r=1}^\infty  t_{p}^r/r=-
\mathrm{ln}(1-t_{p})$. Summing over all
primitive periodic orbits and comparing to
\eqref{eq:tf_counting} then leads to the exact expression
\begin{equation}
  \zetab(k+\I \epsilon)=\prod_{p}
  \left(1-t_{p}\right)
  \label{eq:zeta_po}
\end{equation}
for the
secular function, a truly remarkable identity.
The expression on the left hand side is defined as
the determinant of $\ONE-\evol_B(k)$. As such, when
expanded in periodic orbits it has contributions from
irreducible periodic orbits only\cite{akkermans:2000}. A superficial
look on the right hand side could lead to the conclusion
that it involves all possible multiples of periodic
orbits and their repetions. This apparent paradox
is resolved by observing that there is an 
exact cancellation mechanism which is responsible for
the identity of the left and right hand sides.
From this follows that a truncation of the infinite product
or an approximation of any of the terms $t_p$ leads
to systematic errors and therefore \eqref{eq:zeta_po}
has no practical advantage for graphs.
This contrasts the situation in the analogous semiclassical
trace formulae where shadowing of periodic orbits
is the analogue of the exact cancellations mentioned above.
Nevertheless, for pedagogical reasons mainly, we shall investigate \eqref{eq:zeta_po} further.

The similarity to the Euler product formula for
the Riemann $\zeta$-function $\zeta_\mathrm{Riemann}(s)=
\prod_{\text{primes:}\, n} \left(1-\frac{1}{1-n^s}\right)$
justifies the letter $\zeta$ by which the secular function is denoted.
The product over primitive periodic orbits in
the secular function corresponds to a product over primes in
the Riemann $\zeta$-function.
Like the periodic-orbit sum the
infinite product on the right hand side of \eqref{eq:zeta_po}
is absolutely converging only for complex $k$
beyond the entropy barrier. For real $k$ the product is defined
by analytic continuation. Note that a complex $k$ with positive
imaginary part, orders the product
by the length of the periodic orbit. Thus this ordering is implied in the analytic continuation. Ordering the contributions by the period
$n$ instead of the length is essentially equivalent. Other non-equivalent orderings generally diverge, or worse: they converge to a wrong result
(with false zeros \cite{keating:1987}).
Note, that generally
$|t_p|<1$ for a periodic orbit (if we exclude the line, that is
the trivial graphs with $B=1$). Thus the zeros of the secular function
which define the spectrum of the graph \emph{are not equivalent}
to zeros of the factors $1-t_{p}(k)$ (which have zeros only
for complex $k$ with negative imaginary part). Instead the zeros
of the secular equation for real $k$ are due to the infinite product
over primitive periodic orbits. 

Let us now come back to the exact cancellations on
the right hand side of \eqref{eq:zeta_po} which eventually 
reduce it to the finite polynomial on the left hand side.
One principle behind these cancellations which
is known as \define{shadowing} may
be understood from a closer look at the the
expression $t_{{p}}$ for reducible primitive orbits. Let
$p_a$
and $p_b$ be two irreducible primitive orbits
with codes $a=\overline{\alpha P_a}$
and $b=\overline{\alpha P_b}$
which have exactly one directed bond in common.
Here, $P_{a}$ and $P_{b}$ are the codes for two paths
that close the periodic orbits such that $P_a$
and $P_b$ never visit the same directed bond.
For the reducible primitive periodic orbit $p_{ab}$
with code $ab=\overline{\alpha P_a \alpha P_b}$
one gets the identity $t_{ab}=t_a t_b$
\cite{akkermans:2000}.
Longer reducible primitive orbits may be obtained
from $p_a$ and $p_b$. With an obvious short-hand
notation for the codes, $aab$ and $abb$ are the
only possibilities to create reducible primitive
orbits composed of three irreducible orbits
(with four irreducible orbits one has $aaab$, $aabb$, and
$abbb$; for
five $aaaab$, $aaabb$, $aabab$, $aabbb$, $ababb$, and $abbbb$).
Performing the product over the primitive orbits $\hat{p}$ that
are composed of $p_a$ and $p_b$ one
gets
\begin{equation}
  \begin{split}
    \prod_{\hat{p}} (1-t_{\hat{p}})
    =&(1-t_a)(1-t_b)(1-t_{ab})(1-t_{aab})(1-t_{abb})\dots\\
    =&(1-t_a-t_b+t_at_b)(1-t_at_b)(1-t_a^2t_b)
    (1-t_a t_b^2)\dots\\
    =&(1-t_a-t_b+t_a^2t_b+t_at_b^2+t_a^2t_b^2)
    (1-t_a^2t_b)(1-t_a t_b^2)\dots
  \end{split}
\end{equation}
In the third line all contributions of second order
in $t_a$ and $t_b$ have cancelled. It is
obvious at this stage that the third order
terms will cancel after
the next two factors have been multiplied. Eventually
all higher order terms cancel and one gets the simple
result $\prod_{p} (1-t_p)= 1 - t_a - t_b$ which is
a finite polynomial that only depends
on reducible primitive orbits
(and its zeros are not zeros of one factor
$1-t_p$)\footnote{One may prove, that each finite product
over factors $1-t_{{p}}$ equals $1-t_a-t_b +R$
with $R=\mathcal{O}(t^n)$
if all contributions from periodic orbits that are composed of less than $n$ irreducible periodic orbits appear in the finite
product. It is a signature of the bad convergence of the infinite
product that the modulus of the remainder $|R|$ for real $k$
in generally does not go to zero in the limit of large $n$.}.
If more irreducible orbits are taken
into account a direct multiplication gets much more
involved but the exact {shadowing} 
of reducible periodic orbits remains valid.
Eventually, when all factors $1-t_{{p}}$ are expanded
and the shadowing principle is used to replace
reducible orbits by products of irreducible orbits
the final result
only depends on the \emph{finite} set of irreducible
orbits and their amplitudes $t_{{p}}$. Note, that all these
orbits have a period smaller than (or equal to) the dimension $2B$
of the quantum map\footnote{The final result can also be obtained
with Newton's formulas which express the coefficients $a_n$ of
a secular determinant $\mathrm{det}(z-A)=\sum_{n=0}^N a_n z^n$ through
the first $N$ traces $\mathrm{tr} A^n$ in a recursive way.}.
A more systematic treatment of approximations to
the secular function for calculating the spectrum from
a finite set of primitive periodic orbits
is obtained
within the cycle (or curvature) expansion \cite{webbook,Book:gaspard}.
The cycle expansion also holds for the Ruelle $\zeta$-function  $\zeta_R(z)$ \eqref{eq:Ruelle}
that describes the classical dynamics on the graph.


Cycle expansions are also used
efficiently for semiclassical expansions of 
$\zeta$-functions in Hamiltonian quantum systems \cite{berry:1990d}. 
As soon as semiclassical approximations are used
the analytical properties of the $\zeta$-function
are destroyed.
Also, replacing long orbits by products of short orbits (also
called pseudo-orbits in this context) is a
semiclassical approximation in these systems. For quantum graphs
the expansion of the secular function into products of irreducible
periodic orbits is exact due to the finite dimension 
of the quantum evolution map -- that is, due to the discreteness of
phase space.   \newpage
\thispagestyle{empty}  
\chapter{Spectral statistics}
\label{chapter:spectral_statistics}

Any respectable statistical theory starts by 
specifying the space of
variables over which the probability 
distribution (measure) is defined,
and where the statistical tests and conclusions are drawn. 
Here we would
like to study the spectrum of a single graph or 
of an ensemble of
graphs, from a statistical point of view. To  remain 
within the range of
acceptable respectability we should at least 
state what is our space of
variables.
In other words, what are we averaging over to get a statistical
(probabilistic) description.
The statistical approach to the spectra of graphs can 
be developed at
several levels. The conceptually simplest level is to 
consider the
ensemble of all the graphs which share the same connectivity 
but whose
bond lengths, and possibly also vertex scattering matrices 
are drawn
randomly according to prescribed probability 
distribution functions.
This approach is very similar to disorder averaging common e.g., in the
study of mesoscopic systems. The less obvious approach is 
the  study of
the spectral statistics of a  \emph{single} graph. Here, one can 
consider
non overlapping  spectral domains and study the distribution 
of  the
spectral intervals within each of the domains, assigning 
to the domains
equal probabilities. This approach, which is reffered to as 
\define{spectral
averaging}, makes sense only if one can prove 
that 
the probabilities converge to a well
defined \emph{ limit distribution}
as the spectral
domains and their number  increase. 

\section{Spectral averages}
\label{sec:spectral_averages}

The spectral average over some interval
$k_0 \le k \le k_0+K$ 
of a spectral function
$f(k)$ of a quantum graph is defined by
\begin{equation}
  \langle f(k) \rangle_k =
  \frac{1}{K}
  \INT{k_0}{k_0+K}{k}\ f(k) \ . 
\end{equation} 
In the simplest case one chooses $f(k)=d(k)$
and calculates the mean density of
states 
\begin{equation}
  \langle d(k) \rangle_k = d^\mathrm{Weyl} + 
  \frac{1}{K} 
  \big(N^\mathrm{osc} (k_0+K)
  - N^\mathrm{osc}(k_0)\big).
\end{equation}
Since $|N^{\mathrm{osc}}(k)|\le B$ one may neglect
the second term if $ K \gg
\frac{\pi}{\overline{L}}$. Thus the spectral average
of the density of states equals the Weyl term if
the spectral interval is chosen larger than the
inverse mean bond length (equivalently, when the number of states in that interval is large compared to the dimension
of the quantum evolution map). In most cases we will
take an infinite spectral interval $K \rightarrow \infty$ 
for averaging. For finite $K$ the result explicitly 
depends on $k_0$.
For small $K < 1/d^\mathrm{Weyl}\equiv \overline{\Delta}$ 
where $\overline{\Delta}$ is the \define{mean level spacing}
the $\delta$-peaks 
are broadened but generally
do not overlap. More interesting is the regime
$\overline{\Delta}\ll K < \frac{\pi}{\overline{L}}$ where
pronounced oscillations
modulate the Weyl term.  
Replacing the periodic orbit sum for $N^\mathrm{osc}$ one sees
that the most pronounced of theses oscillations
$\sim \E^{\I L_p k_0}$
stem from the short periodic orbits with length $L_p$. 

Some spectral averages remain very spiky functions even after
the spectral averaging. A simple example for this is the
correlator
\begin{equation}
  \begin{split}
  \langle d(k) d(k+k')\rangle_k =& 
  \left\langle \sum_{m,n} \delta(k-k_n)\delta(k'+k_n-k_m)
  \right\rangle_k\\
  =&
  d^{\mathrm{Weyl}} \delta(k')+ \lim_{K\rightarrow \infty}
  \frac{1}{K}\, \sum_{n\neq m}^{ k_n,k_m <K}
  \delta(k'+k_n-k_m)
  \end{split}
  \label{eq:two-point}
\end{equation}
where we have taken the average over the complete spectrum
and assumed that there are no degeneracies.
In the limit $K\rightarrow \infty$ the second part
becomes an infinite sum over $\delta$-functions with
a weight $\sim 1/K$ which is often equivalent (in a distributional sense)
to some well-behaved function. 
Both in numerical and in analytical approaches an additional
average over a short interval in $k'$ is usually needed
to make a well-behaved function apparent.
In chaotic systems the interesting features of such correlators
occur on the scale of the mean level
spacing $\overline{\Delta}= \frac{1}{d^\mathrm{Weyl}}=\pi/B\overline{L}$ 
and the
additional average over $k'$ has to be on a much shorter
scale.
Equivalently, one may replace the $\delta$-functions by
a Lorentzian or Gaussian function of width $\Delta k'$.
The Lorentzian is natural when the trace formula
is used.
The addition of an imaginary part $\I \epsilon$ to
the wavenumber results in replacing the $\delta$-functions
in the density of states
by Lorentzians of width $\epsilon$.

In practical (numerical) calculations one can circumvent
the difficulties encountered in direct application
of equation \eqref{eq:two-point} by dividing the spectrum
of length $K$ into $M$ subsequent intervals.
For each of the intervals one can use \eqref{eq:two-point}
replacing the $\delta$-functions by Lorentzians or Gaussians
of finite width.
The average over all the 
intervals is a well-behaved numerical approximation.
As a rule of thumb one should use $M\approx \sqrt{K d^{\mathrm{Weyl}}}$.

\section{Disorder averages}
\label{sec:disorder_average}

In the study of spectral statistics averaging over 
an ensemble of systems, usually referred to as `disorder average',
is a usefull theoretical tool which is naturally called
for in many applications. Disorder averaging for graphs
can be implemented in several ways. The most obvious one
is to take a family of graphs with the same connectivity
but with a random distribution of bond lengths. Another
possible variable to average over are the vertex scattering
matrices which can be picked at random from some well-defined
ensemble.
Both methods were used in past works and we illustrate
this method in Section \ref{sec:pot_andreevstars}.
For graphs there is an intimate connection
between disorder and spectral averages which will be explained
in the next section.

\section{Phase ergodicity for incommensurate graphs}
\label{sec:phase_ergodicity}

All spectral functions $f(k)$
that we are going to consider depend on $k$ via
the $B$ phase factors $\E^{\I k L_b}$ in the 
bond propagator $\bondprop (k)$. They are thus
quasiperiodic functions of $k$.
Let us again assume that the bond lengths $L_b$
are \emph{incommensurate}.
Under this condition the flow
\begin{equation}
  k \mapsto \left\{\E^{\I L_1 k},\E^{\I L_2 k},\dots,\E^{\I L_B 
  k}\right\}
  \in T^B
\end{equation}
covers the $B$-dimensional torus $T^B$ ergodically. 
With $\phi_b(k) \equiv k L_b$ and modest conditions on the
function $f(k)\equiv f(\E^{\I \phi_1(k)},\dots,\E^{\I \phi_B(k)})$
we have thus the \emph{exact} equality \cite{kac:ergodicity,barra:2000}
\begin{equation}
  \begin{split}
  \langle f(k)\rangle_k=&\lim_{K\rightarrow \infty}
  \frac{1}{K}\INT{0}{K}{k}\ f(\E^{\I \phi_1(k)},\dots,\E^{\I 
  \phi_B(k)})\\
  =&\frac{1}{(2\pi)^B}\int \D^B \phi\ 
  f(\E^{\I \phi_1},\dots,\E^{\I \phi_B})\\
  \equiv & \langle f(\E^{\I \phi_1},\dots,\E^{\I 
  \phi_B})\rangle_\phi\ .
  \end{split}
  \label{eq:equiv_spectral_phase}
\end{equation}
Thus the spectral average $\langle \cdot \rangle_k$ is equivalent to
a phase (or torus) average $\langle \cdot \rangle_\phi$. We will
make extensive use of this equivalence in the following
section and in Chapter \ref{chapter:susy}.

If only a subset of bond lengths is incommensurate one may 
define ergodic flows on tori of smaller dimension and establish an
equivalence of the phase average with phase averages over these tori.
We refer to the literature \cite{barra:2000} for
details and the application to quantum graphs. 

The equivalence \eqref{eq:spectral_discrete_ff}
between spectral statistics
and eigenphase statistics in large
graphs on one hand and the equivalence between a spectral
average and a phase average \eqref{eq:equiv_spectral_phase}
on the other hand also gives
a connection between the spectral statistics of an individual quantum 
system (with the spectrum $\{k_n\}$) and the spectral
statistics of an \emph{ensemble} of unitary matrices 
$\evol(k)=\scattering \bondprop(k)\mapsto 
\scattering \bondprop(\vec\phi )$
parameterised by $B$ parameters $\phi_b$. 
Thus the spectral statistics of a large graph
(with incommensurable bond lengths and moderate bond length fluctuations)
is equivalent to that of a (special) ensemble of unitary random matrices.

\section{Spectral correlation functions}
\label{sec:correlators}

Let us now define spectral correlation functions. We have already shown
that the spectral average of the density of states reduces to the
Weyl term. Spectral correlations thus appear in the oscillating part.
We define the \define{$n$-point correlation function} as
\cite{Book:mehta}
\begin{equation}
  R_n(s_1,\dots,s_{n-1})=\overline{\Delta}^n
  \langle d^{\mathrm{osc}}(k)d^{\mathrm{osc}}(k+s_1\overline{\Delta})
  \dots d^{\mathrm{osc}}(k+s_{n-1}\overline{\Delta}) \rangle_k
\end{equation}
where the average is over the whole spectrum $0\le k<\infty$. 
In this definition we anticipated
that we will be interested mainly in correlations on the
scale of the mean level spacing $\overline{\Delta}$. 
The corresponding length scale
is the \define{Heisenberg length} $L_H=\frac{2\pi}{\overline{\Delta}}=2 B \overline{L}$
and the corresponding period, the \define{Heisenberg period} is
$n_H=2B$ which is just the dimension of the quantum evolution map.

The two-point correlation
function $R_2(s)$ 
and its Fourier transform, the \define{spectral form factor}
\begin{equation}
  K(\tau)=\INT{-\infty}{\infty}{s} 
  R_2(s)\E^{-2\pi \I 
  s \tau}
\end{equation}
will be at the centre of much of the discussion
in chapters \ref{chapter:periodic_orbit}
and \ref{chapter:susy}.
Though $\tau$ and $s$ are dimensionless quantities 
that have been obtained from rescaling a length and a wave number we
will often follow the convention to call them ``time'' and
``energy'' -- $\tau=1$ corresponds to the Heisenberg time and
$s=1$ to the mean level spacing.

As mentioned above, spectral correlation functions 
(and their Fourier transforms)
are not self-averaging quantities. With the original definition
of the density of states it is obvious that they are usually a sum
over $\delta$-functions.
An additional
average over a small window $\Delta s \ll 1$ 
in the correlators
or $\Delta \tau \ll 1$ in the form factor has to be performed
to smooth the function. A combination of a spectral
average and a subsequent average over $\tau $ or $s$
will be denoted  by $\langle \cdot \rangle_s$ or 
$\langle \cdot \rangle_\tau$.

Instead of 
correlations in the spectrum $\{ k_n \}$ of the graph one may also
be interested in the correlations in the eigenphases of the quantum map
$\evol(k)$ averaged over $k$. The corresponding correlation functions
\begin{multline}
  \tilde{R}_n(s_1,\dots,s_{n-1})=\left(\frac{\pi}{B}\right)^n\times\\
  \langle \tilde{d}^{\mathrm{osc}}(\phi;k)\tilde{d}^{\mathrm{osc}}(\phi
  +s_1 \frac{\pi}{B};k)
  \dots \tilde{d}^{\mathrm{osc}}(\phi+s_{n-1}
  \frac{\pi}{B};k)\rangle_{k}
\end{multline}
are defined analogously\footnote{The $k$-average destroys any dependence on
the value $\phi$ which can thus be set to $\phi=0$ without loss of generality.}. 
For the spectral form factor of the quantum map
this reduces to
\begin{equation}
  \tilde{K}(\tau)=\frac{1}{2B}\sum_{n=1}^\infty 
  \delta\left(|\tau|-\frac{n}{2B}\right)
  \tilde{K}_{n}
  \label{eq:ff_qmap}
\end{equation}
where 
\begin{equation}
  \tilde{K}_{n}=\frac{1}{2B}\left\langle|\mathrm{tr}\, \evol(k)^n
  \right|^2\rangle_k\ .
  \label{eq:ff_discrete}
\end{equation}
The $\delta$-functions in \eqref{eq:ff_qmap} can be smoothed by an 
additional time average over an interval 
$\Delta \tau > \frac{1}{2B}$
\begin{equation}
  \begin{split}
  \left\langle \tilde{K}(\tau=\frac{n}{2B}) \right\rangle_\tau
  \equiv & \frac{1}{\Delta \tau}
  \INT{\tau-\Delta\tau/2}{\tau+\Delta \tau/2}{\tau'}\, 
  \tilde{K}(\tau')\\
  =& \frac{1}{2 B \Delta\tau}\sum_{n': |n'-n|< B \Delta\tau}
  \tilde{K}_{n'}\equiv \left\langle \tilde{K}_n \right\rangle_n \ .
  \end{split}
\end{equation}

\section{Equivalence of spectral correlators and eigenphase 
correlators in large graphs}
\label{sec:periodic_orbit_univ}

For large graphs with moderate bond length fluctuations
spectral and eigenphase correlation functions are 
equivalent. 
We will now discuss this statement 
in a periodic-orbit approach to the two-point
correlation function of a quantum graph with incommensurate bond
lengths.
Replacing $d^\mathrm{osc}$ by the sum over periodic orbits
\eqref{eq:tf_DOS} and performing the spectral average one obtains
\begin{equation}
  R_2(s)=\sum_{p, p^{\prime}}\sum_{r,r^{\prime}=1}^\infty 
  \delta(r L_p, r^{\prime }L_{p^{\prime}})
  \frac{\overline{\Delta}^2 (r_p L_p)^2
  \E^{- 2 r_p L_p \epsilon} }{2 \pi^2 r_p r_{p^{\prime}}} 
  \mathrm{Re}\, \mathcal{A}_p^r {\mathcal{A}_{p^{\prime}}^{r^{\prime}}}^* 
  \E^{\I \overline{\Delta} r_p L_p s}\ .
\end{equation}
Here, the Kronecker
$\delta(r_p L_p ,r_{p^{\prime}}L_{p^{\prime}})$ 
restricts the sum over pairs of periodic
orbits to pairs of the same (full) length 
$r_p L_p = r_{p^{\prime}}L_{p^{\prime}}$.   
This result does not depend on the assumption of incommensurate
bond lengths. The latter has the additional implication that
each bond is visited the same number of times by a pair of
periodic orbits with the same length (thus the full periods $r_p n_p$
and $r_{p^{\prime}}n_{p^{\prime}}$ are also the same). 
Using the notation introduced in Section \ref{sec:length_spectrum}
the two-point correlator of a quantum graph with incommensurate bond lengths
can be written as
\begin{equation}
  R_2(s)=\sum_{\{\vec{q}\}}
  \frac{\overline{\Delta}^2 L_{\vec{q}}^2
  \E^{- 2 L_{\vec{q}} \epsilon} }{2 \pi^2} 
  \sum_{p,p'\in \{\vec{q}\}}
  \frac{1}{r_p r_{p^{\prime}}}
  \mathrm{Re}\, \mathcal{A}_p^r {\mathcal{A}_{p^{\prime}}^{r^{\prime}}}^*
  \E^{\I \overline{\Delta} L_{\vec{q}} s}
\end{equation}
where $\{\vec{q}\}$ is the degeneracy class of periodic
orbits of length $L_{\vec{q}}\equiv \sum_{b=1}^B q_b L_b$.
Interference only takes place between the amplitudes of
periodic orbits in the same degeneracy class. 

The spectral form factor (for $\tau > 0$) 
\begin{equation}
  K(\tau)=\sum_{ \{\vec{q} \}} \tau^2\,\E^{- 2 L_{\vec{q}}\epsilon}
  \delta\left(\tau - \frac{\overline{\Delta} L_{\vec{q}}}{2\pi}\right)
  \sum_{p,p^{\prime}\in \{\vec{q}\}}
  \frac{1}{r_p r_{p'}} 
  \mathcal{A}_p^r {\mathcal{A}_{p^{\prime}}^{r^{\prime}}}^* 
  \ .
\end{equation}
is a sum over $\delta$-functions located at the lengths
of the periodic orbits -- it is thus directly related to
the length spectrum discussed in Section \ref{sec:length_spectrum}.

Finally, 
\begin{equation}
  \tilde{K}_n=
  \sum_{\{\vec{q}\}: \sum_b q_b=n\;\;}
  \frac{n^2\,\E^{- 2 L_{\vec{q}} \epsilon}}{2B }
  \sum_{p,p^{\prime}\in \{\vec{q}\}} 
  \frac{1}{ r_p r_{p^{\prime}}}
  \mathcal{A}_p^{r_p} {\mathcal{A}_{p'}^{r_{p^{\prime}}}}^* 
  \label{eq:periodic_orbit_formfactor}
\end{equation}
is the periodic orbit expression for 
the discrete time form factor obtained from the
Fourier transform of the two-point eigenphase correlator.
Here, the sum 
includes only pairs of orbits of full period
$n=r_p n_p= r_{p^{\prime}} n_{p^{\prime}}\equiv\sum_{b=1}^B q_b$.

An additional time average over a small interval
reveals the equivalence of the
discrete time form factor $\tilde{K}_n$
and the spectral form factor $K(\tau)$ 
for large graphs. To this end, let us write the
spectral form factor as a sum 
\begin{equation}
  K(\tau)=\sum_{n=1}^\infty K_n(\tau)
\end{equation}
where $K_n(\tau)$ contains only those periodic orbits of 
full period $n=n_p r= n_{p^{\prime}} r^{\prime}$.
Next, let us discuss the distribution of the quantity
$\tau_p=\overline{\Delta} r_p L_p/ 2 \pi$ 
(this is the metric length in units of the Heisenberg 
length $L_H=2\pi/\overline{\Delta}$, 
or the traversal ``time'') for
periodic orbits with period $n$. If $n\gg 1$ one may invoke
the central limit theorem 
and set
$\tau_p= n \overline{\Delta} \overline{L}/(2\pi) + \delta \tau_p$
where $n \overline{\Delta} \overline{L}/(2 \pi)$ is the mean 
traversal time. The fluctuations  $\delta \tau_p$ vanish in the mean
over all periodic orbits and have the variance 
$\langle \delta \tau_p^2\rangle_p=  n\overline{\Delta}^2\, 
\overline{\Delta L^2}/(2\pi)^2$ where $
\overline{\Delta L^2}= (\sum_b (L_b -\overline{L})^2)/B$
is the variance of bond lengths. We consider periodic orbits
of large period in a large
graph $n,B \rightarrow \infty$ where $\frac{n}{2B}$ is constant.
Additionally we assume moderate bond length fluctuations in the
sense that $\frac{\overline{\Delta L^2}}{\overline{L}^2}$ remains constant
when $B\rightarrow \infty$.
Thus $K_n(\tau)$ contributes to $K(\tau)$ in a small
time window of width $\sim \sqrt{n}/B \rightarrow 0 $ 
around $\tau= n/2B$. The overall
contribution of $K_n(\tau)$ to $K(\tau)$ can be calculated by the
integral 
\begin{equation}
  \begin{split}
  \INT{0}{\infty}{\tau}\, K_n(\tau)=&
  \sum_{\{\vec{q}\}:\sum_b q_b=n\;\;} 
  \tau_p^2 \E^{-2 L_{\vec{q}} \epsilon}
  \sum_{p,p'\in \{\vec{q}\}} 
  \frac{1}{r_p r_{p^{\prime}}}
  \mathcal{A}_p^{r_p} {\mathcal{A}_{p^{\prime}}^{r_{p^{\prime}}}}^*\\
  =&\frac{1}{2B}\tilde{K}_n\left(1+\mathcal{O}\left(\frac{n}{B^2}
  \right)\right)\ .
  \end{split}
\end{equation}
As a consequence, with a suitable time average over an interval
$\Delta \tau \sim B^{-1/2}$ the spectral form factor is 
well approximated by
the discrete time form factor (averaged over periods which correspond to the time interval) for large graphs and $\tau\gg 1/2B$
\begin{equation}
  \langle K(\tau)\rangle_\tau = \langle \tilde{K}_n \rangle_n.
  \label{eq:spectral_discrete_ff}
\end{equation}

This proves the above statement that the  
statistics of the spectrum
$\{k_n\}$  and of the spectrum of eigenphases of the
quantum map are equivalent for large graphs 
(if bond length fluctuations are moderate).
The discrete time form factor $\tilde{K}_n$
does not explicitly depend on the bond lengths. 
It only depends on the degeneracy classes of
periodic orbits which have the same lengths.
All incommensurate choices of the bond lengths
lead to the same degeneracy classes $\{\vec{q}\}$
and thus to the same value for $\tilde{K}_n$.
As a consequence the 
spectral form factors $\langle K(\tau)\rangle_\tau$ for different
incommensurate choices of the bond lengths are equivelant for
large graphs. 
This is
a first signature of universality: all large graphs with
the same vertex scattering matrices but different incommensurate choices
of bond lengths share the same spectral statistics.

\section[The level spacing distribution for quantum graphs]{The level spacing distribution\\ for quantum graphs}
\label{sec:level_spacing}

The level-spacing distribution $P(s)$ is one of the most well-known
signatures of quantum chaos. For a graph with the ordered spectrum
$\{k_n\}$, $k_{n+1}\ge k_n$  it is defined by
\begin{equation}
  P(s)=\lim_{N\rightarrow \infty} \frac{\overline{\Delta}}{N}\sum_{n=1}^N 
    \,\delta_\epsilon
      \left(\overline{\Delta}s-(k_{n+1}-k_n)\right)
\end{equation}
where $\delta_\epsilon(x)$ is some continuous approximation to a $\delta$-function with width $\epsilon$. 
The limit $\epsilon \rightarrow 0$ is implied
at the end after $N\rightarrow \infty$ and we will omit $\epsilon$ in the following. 

The level-spacing distribution depends on spectral correlators
of all orders. For this reason it is desirable to express
the level spacing distribution of a quantum graph directly. 
Such an expression has been derived
by Barra and Gaspard \cite{barra:2000}
and we will outline their approach in this section.

The starting point is the observation that
the  spectrum is given by the (positive) zeros
of a quasi-periodic function $f(k)$. For every graph there are many
such functions. For example, the secular function $f(k)=\zetab(k)$ will
do the job but $\zetab(k) \alpha(k)$ where $|\alpha(k)|>0$ will
equally be valid. We will not specify the choice of $f(k)$ in the
following and the approach can be generalised 
beyond quantum graphs to any spectrum
which is given by such a function. 
By definition quasi-periodicity
implies that there is a function $\tilde{f}(\phi)\equiv\tilde{f}(\phi_1,\dots,\phi_B)$ 
of $B$ variables $\phi=(\phi_1,\dots,\phi_B)$ 
which is periodic in each argument
\begin{equation}
  \tilde{f}(\phi_1,\dots,\phi_b+2\pi,\dots,\phi_B)=
  \tilde{f}(\phi_1,\dots,\phi_b,\dots,\phi_B)
\end{equation}
such
that 
\begin{equation}
  f(k)=\tilde{f}(\phi_1=L_1 k,\dots,\phi_B=L_B k)
\end{equation}
where $L_1,\dots,L_B$ is a set of rationally independent frequencies.
In the case of quantum graphs these frequencies are the bond lengths\footnote{
If not all bond lengths of the quantum graphs are rationally independent
the number of variables has to be reduced to some number
smaller than $B$.
The generalisation of the following argument is straight forward. However,
we will keep a notation that implies incommensurability of all bond lengths.}.
For the secular function $\zetab(k)$ of a graph 
quasiperiodicity is easily seen since it 
depends on $k$ only through the
bond propagator $\bondprop (k)$. Replacing
$L_b k \mapsto \phi_b$ in $\bondprop(k)\mapsto \bondprop(\phi_1,\dots,\phi_B)$
leads to a periodic function $\tilde{\zeta}_B(\phi_1,\dots,\phi_b)$ which shows
that $f(k)=\zetab(k)$ is a quasiperiodic function. 
We have already encountered this in Section
\ref{sec:phase_ergodicity}. 
The variables $\phi_b$ live on a B-dimensional torus $\mathbbm{T}^B$ and
\begin{equation}
  \Phi^k(\phi)=(\phi_1 + k L_1,\dots,\phi_B+kL_b)
\end{equation}
defines an ergodic flow in `time' $k$ on this torus. When the initial
condition of the flow is set to $\phi_1^{(0)}=\dots=\phi_B^{(0)}=0$
one has
\begin{equation}
  f(k)=\tilde{f}\left(\Phi^k(\phi^{(0)})\right).
\end{equation}
Thus the intersections of the flow with the Hyperplane $\Sigma$
defined by
\begin{equation}
  \tilde{f}(\phi)=0
\end{equation}
give the spectrum. Denoting a point on the $B-1$-dimensional
hypersurface $\Sigma$ by $\chi$ the flow defines a Poincar\'e map
\begin{equation}
  \begin{split}
    \chi_{n+1}=&\tilde\Phi(\chi_n)\\
    k_{n+1}=&k_n+ \tau_R(\chi_n)
  \end{split}
\end{equation}
where $\chi_n$ is a point on $\Sigma$ that is mapped by the flow to
$\chi_{n+1}$ on $\Sigma$. The two points are the successive intersections
of the trajectory $\phi_b=k L_b$  with $\Sigma$ at times $k_n$ and $k_{n+1}$.
The difference $\tau_R(\chi_n)\equiv k_{n+1}-k_n$ is the first 
return time to the
surface of section. 

We can now write the level spacing distribution as
\begin{equation}
  \begin{split}
  P(s)=&
  \lim_{N\rightarrow \infty}\frac{\overline{\Delta}}{N}\sum_{n=1}^N\,
  \delta(\overline{\Delta}s-\tau_R(\chi_n))\\
  =&
  \lim_{N\rightarrow \infty}\frac{\overline{\Delta}}{N}\sum_{n=1}^N\,
  \delta\left(\overline{\Delta}s-\tau_R\left(\tilde\Phi^{n-1}(\chi_1)\right)
  \right)
  \end{split}
\end{equation}
As a consequence of ergodicity this distribution is independent of the
initial conditions $\phi_b^{(0)}=0$ of the flow.
Thus, the initial conditions
can be taken anywhere on the torus without changing the limiting distribution
for $N\rightarrow \infty$. Moreover, ergodicity implies that the initial
condition can be taken anywhere on the surface $\Sigma$ and
that there exists an appropriate
measure $\D^{B-1}x\, \nu(x)$ 
such that
\begin{equation}
  P(s)=\int_\Sigma \D^{B-1} x \, \nu(x) \overline{\Delta}\, 
  \delta\left(\overline{\Delta}s-\tau_R(\chi(x))\right)\ ,
\end{equation}
where $x=(x_1,\dots,x_{B-1})$ is a
set of $B-1$ variables that parameterise $\chi(x) \in \Sigma$.

We now turn to the calculation of the function $\nu(x)$ that determines
the invariant measure on the hypersurface. 
A good starting point is the known invariant measure on the complete
torus. That is for every (measurable) function $g(\phi_1,\dots,\phi_B)$
on the torus ergodicity implies
\begin{equation}
  \lim_{K\rightarrow \infty} 
  \frac{1}{K} 
  \INT{0}{K}{k}\,
  g(\Phi^k(\phi^{(0)}))=
  \frac{1}{(2\pi)^B}
  \int \D^B \phi\, g(\phi)\ .
  \label{eq:torus_ergodicity}
\end{equation}
where $\frac{1}{(2\pi)^B}\D^B \phi$ is the invariant measure on the
torus implied by the homogeneous ergodic flow.
Defining $\tau[\Phi^k(\phi^{(0)})]$ as the time of flight
after the last intersection  with the surface of 
section $\Sigma$ (thus $\tau[\Phi^k(\phi^{(0)})]=k-k_n$ if
the last intersection happened at time $k_n$) 
and setting $\phi^{(0)}=\chi_1$\footnote{This implies a shift
$k \mapsto k-k_1$ without changing the
time difference between the last intersection. Thus, $\tau[\Phi^{k-k_1}(\chi_1)]=k-k_n$.}
to the first point of
intersection we replace the function
$g(\phi)$ by
\begin{equation}
  g[\Phi^{k-k_1}(\chi_1)]=\overline{\Delta}
  \Theta(\overline{\Delta}s-\tau[\Phi^{k-k_1}(\chi_1)])
  \sum_{n=1}^\infty \delta(k-k_n).
\end{equation}
Let us first show that the left hand side of \eqref{eq:torus_ergodicity}
is the cumulative function
of the level spacing distribution, that is
\begin{equation}
  \begin{split}
  I(s)=&\INT{0}{s}{s'}\, P(s')\\
  =&
  \lim_{K\rightarrow \infty}\frac{\overline{\Delta}}{K}
  \INT{0}{K}{k}\,\Theta(\overline{\Delta}s- \tau[\Phi^{k-k_1}(\chi_1)])
  \sum_{n=1}^\infty \delta(k-k_n)\ .
  \end{split}
\end{equation}
We assume that there are $N$ intersections at times $k_2,\dots,k_{N+1}$
of the trajectory $\phi^k(\chi_1)$
with $\Sigma$ in the interval $0<k\le K$. The integral over $k$ then gives
\begin{equation}
  \begin{split}
  \INT{0}{K}{k} g[\Phi^{k-k_1}(\chi_1)]=&\sum_{n=1}^N \overline{\Delta}
  \Theta(\overline{\Delta}s - (k_{n+1}-k_n))\\
  =&
  \sum_{n=1}^N \overline{\Delta}
  \Theta(\overline{\Delta}s - \tau_R[\tilde{\Phi}^{n-1}(\chi_1)])\ .
  \end{split}
\end{equation}
For large values one may replace $K=N \overline{\Delta}$ and arrives
at
\begin{equation}
  I(s)=\lim_{N\rightarrow\infty}\frac{1}{N}
  \Theta(\overline{\Delta}s - \tau_R[\tilde{\Phi}^{n-1}(\chi_1)])
\end{equation}
which shows that $I(s)$ is indeed the cumulative function of
the level spacing distribution.

For the calculation of the right hand side of \eqref{eq:torus_ergodicity}
one first needs to rewrite the sum over $\delta$-functions such that
it depends explicitly on the point $\phi$ on the torus.
Using that the values $k_n$ are given as the zeros of $f(k)$
and thus
\begin{equation}
  \sum_{n=1}^\infty \delta(k-k_n)=\left|\frac{\D f(k)}{\D k}\right|
  \delta(f(k)).
\end{equation}
Now  $f(k)=\tilde{f}(\Phi^{k-k_1}(\chi_1))$ and
$\frac{\D f}{\D k}= \sum_{b=1}^B \frac{\partial \tilde{f}}{\partial \phi_b} L_b$
lead to 
\begin{equation}
  \begin{split}
  g[\Phi^{k-k_1}(\chi_1)]=&\overline{\Delta}\,\Theta(\overline{\Delta} s-
  \tau[\Phi^{k-k_1}(\chi_1)]) \times\\
  &\left|\sum_{b=1}^B\frac{\partial 
  \tilde{f}(\Phi^{k-k_1}(\chi_1))}{\partial \phi_b} L_b \right|
  \delta(\tilde{f}(\Phi^{k-k_1}(\chi_1)))\ .
  \end{split}
\end{equation}
As a next step we change the variables $\phi\mapsto(x,\tau)$ 
where the $B-1$ variables
$x=(x_1,\dots,x_{B-1})$ parameterise
the surface of section $\Sigma$ for $\tau=0$ and $\tau$ 
is the time of flight of the flow
since the last intersection with $\Sigma$.
That is
\begin{equation}
  \phi_b=L_b \tau + s_b(x)
\end{equation}
where the $B$ functions $s_b$ satisfy $\tilde{f}(s_1(x),\dots,s_B(x))=0$
such that $\chi(x)\equiv (s_1(x),\dots,s_B(x))\in \Sigma$ 
In the new variables we have
$\tilde{f}(\Phi^{k-k_1}(\chi_1))\equiv\tilde{f}(x,\tau)$
and a measure
\begin{equation}
  \D^B\phi= \D^{B-1}x \D\tau\, J(x)
\end{equation}
with the Jacobean determinant
\begin{equation}
  J(x)=
  \begin{vmatrix}
    L_1 & \dots & L_B\\
    \frac{\partial s_1}{\partial x_1}& \dots 
    & \frac{\partial s_B}{\partial x_1}\\
    \dots & \dots &\dots\\
    \frac{\partial s_1}{\partial x_{B-1}}& \dots 
    & \frac{\partial s_B}{\partial x_{B-1}}
  \end{vmatrix}\ .
\end{equation}
The cumulative function becomes
\begin{equation}
  \begin{split}
  I(s)=&\frac{\overline{\Delta}}{(2\pi)^B}
  \int_\Sigma \D^{B-1}x\, J(x)\, \INT{0}{\tau_R[\chi(x)]}{\tau}\,
  \Theta(\overline{\Delta}s-\tau)\times\\
  &\left|
  \sum_{b=1}^B \frac{\partial\tilde{f}}{\partial \phi_b}L_b\right|
  \delta(\tilde{f}(x,\tau))\ .
  \end{split}
\end{equation}
The integral over $\tau$ can be performed explicitly by
introducing the new variable
\begin{equation}
  u(\tau)=\tilde{f}(x,\tau)
\end{equation}
where $x$ is fixed. With $\frac{\D u}{\D t}=\sum_{b=1}^B 
\frac{\partial \tilde{f}}{\partial \phi_b} L_b$ one
arrives at
\begin{equation}
  I(s)=\frac{\overline{\Delta}}{(2\pi)^B}\int_\Sigma \D^{B-1}x\
  J(x)\int du \Theta(\overline{\Delta}s - \tau(u))\delta(u)\ .
\end{equation}
The integral over $u$ picks up the value of $\tau(u)$ at
$u=0$. This has the two solutions $\tau=0$ and $\tau=\tau_R(\chi(x))$.
Since $\tau$ is the time of flight since the last intersection only
the second solution is valid and we finally arrive at
\begin{equation}
  \begin{split}
    I(s)=&\frac{\overline{\Delta}}{(2\pi)^B}\int_\Sigma \D^{B-1} x
    \, J(x) \theta(\overline{\Delta})\Theta(\overline{\Delta}s - 
    \tau_R(\chi(x)))\\
    =&\frac{1}{\int_\Sigma \D^{B-1} x J(x)}\int_\Sigma \D^{B-1} x
    \, J(x) \theta(\overline{\Delta})\Theta(\overline{\Delta}s - 
    \tau_R(\chi(x)))
  \end{split}
\end{equation}
where the last line follows from
\begin{equation}
  \frac{\overline{\Delta}}{(2\pi)^B}\int_\Sigma \D^{B-1}x J(x)=1\ .
  \label{eq:meanlevel}
\end{equation}
Equation \eqref{eq:meanlevel} is consistent with
the normalisation of the level spacing distribution
\begin{equation}
  P(s)=\frac{\D I(s)}{\D s}=
  \frac{\overline{\Delta}\int_\Sigma \D^{B-1} x J(x) 
  \delta(\overline{\Delta}s-\tau_R(\chi(x)))}{\int_\Sigma \D^{B-1}x J(x)}
  \label{eq:levelspacing_integral}
\end{equation}
and can be derived from
\begin{equation}
  1=\lim_{K\rightarrow \infty}\frac{\overline{\Delta}}{K}
  \INT{0}{K}{k} \sum_{n=1}^\infty \delta(k-k_n)
\end{equation}
in analogy to the above derivation.
The invariant measure on the surface of section can now be read off
easily as
\begin{equation}
  \D^{B-1}x\, \nu(x)= \D^{B-1}x\, \frac{J(x)}{\int_\Sigma \D^{B-1}x J(x)}.
\end{equation}

The integral \eqref{eq:levelspacing_integral} for a large quantum graph
is in general too complex to be performed analytically. 
Barra and Gaspard \cite{barra:2000}
have used this integral to investigate the level-spacing distribution of small graphs.
For large generic graphs approximations have to be used and it is not known
how to proceed from this expression to understand the universal behaviour
of generic large graphs that we will discuss in the remaining chapters.\newpage
\thispagestyle{empty}  
\chapter{Quantum chaos and universal spectral statistics}
\label{chapter:universality}

More than twenty years ago it has been observed
that spectral fluctuations in individual
complex (chaotic) quantum systems
are universal \cite{haq}. That is, all the spectral correlators 
defined in Chapter \ref{chapter:spectral_statistics}
(and also the
level spacing distribution) of any chaotic quantum system
are described by non-trivial system-independent functions.
Once rescaled by the (system dependent) mean level spacing, 
they only depend on some general symmetry properties
that follow the classification scheme of Wigner and Dyson
\cite{wigner:1958,dyson:1962,dyson:1962a,dyson:1962b}. This symmetry
classification scheme is based on the behaviour of the system under
a time-reversal operation. Systems either violate
time-reversal invariance (symmetry class $A$ in the notation
of \cite{zirnbauer:symmetry}) or they are time-reversal invariant.
In the latter case there are two symmetry classes which,
in fermionic systems, are realised by time-reversal invariant
dynamics with conserved spin (symmetry class $A$I) or
with broken spin rotational invariance (symmetry class $A$II).
For a general overview of symmetry classes and their algebraic properties,
see Appendix \ref{app:symmetry_classes}.
The spectral statistics in integrable systems is different
from the chaotic case. Spectral statistics
in integrable systems with at least
two degrees of freedom is generically Possonian \cite{berry:1977b}. 
The above statement of universal spectral
statistics is based on an overwhelming basis of
experimental and numerical evidence and has been
promoted to a conjecture for chaotic (hyperbolic)
Hamiltonian systems in the semiclassical regime
by Bohigas, Giannoni and Schmit \cite{bohigas:1984} (see also \cite{berry:1981,casati}). Since, proving the Bohigas-Giannoni-Schmit
conjecture and understanding the physical basis of 
this type of universality has been one of the major
challenges in the field of Quantum Chaos \cite{Book:Haake,Book:Stockmann,Book:Gutzwiller,Book:Ozorio,berry_LesHouches:1991}.
The dominant method that has been applied is semiclassical
periodic orbit theory based on Gutzwiller's trace formula
\cite{Book:Gutzwiller,gutzwiller:trace}
which expresses (the oscillatory part of) the
density of states as a sum over periodic orbits of the
corresponding classical Hamiltonian dynamics.
In spite of significant recent progress, universal spectral correlations
in \emph{individual} Hamiltonian systems are not fully understood.
Moreover, some counterexamples of classically chaotic quantum systems
have been identified which do not show universal spectral
statistics \cite{keating:1991a,bogomolny:1997,zakrzewski}.
While in these systems the strong deviation from universality
is fairly well understood it raises the question what are
the precise conditions to find universality in the
spectra of classically chaotic Hamiltonian systems.
Today, large quantum graphs is the only class of individual 
quantum systems for
which universality has been proven and precise sufficient
conditions can be stated \cite{gnutzmann:2004c,gnutzmann:2005}. We will give an outline of this proof later
in Chapter \ref{chapter:susy}.

\section[Universal correlators and random-matrix theory]{Universal correlators\\ and random-matrix theory} 

It has been known for some time that universality can 
be proven with an (\emph{additional}) average
over \emph{ensembles} of systems. The first successful
approach was random-matrix theory 
\cite{Book:mehta,guhr:1998,fyodorov:2004b} where a complex quantum system
is described by a hermitian matrix with random entries.
Random-matrix theory has become an important
tool to predict spectral statistics, wave function statistics
and transport statistics in complex quantum systems (e.g. 
disordered mesoscopic systems) \cite{susy,beenakker:review,brouwer:diss}.
Wigner and Dyson \cite{wigner:1958,dyson:1962,dyson:1962a,dyson:1962b}
proposed three Gaussian random-matrix ensembles,
one for each symmetry class: the
Gaussian Unitary Ensemble (GUE, symmetry class $A$), the Gaussian Orthogonal Ensemble
(GOE, symmetry class $A$I) and the Gaussian Symplectic Ensemble 
(GSE, symmetry class $A$II).  
The universal spectral correlators for these ensembles have been calculated
analytically in the limit of infinite matrix dimension \cite{Book:mehta}.
See appendix \ref{app:random_matrix_theory} for a definition of the
ensembles and for their correlation functions. 
The success of random-matrix theory in predicting
universal correlators is reflected in the physical
literature where `universality' is frequently replaced
by `random-matrix behaviour'.

\begin{figure}[ht]
  \begin{center}
    \includegraphics[width=12cm]{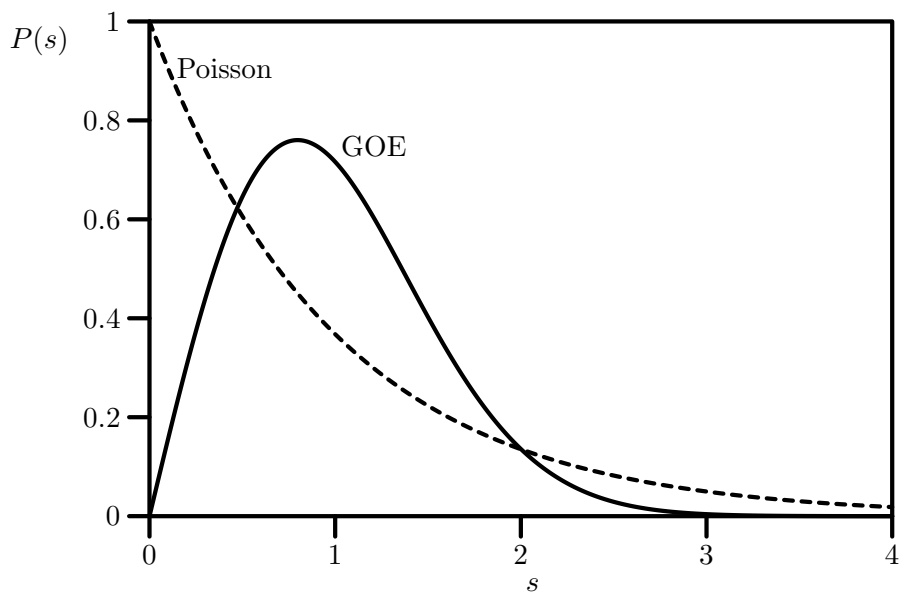}
  \end{center}
  \caption{Level spacing distributions of a complex quantum system
  (Wigner surmise for the GOE, full line) 
  and of an uncorrelated spectrum (Poisson
  distribution, dashed line).}
  \label{fig:levelspacing}
\end{figure}

If the spectrum of a quantum system is known (either by experimental
measurements or by a numerical calculation) a simple
way to see, if its spectral statistics follows the universal
predictions of random-matrix theory is to plot a histogram of
its level-spacing distribution against
the Wigner surmises (an energy interval with a few hundred eigenvalues is usually sufficient). 
The Wigner surmises \eqref{eq:Wigner_surmises} are the random-matrix
results for the level-spacing distributions of 
the Gaussian random-matrix ensembles of dimension $2\times 2$
(GUE and GOE) or $4\times 4$ (GSE). They only deviate very
slightly from the exact universal result for infinite matrix dimension
\cite{Book:Haake} and the difference can only be resolved
with very large data sets. 
In figure \ref{fig:levelspacing} the level spacing distribution
of a completely uncorrelated spectrum is plotted against the
Wigner surmise of the GOE. 
Uncorrelated spectra are
realisations of a Poisson
process. Poissonian eigenvalue statistics is
generic for integrable systems (with at least two freedoms)
\cite{berry:1977b,Book:Haake}
and the level spacing distribution is just an exponential
decay 
\begin{equation}
  P(s)= \E^{-s}\ .    
\end{equation}
The strong deviation of the Wigner surmise from the Poisson
distribution is a signature of the non-trivial
correlations in complex quantum systems. The most obvious
difference of the two distributions is that small values
of the level spacing are 
favoured in a Poisson spectrum (level clustering)
but are strongly suppressed in the Wigner surmise. This 
\define{level repulsion} in complex quantum systems can be understood
using a perturbative approach \cite{Book:Haake}.
Also large values of the level spacing occur much more frequently
in a Poisson spectrum than in a universal complex quantum system.
That is universal spectra are much more rigid than Poisson spectra.

\section{Analytical approaches to universality}

Apart from random-matrix theory there have been two
other successful attempts to understand universality for
\emph{ensembles} of quantum systems: a particle in a
random potential (disordered system) \cite{Book:Efetov,mirlin} and the 
so-called
Pechukas-Yukawa gas where the parametric dependence
of eigenvalues and eigenvectors
on a parameter $\lambda$ in a Hamiltonian
of the form $H= H_1 + \lambda H_2$ is investigated 
\cite{braun,pechukas,yukawa,Book:Haake}.
In spite of
their success and significance these studies 
(and also random-matrix theory)
cannot explain the deep quantum-to-classical correspondence
between spectral correlations in a Hamiltonian systems
and the dynamic properties of the underlying classical
flow. 

This correspondence is the main focus in quantum chaos
where semiclassical periodic-orbit theory
has been used
\cite{argaman:1993a,berry:1985a,berry:1990d,bogomolny:1996,cohen:1998,smilansky:2003}. 
We will not go into the details of the semiclassical
periodic orbit approach here but only summarise 
qualitatively some of 
the main results and ideas. For an introduction
we refer to textbooks on Quantum Chaos \cite{Book:Gutzwiller,Book:Haake,Book:Stockmann,Book:Ozorio}  
and to the cited literature.
The periodic-orbit
expansions for quantum graphs 
which we will present in the next chapter are in most parts 
analogous to semiclassical periodic orbit theory. 
While periodic-orbit theory on graphs is not always trivially
generalised to Hamiltonian systems, its main advantage
is that it allows for exact analytical treatments with
controlled approximations.

The semiclassical theory distinguishes between a number of relevant
time scales which have either classical or quantum mechanical
origin. The classical scale $t_{\mathrm{erg}}$
is determined by the inverse Lyapunov exponent. The shortest
quantum time scale is the Heisenberg time 
$t_H=2\pi \hbar / \overline{\Delta}$. 
This is the time scale on which the discreteness of the spectrum
is resolved.
In the semiclassical limit
$t_H \gg t_{\mathrm{erg}}$.

\begin{figure}[ht]
  \begin{center}
    \includegraphics[width=12cm]{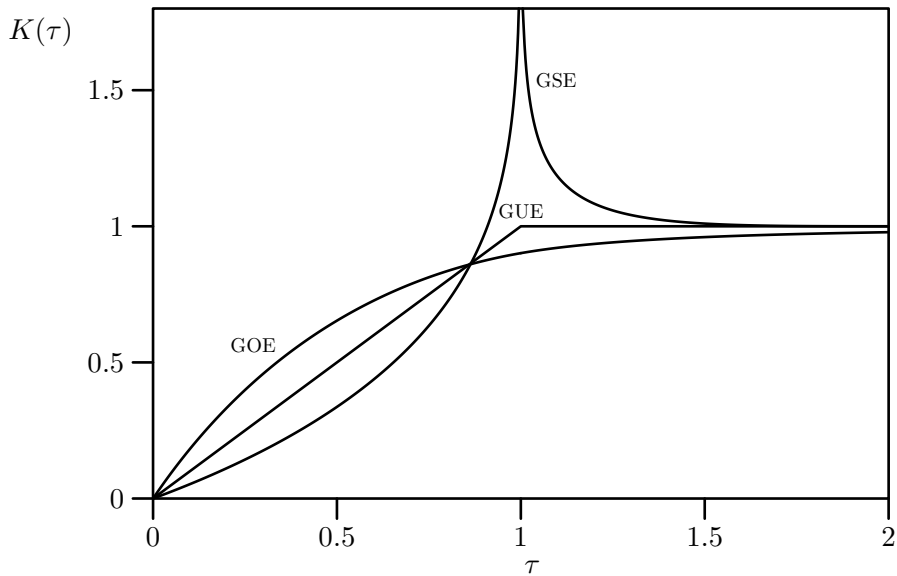}
  \end{center}
  \caption{The spectral form factors for the three universality classes
  described by the GUE, GOE, and GSE.}
  \label{fig:formfactor}
\end{figure}

The spectral form factor 
has been the main focus of semiclassical
periodic orbit theory of spectral statistics.
The universal spectral form factors for
the three Wigner-Dyson classes are shown
in Fig. \ref{fig:formfactor}. 
The non-trivial spectral correlations in the form factor
are apparent as deviations from $K^{\mathrm{Poisson}}(\tau)=1$.
 
Using Gutzwiller's trace formula \cite{gutzwiller:trace,Book:Gutzwiller}
the form factor $K(\tau)$ 
for a classically chaotic Hamiltonian system 
can be approximated
semiclassically as a sum over pairs of periodic
orbits with period $t_p=\tau t_H$
\begin{equation}
  K(\tau) \sim \sum_{p,p'} \frac{t_p^2}{r_p r_{p^{\prime}}}
  \mathcal{A}_p \mathcal{A}_{p^{\prime}}^*
  \label{eq:formfactor_semiclassical}
\end{equation}
where $\mathcal{A}_{p/ p^{\prime}}$ are amplitudes corresponding
to the periodic orbits and $r_{p,p^{\prime}}$ are repetition
numbers. The magnitude of $\mathcal{A}$
depends on the classical stability, and its phase on the classical
action of the orbit $p$. 

The 
contribution of 
periodic-orbits in the range $t_{\mathrm{erg}}\ll t_p\ll t_H$ 
to the spectral correlators 
is quite well understood and it reproduces the
universal form factor. 
In a seminal paper Berry 
\cite{berry:1985a} showed
that a reduction of the double sum in \eqref{eq:formfactor_semiclassical}
to pairs which are either the same or time reversed 
(the \define{diagonal approximation}) can be performed
on the basis of the sum rules \cite{ozorio:sumrule} 
and leads to an agreement with the linear universal behaviour 
of the form factor for short times.
Since, it has been a challenge to understand the next-to-leading order
in the short time expansion
\begin{equation}
  \begin{split}
    K_{\mathrm{GUE}}(\tau)=& \tau\\
    K_{\mathrm{GOE}}(\tau)=& 2 \tau-2\tau^2 +\mathcal{O}(\tau^3) \\
    K_{\mathrm{GSE}}(\tau)=&
     \frac{\tau}{2}+\frac{\tau^2}{4}+\mathcal{O}(\tau^3) 
  \end{split}
  \label{eq:formfactor_short_time}
\end{equation}
of the form factor or to understand the long-time behaviour $\tau>1$.
Many interesting results on action correlations 
\cite{smilansky:2003,argaman:1993a,cohen:1998} in terms of 
so-called `pseudo-orbit' expansions have been
obtained \cite{berry:1990d,bogomolny:1996}. 
Sieber and
Richter made a crucial progress by identifying pairs of 
correlated periodic orbits 
whose contribution gives the next-to-leading
order term in \eqref{eq:formfactor_short_time} 
\cite{richter,sieber,sieber:2002} (see also
\cite{aleiner,aleiner:1997,aleiner:1997a}). 
This approach has been developed further
in a number of recent articles \cite{heusler:2004,muller:2003,turek,turek:2003,spehner}.
The present status of
semiclassical periodic orbit theory
singles out in a systematic way the pairs
of periodic orbits contribute to
progessivly higher terms in the short-time expansion
of the form factor \cite{muller:2004,muller:2005}. 
However, due to the essential singularity of the form factor
at the Heisenberg time $\tau=1$
this result cannot be extended beyond $\tau>1$.
Another drawback of the present theory
is that not all periodic orbits
are considered. An estimate
of the contribution of the omitted orbits to the form factor is not given.

There exists yet another attempt to explain
universality in individual chaotic quantum systems.
It relies on a mapping to a supersymmetric
field theory. 
It was hoped that 
these so-called `ballistic $\sigma$-models'
would allow a direct 
access to universality via their
mean field approximations 
\cite{agam:1995,agam:1997,andreev,muzykantskii}. 
However, there are serious problems
in regularising these theories, and so far they could only be established
under the protection of an additional average over (very weak) disorder
\cite{muller:2005a}.

For \emph{individual} quantum graphs periodic-orbit methods and methods
from random-matrix theory and disordered systems like mappings
to supersymmetric field theories can all be applied rigorously. 
The inherent disorder introduced
into the quantum graph by a fixed choice of rationally independent
bond lengths is sufficient to treat a quantum graph in a similar
way as an ensemble of disordered systems or as an ensemble of 
unitary matrices.
In the following chapters we will give a detailed account
of the periodic-orbit theory and its supersymmetric
counterpart to quantum graphs.   

\newpage
\thispagestyle{empty}
\chapter{Periodic orbit theory for spectral statistics}
\label{chapter:periodic_orbit}

In this chapter we describe the periodic-orbit theory for
the spectral form factor for large graphs. We will not try to give
the most general account but will restrict our discussion to star graphs.
While the results can be generalised
 to other topologies,
the main ideas can be explained in a clearer way
for star graphs, where one avoids a few technical problems 
that arise in the general case.
For the same reason we will only consider the
discrete version of the form factor which 
corresponds to two-point correlations in
the eigenphases of the quantum map.
As shown in Section \ref{sec:periodic_orbit_univ} eigenphase
correlations are equivalent to spectral correlations
for large graphs with bond lengths which are incommensurate
and which are confined to a moderately narrow
interval of lengths.   

The quantum dynamics on a star graph is not affected
by adding a magnetic field \eqref{eq:zetastar}. The only
way to violate time-reversal invariance is by choosing
a central vertex scattering matrix in an appropriate way.
Similarly one can add a spin degree of freedom which
enables us to study the resulting spectral statistics
and compare them with the predictions of random-matrix
theory. 

One of the important features of random-matrix theory
is that the statistics derived for ensembles of $N\times N$
matrices tend to well-defined limiting distributions in
the limit $N\rightarrow \infty$, provided that the
spectral parameters are properly scaled. In 
periodic-orbit theory the analogous limit is $t_H\rightarrow
\infty$ which is the semiclassical limit for Hamiltonian flows.
For graphs the limit $\hbar \rightarrow 0$ has no consequence.
However, one can make the Heisenberg time $t_H$ arbitrarily
large by increasing the number of bonds $B$. 
The disadvantage of graphs with respect to Hamiltonian flows
is that there is no way
to increase the number of bonds and at the same time
preserve the classical dynamics on the graph.
As a matter of fact 
there is no unique way to reach the limit $B\rightarrow \infty$. 
In the case of star graphs the limit $B\rightarrow \infty$
is topologically straight forward. However, one has to
prescribe a sequence of central vertex scattering matrices.
For this reason it is more appropriate to consider fixed
large graphs and expand the spectral functions as powers series
in $B^{-1}$. In the cases of Neumann or
DFT vertex scattering matrices which we study here as examples
one can perform the limit $B\rightarrow \infty$ in
a consistent way.

\section{Spectral form factor for star graphs}
\label{sec:stars}

One of the main technical advantages of star graphs
is the fact that one may reduce the dimension of the
quantum evolution operator from $2B$ to $B$.
This follows from the observation that
a wave which scatters into a bond is totally reflected
at the peripheral vertex.
This idea can be implemented in the following way.
The $2 B \times 2 B$ graph scattering matrix 
for star graphs has the form
\begin{equation}
  \scattering=
  \begin{pmatrix}
  0 & -\scattering_\star\\
  -\ONE_{B} & 0
  \end{pmatrix}
\end{equation}
where $\scattering_\star$ is the $B \times B$
\define{star scattering matrix}. 
We assume only dynamically connected star graphs for
which $\scattering_\star$ cannot be brought into block-diagonal form
by a permutation of bond indices.
It is equal to the
central vertex scattering matrix up to a minus sign which we introduce
for convenience. We have chosen Dirichlet boundary conditions at
the peripheral vertices. 
These appear as minus the identity matrix in
the graph scattering matrix.
The secular function of a star graph can be written as
\begin{equation}
  \zetab(k)=\mathrm{det}\left( 
  \ONE_{2B}-\evol(k)\right)=\mathrm{det}\left(
  \ONE_{B}-{\evol}_\star(k)\right) 
\end{equation}
where  
\begin{equation}
  {\evol}_\star(k)=\bondprop_\star(k)\scattering_\star\ .
\end{equation} 
with $\bondprop_\star(k)_{b,b'}=
\delta_{bb'}\E^{2\I k L_b}$ is the $B\times B$ star quantum
evolution map.
The star bond propagation matrix $T_\star(k)$ 
describes the phase accumulated during
the propagation from the centre to the peripheral vertices and back.
The minus signs picked up during the scattering at the peripheral vertices
are cancelled by the minus sign of the central
scattering matrix $- \scattering_\star$. Thus
the star quantum evolution map ${\evol}_{\star}(k)$ describes the process of scattering at the centre
and subsequent propagation along the bonds until coming back to the centre.
This way we have removed the direction index $\omega$ from the
description. 

The topoligical length $n$ of a trajectory  in this reduced picture
is equal to the number of peripheral vertices (or undirected bonds)
visited. This
corresponds to a length $2n$ in the general approach
based on directed bonds. Thus the Heisenberg period is now $n_H=B$.
The discrete form factor at time $\tau=\frac{n}{B}$ 
is 
\begin{equation}
  \tilde{K}_{\star,n}=\frac{1}{B}\langle |\mathrm{tr}\, 
  \scattering_\star^n|^2\rangle_k=
  \sum_{p, p'\in \mathcal{P}(n)} 
  \delta(r_p L_p,r_{p^{\prime}}L_{p'})
  \frac{n^2}{B \,r_p r_{p'}} 
  \mathcal{A}^{r_p}_p {\mathcal{A}^{r_{p^{\prime}}}_{p^{\prime}}}^* 
  \label{eq:star_formfactor}
\end{equation}
where a periodic orbit $p=\overline{b_1,b_2,\dots,b_{n_p}}$ 
is now specified by the sequence of undirected bonds $b=1,\dots,B$ and
its amplitude is the
product of the amplitudes of the star scattering matrix
$A_p=\scattering_{\star\, b_1,b_{n_p}}\dots\scattering_{\star\, b_3,b_2}
\scattering_{\star\, b_2,b_1}$. Another way of writing
the same expression is 
\begin{equation}
  \tilde{K}_{\star,n}=\sum_{\{ \vec{q}\}:\sum q_b=n} \frac{n^2}{B}
    \left|\sum_{p \in \{\vec{q}\} }
    \frac{1}{r_p } 
    \mathcal{A}^{r_p}_p\right|^2
    \label{eq:star_formfactor2}  
\end{equation}
where $\{\vec{q}\}$ is the degenercy class
of periodic orbits of length $L=2 \sum_{b=1}^\infty q_b L_b$.

A 
sufficient condition for a star graph 
to be  invariant under time-reversal  (symmetry class $A$I)
is that the star scattering matrix is symmetric
\begin{equation}
  \scattering_{\star}=\scattering_{\star}^T.
\end{equation}
Otherwise, time-reversal symmetry is in general
broken (symmetry class $A$). More precisely, any
star graph with a scattering matrix of the form $\scattering_\star=S_0 D$ where
$S_0$ is symmetric $S_0=S_0^T$ and $D=\mathrm{diag}(\E^{\I \beta_b})$
is time-reversal invariant. Under this condition 
every periodic orbit has the same amplitude as its time-reversed partner. 
Note also, that the secular equations for
a graph with scattering matrix $\scattering_\star=S_0 D$ and a
graph with the scattering matrix $\scattering'_\star=D^{1/2}S_0 D^{1/2}$
are completely equivalent. 

Time reversal invariant star graphs of symmetry class
$A$II can be constructed by adding a spin freedom to the
wave function \cite{gnutzmann:2004a}. This will be implemented in Section
\ref{sec:spin_GSE}.

We will use two examples of scattering matrices
for star graphs
to illustrate our results: \textit{i.} \define{Neumann star graphs}
\cite{berkolaiko:1999,berkolaiko:diss,berkolaiko:2001,berkolaiko:2003,berkolaiko,stargraphs} 
\begin{equation}
  \scattering_{\mathrm{Neumann} \star\, bb'}= \delta_{bb'}- \frac{2}{B}\ ,
\end{equation}
and \textit{ii.} \define{DFT star graphs}
\cite{gnutzmann:2004a,gnutzmann:2004b} 
\begin{equation}
  \scattering_{\mathrm{DFT} \star\, bb'}=
    -\frac{1}{\sqrt{B}} \E^{\I \frac{2\pi bb'}{B}}
\end{equation}
for which the
the central vertex scattering matrix is taken as the discrete Fourier transform matrix (see Section \ref{sec:vertexscat}). 
Both scattering matrices are symmetric,
thus both families of star graphs are time-reversal invariant. 
The main difference between the two families is revealed by comparing the
probability of backscattering $b \rightarrow b$ to the probability to
be scattered into any other bond $b \rightarrow b'\neq b$.
For DFT star graphs an incoming wave packet is scattered
into any bond with the same probability
$P_{\mathrm{DFT} \star\, b\leftarrow b}=
P_{\mathrm{DFT} \star\, b\leftarrow b'}=1/B$. For Neumann star graphs
with large $B$
back scattering $ P_{\mathrm{Neumann} \star\, b\leftarrow b}=(1-\frac{2}{B})^2$ is favoured, 
while scattering into any other bond has a small probability
$P_{\mathrm{Neumann} \star\, b\leftarrow b'}=4/B^2$. In the limit $B\rightarrow \infty$ this difference
has a strong impact on spectral statistics. While DFT star graphs
have the canonical universal spectral statistics 
of the GOE
the Neumann star graphs belong to
a different universality class \cite{berkolaiko:1999,berkolaiko:2001,stargraphs}.

\section{The diagonal approximation}
\label{sec:diagonal_approximation}

We shall start the discussion of the periodic-orbit
theory by considering the first leading term in the
short time expansion of the form factor $K(\tau)$.
This limit is defined by the requirement that
$\tau=n/B$ is
much smaller than the Heisenberg time
$\tau \ll \tau_H \equiv 1$. 

In \eqref{eq:star_formfactor2}
we have expressed
the form factor $\tilde{K}_{\star\, n}$ as a sum over 
degeneracy classes $\{\vec{q}\}$ of periodic orbits which share
the same length $L_{\vec{q}}$.
Each degeneracy class $\{\vec{q}\}$ gives a contribution
\begin{equation}
  \frac{n^2}{B}\left|\sum_{p\in\{\vec{q}\}} 
  \frac{\mathcal{A}_p}{r_p}\right|^2
  =\frac{n^2}{B}\left( 
   \sum_{p\in\{\vec{q}\}} \frac{|\mathcal{A}_p|^2}{r_p^2}
   +\sum_{p,p'\in\{\vec{q}\}: p\neq p'} 
   \frac{\mathcal{A}_p^*\mathcal{A}_{p'}}{r_p r_{p'}}
  \right)\ .
  \label{eq:degeneracy_class}
\end{equation}
If the phases of the amplitudes $\mathcal{A}_p$ were
completely random one would expect that the 
purely diagonal term which is the
first term
on the right hand side of \eqref{eq:degeneracy_class}
dominates the form factor.
However, the phases of the terms are not random and
the extraction of the form factor as a power series in $\tau$
amounts to unraveling the phase correlations in a systematic way.

Let us now consider the time-reversal invariant case where
$\scattering_\star=\scattering_\star^T$. In general each periodic orbit 
has then the same amplitude as its time-reversed
partner. Therefore time-reverse pairs add coherently to
the form factor which results in doubling the contribution of
the pure diagonal approximation.
This is not exactly true because there always exist
\define{self-retracing orbits}, for which
by definition 
the time-reversed partners are identical.

It is natural to devide the form
factor into diagonal and off-diagonal
contributions  
\begin{equation}
  \tilde{K}_{\star\, n} 
  =\tilde{K}_{\star\, n}^\mathrm{diag}+
  \tilde{K}_{\star\, n}^\mathrm{off-diag}
\end{equation}
where the diagonal part 
\begin{equation}
  \tilde{K}_{\star\, n}^\mathrm{diag}=\frac{2 n^2}{B}
  \sum_{p\in\mathcal{P}(n)} \frac{|\mathcal{A}^{r_p}_p|^2}{r_p^2} 
  - \frac{n^2}{B} 
  \sum_{p\in\mathcal{P}(n): p= \hat{p}}
   \frac{|\mathcal{A}_p^{r_p}|^2}{r_p^2}
  \label{eq:diag1}
\end{equation}
consists of all pairs $p,p' \in \mathcal{P}(n)$ 
of primitive periodic orbits 
that are either equal $p'=p$ or time-reversed to each other $p'=\hat{p}$. 
The second term is a sum over
 \define{self-retracing} periodic orbits which have been counted
 twice in the first term.
 
The off-diagonal
contribution 
\begin{equation}
  \tilde{K}_{\star\, n}^\mathrm{off-diag}=
  \frac{n^2}{B}\sum_{p,p'\in \mathcal{P}(n):p'\neq p,p'\neq
  \hat{p}} \delta(r_pL_p, r_{p^{\prime}}L_{p^{\prime}})
  \frac{\mathcal{A}_p^{r_p} 
  {\mathcal{A}_{p^{\prime}}^{r_{p^{\prime}}}}^*}{r_p r_p'}
\end{equation}
to the form factor amounts to all remaining pairs of periodic orbits.
It is responsible for the quantum interference effects 
which are not present
in the diagonal part. 

The diagonal approximation \cite{berry:1985a} 
consists of writing the form factor in terms of the diagonal
part only. This is justified for short
times $n\ll B$ because of the following argument.
For $n \ll B$ the majority
of  periodic orbits visit any bond at most once.
In that case only the orbits which contribute
to the diagonal approximation contribute
coherently to the sum irrespectively of the chosen
star scattering matrix $\scattering_\star$
and one may expect that the off-diagonal
terms are suppressed due to the remaining phase factors.

At larger times the probability to visit a bond twice is non-negligible
-- this probability grows proportional to the time for $n\ll B$.  
When a bond is visited  twice (or more often) some off-diagonal orbit
pairs have correlated phases and add up coherently. 
Eventually these give higher order corrections in $\tau=n/B$. 

We will come back to a discussion
of the off-diagonal part later
in Section \ref{sec:offdiagonal_contributions}. 
For the rest
of this section we will discuss the diagonal approximation
to the form factor in detail.
Note that
\begin{equation}
  \left|\mathcal{A}_p\right|^2=\prod_{j=1}^{n_p}
  \left|\scattering_{\star\, b_{j+1},b'_{j}}\right|^2
  =\prod \mathcal{M}_{\star\, b_{j+1}b_j}= W_p   
\end{equation}
is just the classical weight of a periodic orbit \eqref{eq:classical_amplitude}
for the classical evolution map
\begin{equation}
  \mathcal{M}_{\star\, bb^{\prime}}=
  \left|\scattering_{\star\, bb'}\right|^2\ .
\end{equation}
Replacing $\left|\mathcal{A}_p\right|^2\rightarrow
W_p$ in the diagonal part \eqref{eq:diag1} and comparing it
to the classical periodic orbit expansion
\eqref{eq:classreprb} of the return probability
$u_n=\mathrm{tr} \mathcal{M}^n$ 
one obtains
the identity
\begin{equation}
  \tilde{K}^{\mathrm{diag}}_{\star \, n}=
  \frac{2 n}{B} u_n
  -\frac{2 n^2}{B} \sum_{p\in \mathcal{P}(n): r_p\ge 2} 
    \frac{r_p-1}{r_p^2} W_p^{r_p}
  -\frac{n^2}{B} \sum_{p\in\mathcal{P}(n):p=\hat{p}} 
  \frac{1}{r_p^2} W_p^{r_p}
  \label{eq:diag_class}
\end{equation}
in terms of the classical weights. In the classical periodic
orbit expansion of the return probability
\eqref{eq:classreprb}
repetition numbers $r_p\ge 2$ have a different factor in front
of the weight $W_p^{r_p}$ compared to the form factor.
This difference is neglected in the first term 
but corrected in 
the second term of \eqref{eq:diag_class}.
Below we will show that only a very small fraction of orbits
$p\in \mathcal{P}(n)$ has $r_p\ge 2$.
The last summand in \eqref{eq:diag_class} 
is the sum over self-retracing
periodic orbits.

The first term in \eqref{eq:diag_class} is the graph analogue
of a general semiclassical relation which expresses
the short-time form factor as the product of
$2\tau$ with the classical probability to return
\cite{dittrich:returnprob,argaman:1993a}.

The total number of periodic orbits of period $n$ grows
exponentially $\propto B^n/n = \E^{\lambda_T n}/n$
with the toplogical entropy $\lambda_T = \mathrm{ln}\, B$ 
for a star graph. 
The number of periodic orbits $p\in \mathcal{P}(n)$
which are repetitions of shorter periodic orbits, that is
$r_p\ge 2$,  is only proportional to $B^{n/r_p} r_p/n \propto
\E^{\lambda_T n/r_p}r_p/n$. Though this number also grows
exponentially it is only an exponentially small fraction
of all periodic orbits. For self-retracing orbits a similar
argumentation is straight forward because one half of the
corresponding code is just the (time) reverse of the other half. 
Altogether the total number of
periodic orbits that contribute in the second and third summand
of \eqref{eq:diag_class} only grows as
$\E^{\lambda_T n/2}/n$
with half the topological entropy. 

Let us now use the classical sum rule \eqref{eq:classicalsum}
to estimate the diagonal approximation.
The sum rule states $u_n \rightarrow 1$ for periods
$n \gg n_{\mathrm{mix}}$. On the same time scale
the second and third part of \eqref{eq:diag_class} 
which contain the
contributions from repetitions of shorter orbits
and self-retracing orbits are exponentially small
because the  number 
of these periodic orbits 
cannot compensate the small
weight of the repeated orbit,
which is $W_p^{r_p} \propto (1/{B})^n$ in the mean.
The largest contributions to the second and third parts come
from the finite number of short primitive orbits $p$ with
a period smaller than the mixing time 
$n_p < n^{\mathrm{mix}}$. Their repetitions have a weight $W_p^{r_p} \lesssim \E^{-n/n^{\mathrm{mix}}}$ which can also be neglected.
In summary, a mixing classical evolution implies
the quantum mechanical sum rule
\begin{equation}
  \frac{B}{n} \tilde{K}_n^{\mathrm{diag}} \xrightarrow{n\rightarrow \infty} 2
  \label{eq:diag_sumrule}    
\end{equation}
for the diagonal part of the form factor for
time-reversal invariant quantum graphs.
The
deviations die out on the mixing time scale $n^{\mathrm{mix}}$.
This time scale can be replaced by the (in general shorter)
ergodic time scale if an additional time average over
a small interval $\Delta n$ is applied.

Since the diagonal approximation is only valid
for quantum mechanically short times $n\ll B$
the sum rule \eqref{eq:diag_sumrule}
can only be used effectively to predict
the form factor if there is a
 time scale separation such that
mixing sets in much faster than the
Heisenberg time $n^{\mathrm{erg}}\ll n_H=B$.
In that case there is a time
regime, $n^{\mathrm{erg}}/B \ll \tau \ll 1$, where
the spectral form factor  
\begin{equation}
  \langle K(\tau=\frac{n}{B}) \rangle_\tau \approx \langle \tilde{K}\rangle_n
  \approx 2 \tau
  \label{eq:diag_universal}  
\end{equation}
follows
the universal prediction from random-matrix theory up to
corrections of order $\mathcal{O}(\tau^2)$. 

In the limit of large graphs $B\rightarrow \infty$
at a fixed time 
$\tau=\frac{n}{B}$
a more detailed account is necessary because
the number of classical modes (which equals the number of bonds)
grows.
The
spectral gap $\Delta_g=1/n^{\mathrm{erg}}$ 
defined in \eqref{eq:spectral_gap} may also depend on $B$.
If the 
spectral gap is bounded from below by a constant $\Delta_g>\delta>0$
equation \eqref{eq:diag_universal} becomes exact for $B\rightarrow \infty$
for every fixed time $\tau$. 
If the spectral gap approaches zero more care has to be taken
for those classical modes with eigenvalues $\nu_\ell\rightarrow 1$
\cite{tanner:2001}. In that case the corrections 
\begin{equation}
  u_n -1=  \sum_{\ell=2}^B \ \nu_\ell^n 
\end{equation}
to the classical sum rule
\eqref{eq:classicalsum} have to be considered in more detail.
One may estimate these corrections by 
\begin{equation}
  \langle|u_n -1|\rangle_n \lesssim (B-1) (1-\Delta_g)^n \sim B \
  \E^{- \Delta_g n} 
\end{equation}
which shows that the corrections in the limit $B\rightarrow \infty$
can only be neglected if $\Delta_g B \rightarrow \infty$.
Otherwise deviations from the universal behaviour remain in the
diagonal approximation for the short-time form factor.
Based on this observation, Tanner conjectured \cite{tanner:2001} that
large quantum graphs have universal
spectral statistics (also beyond the diagonal approximation)
if the spectral gap either remains finite or decays
as 
\begin{equation}
  \Delta_g \sim B^{-\alpha} \qquad \text{with}
  \qquad 0\le \alpha <1
  \label{eq:gap_condition}
\end{equation}
in the limit $B\rightarrow \infty$.
We will discuss this
condition again in connection with the supersymmetry approach to 
quantum graphs in Chapter \ref{chapter:susy}. 
We shall conclude this general discussion with the following
remarks.\\
\noindent
\textit{i.} 
If the gap condition is violated, the contributions of repetitions and
self-retracing orbits 
\eqref{eq:diag_class}
to $\tilde{K}_n^{\mathrm{diag}}$
should be examined because there is no a priori reason
to neglect them. We shall give an example below.\\
\noindent
\textit{ii.}
The gap condition is consistent with the existence of the
time regime
$1 \gg \tau \gg n^{\mathrm{erg}} /B$.
It has been shown that the gap conditions
holds generically for large star graphs
\cite{berkolaiko:2001,zyczkowski:2003}.
 
Let us illustrate the general discussion by two examples --
the DFT star graph and the Neumann star graph.
The DFT star graph is known to
show universal spectral statistics \cite{gnutzmann:2004a} -- 
the classical map is
simply $M_{\mathrm{DFT}\star \, bb'}=1/B$ for which all eigenvalues apart
from $\nu_1=1$ vanish exactly. Thus the gap condition is fulfilled maximally. 
The Neumann star graph is an example with
spectral statistics which belongs to a universality class that is not described by the GOE \cite{kottos:1998,stargraphs,berkolaiko:1999}, here 
$M_{\mathrm{Neumann}\star\, bb'}=(4+\delta_{bb'}(B^2-4B))/B^2$ for which
all eigenvalues (except for $\nu_1=1$) have the same value
$\nu_\ell=(B-4)/B$  such that the gap is $\Delta_g=4/B$ which
does not fulfil the gap condition. 
Thus we also have to consider repetitions of short orbits and
self-retracing orbits.
In the case of Neumann star graph
it suffices to consider 
the repetitions of the shortest periodic orbits
with primitive period $n_p=1$. These are at the same time
self-retracing orbits.
Altogether one arrives at \cite{kottos:1998}
\begin{equation}
\begin{split}
  \tilde{K}_{\mathrm{Neumann}\star\, n}^\mathrm{diag}=&
  2\frac{n}{B} u_n +\frac{1}{B}
  \left(1-2n\right) 
  \sum_{b=1}^B (\mathcal{M}_{\mathrm{Neumann \star}\, bb})^n \\
  \longrightarrow&\;\E^{-4\tau}+ 2\tau\left(1-\E^{-4\tau}\right) 
  =1-4 \tau + \mathcal{O}(\tau^2)
\end{split}
\end{equation}
which shows that
the contributions from repetitions 
and self-retracing orbits dominate the short-time behaviour.

We will now study the form factor in the case
of a weakly broken time-reversal symmetry such that $\scattering_\star-\scattering_\star^T\neq 0$
is small. The contributions from time reversed pairs
to the diagonal part of the form factor aquire phases and
they do not add up coherently any more. 
In that case \eqref{eq:diag_class} is replaced by
\begin{equation}
  \tilde{K}_{\star\, n}\approx\tau\left(\mathrm{tr}\, 
  \mathcal{M}_\star^n + \mathrm{tr}\, 
  \mathcal{R}_\star^n \right) 
  \label{eq:diag_A}
\end{equation}
where
\begin{equation}
  \mathcal{R}_{\star\, bb'}=
  \scattering^*_{\star\, b'b}\scattering_{\star\, bb'}
  \label{eq:timerev_evol}
\end{equation}
and we have neglected the contributions from repetitions and self-retracing orbits. 
The first term in \eqref{eq:diag_A} 
is the contribution of equal pairs of periodic orbits.
For these the above discussion remains valid
and we may replace $\mathrm{tr}\,\mathcal{M}_\star^n\rightarrow 1$
for $n\gg n_{\mathrm{erg}}$ if we  
assume that the gap condition holds for
the classical map $\mathcal{M}_\star$.
The second term in \eqref{eq:diag_A}
is the contribution from time-reversed pairs of
periodic orbits. 
The matrix $\mathcal{R}_\star$ has generally complex entries.
This reflects the fact that broken time-reversal invariance leads
to destructive interference of counter-propagating orbits.
In the present context only the leading eigenvalue
the matrix $\mathcal{R}_\star$ has to be considered. This
can be estimated using perturbation theory.
As a result the contribution of the time-reversed
pairs decays exponentially
\begin{equation}
\mathrm{tr}\, \mathcal{R}^n 
\propto \E^{-n/n^{\mathrm{crit}}}
\end{equation}
 in $n$ on a time scale 
set by
\begin{equation}
  1/n^{\mathrm{crit}} \approx
  \frac{1}{B}\left|\mathrm{tr}\scattering_\star^*(\scattering_\star
  -\scattering_\star^T)\right|\ . 
\end{equation}
The inverse time scale $1/n^\mathrm{crit}$ is an obvious measure for
the violation of time-reversal invariance. 
For large graphs $B\rightarrow \infty$ and fixed time $\tau=n/B$
the contribution of counter-propagating orbits vanishes if
$1/n^{\mathrm{T}} \propto B^{\alpha^{\prime}}$ where
$0 \le \alpha^{\prime} < 1$. This leaves
\begin{equation}
  \langle{K(\tau)}^{\mathrm{diag}}\rangle_{\tau}
  =\langle K_{\star\, n}^\mathrm{diag}\rangle_n\rightarrow \tau
\end{equation}
which is the leading order of the universal result
\eqref{eq:formfactor_short_time} for broken time-reversal
invariance.

\section{Off-diagonal contributions}
\label{sec:offdiagonal_contributions}

The diagonal approximation only gives the leading order
in the short-time expansion of the form factor. We will now discuss
how universality is built up order by order when more and more
off-diagonal pairs of periodic orbits are included 
in the expansion of the form factor.
We will discuss the limit of large star graphs $B\rightarrow \infty$
at a fixed time $\tau=n/B\ll 1$. 
The leading off-diagonal correction to the diagonal approximation
for general large graphs has been given by Berkolaiko, Schanz, and Whitney
\cite{berkolaiko:2002} following the method by Sieber and Richter \cite{sieber,sieber:2002} 
for Hamiltonian systems.

For simplicity we will assume that the elements of 
the quantum map of the star graph are all
of the same order $|\evol(k)_{\star\, bb'}|\sim B^{-1/2}$ as
for example in the DFT star graph. As a consequence
the gap condition
\eqref{eq:gap_condition} holds with $\alpha=0$. With some
extra effort the following discussion can be generalised
to graphs where the gap condition holds with $0 \le \alpha <1$.

For time-reversal invariant graphs random-matrix theory
predicts a leading off-diagonal correction
$K(\tau)-K^{\mathrm{diag}}(\tau)=-2 \tau^2 +\mathcal{O}(\tau^3)$
to the diagonal
approximation of the form factor.
For broken time-reversal symmetry random-matrix theory predicts
that there are no corrections to the diagonal part
for short times $\tau<1$. We will start
with the time-reversal invariant case
and come back to broken time-reversal at the end.

A family of pairs of periodic orbits can conveniently
be described in terms of diagrams which show the geometry
of the two orbits. A diagrammatic language has first been introduced
in \cite{berkolaiko:2002,berkolaiko:2003a}. We will follow closely \cite{gnutzmann:2004a,gnutzmann:2004b}
where a variant which is suitable for star graph dynamics has been developed.
The art of finding the diagrams 
which contribute to a given order 
in $\tau$ to the form factor is significantly simplified by our assumptions.

To introduce diagrams let us remind
that a pair of periodic orbits $p$ and $p^{\prime}$ only contributes
to the (discrete) form factor $\tilde{K}_n$ if both orbits
have the same length $L_p= L_{p^{\prime}}$. Rationally independent
bond lengths imply that the two orbits visit the
same bonds with equal multiplicity in a permuted order. 
That is the code of one orbit is a permutation of the code
of the other orbit.
All periodic orbit pairs for which this permutation is equal
(up to a cyclic permutation) share a similar geometry which can be expressed
as a diagram. 

The diagrammatic language is most easily introduced
by first considering the $n$-th trace
\begin{equation}
  s_n=\mathrm{tr}\, \scattering_\star^n = \sum_{p\in \mathcal{P}(n)}
  \frac{n}{r_p} \mathcal{A}_p^{r_p}
\end{equation}
as a sum over periodic orbits of period $n$.
For a fixed period, say $n=4$, one may write this sum in terms
of the following diagram
\begin{equation}
 s_4=\sum_{b_1,b_2,b_3,b_4=1}^B \scattering_{\star\, b_1b_4}
 \scattering_{\star\, b_4b_3}\scattering_{\star\, b_3 b_2}
 \scattering_{\star\, b_2b_1}=
 \begin{array}{c}
  \includegraphics[width=2.5cm]{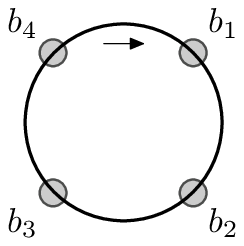}\ .
 \end{array}
\end{equation}
The diagram on the right hand side 
represents the sum over all periodic orbits
of period $n=4$. The sequence of bonds visited
by the periodic orbit in the star graph is represented
by a sequence of vertices in the diagram (these should not be confused
with the vertices of the graph) which is traversed in the direction
indicated by arrow. Each vertex in the diagram on the right hand
side
is translated on the left hand side 
as one summation index $b=1.\dots,B$ which is summed over.
Each
(directed) line $b \rightarrow b'$ is translated as a factor
$\scattering_{\star\, b'b}$. A diagram needs not be labelled
with the summation indices $b_j$ as above. We will do so
sometimes to facilitate the translation. 
In the diagrams for
the traces $s_n$ with higher period $n$
it is useful to define a line element for 
matrix elements of the $n$-th
power $\scattering_\star^n$ by writing the power $n$ explicitly
next to the line. This leads to
the diagram
\begin{equation}
  s_n =\sum_{b=1}^B \left(\scattering_\star^n\right)_{bb}=
  \begin{array}{c}
    \includegraphics[width=1.5cm]{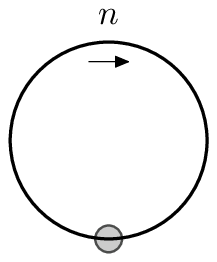}
  \end{array}
\end{equation}
which completes the introduction to diagrams for the trace $s_n$.

Now we shall introduce diagrams that
contribute to the form factor $\tilde{K}_n$. 
Instead of immediately giving
a full set of rules how to draw and translate diagrams for the off-diagonal
part we will first consider the diagonal part where the partner
orbits are either the same or time-reversed. 
Neglecting the overcounting of repetitions of shorter
orbits and the double counting of self-retracing orbits
the following two diagrams give the
dominant contribution to the diagonal approximation
of the form factor (here explicitly for $n=4$)
\begin{equation}
  \begin{split}
    \tilde{K}^{\mathrm{diag}}_4=&
    \begin{array}{c}
      \includegraphics[width=2cm]{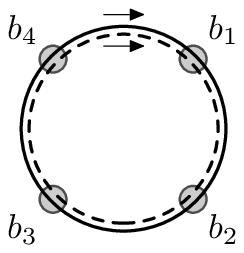}
    \end{array}
    +
    \begin{array}{c}
      \includegraphics[width=2cm]{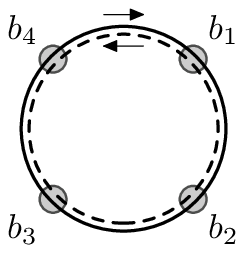}
    \end{array}
    \\
    =&
    \frac{4}{B}\!\!
    \sum_{b_1,b_2,b_3,b_4=1}^B\!\!\!\!\!
    \scattering_{\star b_1b_4}
    \scattering_{\star b_4b_3}
    \scattering_{\star b_3b_2}
    \scattering_{\star b_2b_1}\!\!
    \left(
    \scattering_{\star b_1b_4}
    \scattering_{\star b_4b_3}
    \scattering_{\star b_3b_2}
    \scattering_{\star b_2b_1}\right)^*+\\
    &
    \frac{4}{B}\!\!
    \sum_{b_1,b_2,b_3,b_4=1}^B\!\!\!\!\!
    \scattering_{\star b_1b_4}
    \scattering_{\star b_4b_3}
    \scattering_{\star b_3b_2}
    \scattering_{\star b_2b_1}\!\!
    \left(
    \scattering_{\star b_4b_1}
    \scattering_{\star b_3b_4}
    \scattering_{\star b_2b_3}
    \scattering_{\star b_1b_2}\right)^*\\
    =&
    \frac{8}{B}\!\!
    \sum_{b_1,b_2,b_3,b_4=1}^B\!\!\!\!\!
    \mathcal{M}_{\star b_1b_4}
    \mathcal{M}_{\star b_4b_3}
    \mathcal{M}_{\star b_3b_2}
    \mathcal{M}_{\star b_2b_1}\ .
  \end{split}
  \label{eq:diag_diag4}
\end{equation}
The first diagram 
represents the sum over all equal pairs and the second the
sum over time-reversed pairs of periodic orbits.
As before each vertex in the diagram is translated as a summation
index $b_j$ (the explicit labels in the diagram are not necessary).
A full line from vertex $b$ to $b'$ is translated as the corresponding
matrix element $\scattering_{\star b'b}$ and a dashed line
as its complex conjugate. 
If a full and a dashed line are parallel (or antiparallel)
one may translate both lines as one matrix element of
the classical evolution map $\mathcal{M}_{\star b'b}$ 
as in the last line \eqref{eq:diag_diag4}.
An overall prefactor $\frac{4}{B}$
is added where $1/B$ stems from the definition of the form
factor and $4$ is the number of cyclic permutations of one orbit
with respect to the other.

In general the two diagrams for the diagonal
approximation to the form factor $\tilde{K}_n^{\mathrm{diag}}$
has $n$ vertices. The translation rules for lines and vertices
are the same as in the example above and the prefactor
is $\frac{n}{B}$.
For drawing the diagrams it is useful to replace a long stretch
of parallel (antiparralel) full and dashed lines 
by a single
pair of parallel (antiparallel) lines and put a label
next to the line which gives the length of the stretch.
For the diagonal approximation this leads to the
diagrams
\begin{equation}
  \begin{split}
    \tilde{K}^\mathrm{diag}_{\star\, n}=& 
    \begin{array}{c}
      \includegraphics[width=2cm]{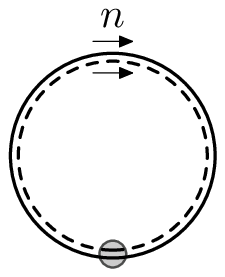}
    \end{array} 
    +
    \begin{array}{c}
      \includegraphics[width=2cm]{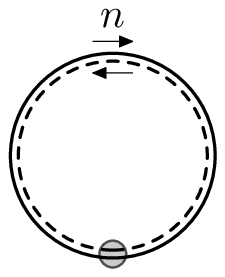}
    \end{array}\\
    =&
    \frac{2n}{B}
      \sum_{b=1}^B \left(\mathcal{M}_{\star}^n\right)^*_{bb}\ .
  \end{split}
  \label{eq:diagonal_diagrams}
\end{equation}
In general a stretch of parallel (antiparallel) lines
from vertex $b$ to $b'$ is translated as a matrix
element $(\mathcal{M}^n)_{\star b'b}$ of the $n$-th power
of the classical map.
Ergodicity sets in for $n\gg n^{\mathrm{erg}}$
and one may replace $(\mathcal{M}^n)_{\star b'b}\rightarrow \frac{1}{B}$
in the limit $n,B\rightarrow \infty$.
The corrections
due to classical decaying modes 
vanish exactly
in that limit and one obtains the form factor
$\tilde{K}^{\mathrm{diag}}_n \rightarrow 2\tau$ as predicted
by random-matrix theory.

We will now introduce diagrams that contribute to the
off-diagonal part of the form factor. The diagrams have
to be ordered by the power of $\tau$ to which they contribute
in the short-time expansion of the form factor. A lot of care has
to be taken when more and more diagrams are added to the form factor
since certain pairs of periodic orbits occur in more than one diagram.
Off-diagonal diagrams and the carefull considerations
that have to be taken when they are added to the form factor are most easily
introduced by considering the diagram 
\begin{equation}
  \begin{array}{c}   
    \includegraphics[width=7cm]{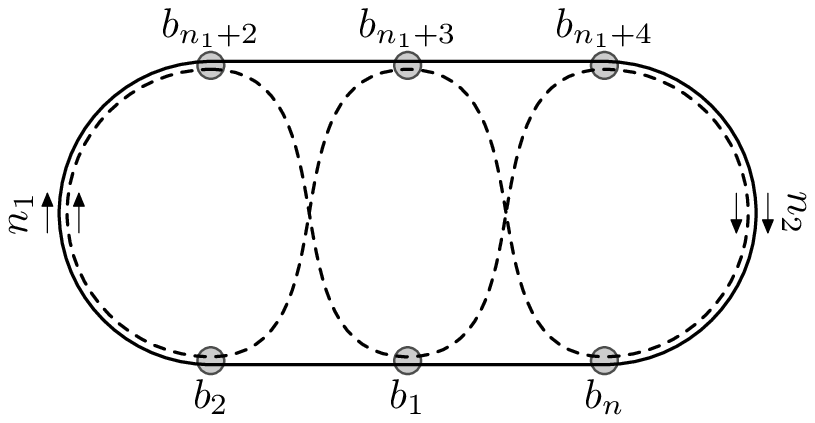}\ ,
  \end{array}
\end{equation}
as a pedagogical example. 

We will first describe how this diagram is
translated into a sum over pairs of periodic orbits. 
Both periodic orbits in the diagram have period $n=n_1+n_2+4$.
The full line is a periodic orbit
with code $\overline{b_1,\dots,b_n}$. The dashed line is
parallel to the full line along the long stretches of
length $n_1$ and $n_2$ on the left and right.
In the code of the partner orbit two indices have interchanged
$b_1 \leftrightarrow b_{n_1+3}$ and, for this reason, 
we will call this diagram a
\define{transposition diagram}.
The contribution of this diagram to
the discrete form factor $\tilde{K}_n$ is the sum over all
pairs of periodic orbits with the given geometry. For $n_1\neq n_2$
the diagram is translated
as the following sum
\begin{equation}
  \begin{split}
    \begin{array}{c}   
    \includegraphics[width=4cm]{./figures/transposition_diagram.eps}
    \end{array}=&\frac{n^2}{B}\!\!\sum_{
      \genfrac{}{}{0pt}{}{b_1,b_2,b_{n_1+2},}{b_{n_1+3},b_{n_1+4},b_n}}
    (\mathcal{M}^{n_1}_\star)_{b_{n_1+2}b_2} 
    (\mathcal{M}^{n_2}_\star)_{b_{n_1+4}b_n}
    \times\\
    &\scattering_{\star\,b_{n_1+4} b_{n_1+3}}      
    \scattering_{\star\,b_{n_1+3} b_{n_1+2}}
    \scattering_{\star\,b_2 b_1}      
    \scattering_{\star\,b_1 b_n}\times\\
    &\scattering_{\star\,b_{n_1+4} b_1}^*
    \scattering_{\star\,b_1 b_{n_1+2}}^*
    \scattering_{\star\,b_{2} b_{n_1+3}}^*
    \scattering_{\star\,b_{n_1+3} b_n}^*\ .
    \label{eq:transpos_diag}
  \end{split}
\end{equation}
Apart from an additional factor $n$ in the prefactor the translation
follows the same rules as in the diagonal approximation.
The additional factor $n$
is explained by the obeservation that interchanging
the indices $b_2 \leftrightarrow b_{n_1+4}$ in the
code $\overline{b_1\dots b_n}$ 
describes a different partner orbit
than interchanging $b_1 \leftrightarrow b_{n_1+3}$. The diagrams
for both transpositions lead to the same diagram -- only the labels
at the vertices are changed. For $n_1 \neq n_2$ 
all transpositions
$b_j \leftrightarrow b_{j+n_1+2}$ with $j=1,\dots,n$ 
lead to different partner orbits with the same contribution
which results in the factor $n$.
For $n_1=n_2$ (thus $n=2n_1+4$ is even) only $b_j \leftrightarrow b_{j+n_1+2}$ with $j=1,\dots,n/2$
lead to different partner orbits due to the symmetry of the diagram.
In that case the additional factor is $\frac{n}{2}$ instead of $n$.
In general the prefactor of any diagram (including the diagonal diagrams)
is given by $\frac{n^2}{sB}$ for a diagram with $s$-fold symmetry (e.g.
$s=n$
for the diagonal diagrams).

The total contribution of the transposition diagram is now computed
using the unitarity of the graph scattering matrix $\scattering_\star$
and replacing ${\mathcal{M}^{n_{j}}_{\star}}_{bb'}
\rightarrow
\frac{1}{B}$ for $j=1,2$.  
For fixed lengths $n_{1}$, $n_2=n-n_1-4$ of the two loops
one gets  
\begin{equation}
  \begin{array}{c}   
  \includegraphics[width=5cm]{./figures/transposition_diagram.eps}
  \end{array}=
  \frac{n^2}{s B^2}=\frac{1}{s}\tau^2\ ,
  \label{eq:transposition_value}
\end{equation}
where $s=2$ for $n_1=n_2$ and else $s=1$. 

This shows that a transposition diagram contributes to order $\tau^2$ in
the form factor which is the leading correction to the diagonal part.
We will now add all diagrams of this type to the two diagonal
diagrams. Eventually it will turn out that their total contribution
vanishes once pairs of periodic orbits that occur in more
than one diagram have been considered correctly. Later we will
consider other diagrams that contribute to the same order $\tau^2$
in the form factor and give a non-vanishing contribution which is
consistent with the predictions of random-matrix theory.  

Taking into account
all different transposition diagrams which differ by the size of
the loops one gets an overall contribution of $\frac{(n-1)}{2}\tau^2$
which diverges as $n,B\rightarrow \infty$ \footnote{Transpositions of an element with its neighbour (formally $n_1=-1$) or next-neighbour ($n_1=0$) leads to diagrams which look a little bit different but have the same
value.}. 
This apparent divergence is resolved due to the fact that certain pairs
of periodic orbits occur in different diagrams and are thus
double (or multiply) counted.
In particular,
each of the transposition diagrams contains diagonal periodic
orbit pairs which have to be subtracted before performing
the limit of large graphs. 
In the explicit summation \eqref{eq:transpos_diag} 
diagonal pairs of periodic orbits appear 
whenever the two transposed bond
indices are the same
($b_1=b_{n_1+3}$). 
The subset of pairs of periodic orbits which
appears both, in the left diagonal diagram in \eqref{eq:diagonal_diagrams} and in the transposition diagram \eqref{eq:transpos_diag} 
can be expressed by the diagram
\begin{equation}
\begin{split}  
   \begin{array}{c}
     \includegraphics[width=5cm]{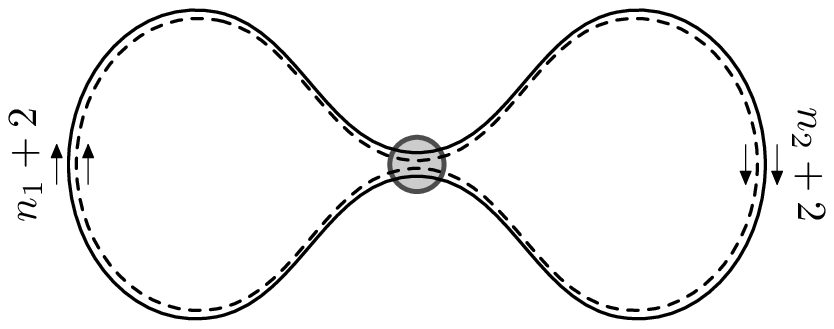}
   \end{array}=&
   \frac{n^2}{s B} \sum_{b=1}^B \frac{1}{B^2}
   \\=& \frac{1}{s}\tau^2
  \end{split}
\end{equation}
which has the same value as the transposition diagram \eqref{eq:transposition_value}.
Each transposition diagram is thus cancelled by subtracting
the doubly counted orbit pairs. As a consequence the transposition
diagrams do not contribute at all to the form factor in spite of the fact
that each diagram is of order $\tau^2$.
A transposition of two indices 
to the right diagram of the diagonal approximation in \eqref{eq:diagonal_diagrams} 
(thus changing the
direction of the dashed arcs in the transposition diagram) also leads
to new diagrams which are cancelled in the same way. 
This observation can be generalised. 
Adding a diagram which differs from the transposition diagram
by a further interchange of two bonds inside one of the two long stretches
of parallel lines gives a contribution $\frac{n^2}{B^3}$ which
is a factor $\frac{1}{B}$ smaller than the transposition diagram.
Subtracting all periodic orbits which have been counted twice --
once in the transposition diagram and once in the new one --
cancells the overall contribution of these orbits.

In the transposition diagrams discussed
above the two periodic orbits are parallel in two long 
stretches on the left and right of the diagram.
We will call such long stretches \define{loops}
because they describe a parallel propagation of the two
trajectories in phase space.
On these the motion
is free of interference and described by the classical map. 
A long loop of length $n_1$ in a diagram just gives a factor $\left(\mathcal{M}^{n_1}\right)_{bb'}\rightarrow 1/B$.
Interference occurs only in the small
region (composed of six bond indices) near the transposed
indices. It seems natural to consider diagrams
with long parallel stretches of classical motion with a few
regions where quantum interference is active in order to
find quantum corrections to the diagonal approximation. This
leads directly to the so-called loop expansion which orders
the diagrams according to the number of classical loops.
The diagonal approximation is represented by a single loop. We will
show that the universal $\tau^2$
correction to the diagonal approximation is obtained by 
all diagrams with two loops (including, of course, the correction diagrams for multiply counted pairs). We have already discussed two
types of diagrams with two loops: the transposition diagrams
where the periodic orbits traverse the loops in the same direction,
and the same type of diagram for loops which are traversed in opposite (time-reversed) direction. None of these gave any contribution
after the doubly counted orbits have been subtracted. There is 
a third type of diagram
with two loops. For these one loop contains time-reversed trajectories
and the other parallel trajectories. The diagram is easily 
drawn and calculated
\begin{equation}
  \begin{array}{c}
    \includegraphics[width=5cm]{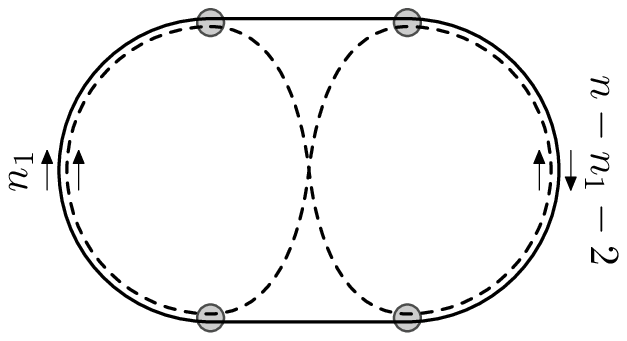}
  \end{array}=\frac{n^2}{B^2}=\tau^2.
\end{equation}
This eight-shaped diagram is the analog of the semiclassical
Sieber-Richter \cite{richter,sieber,sieber:2002} pairs for 
Hamiltonian systems.
The left loop has a minimal length $n_1\ge2$ since the cases $n_1=0,1$
have already been included as transpositions in the time-reversed
diagonal diagram.
For the same reason the right loop has a minimal length $n-n_1-2\ge 2$.
These minimal lengths of the loops are the graph analog
of the minimal length for a classical loop in
Hamiltonian flows.
Altogether there are $n-5$ new diagrams of the same value
$\frac{n^2}{B^2}$. Let us now consider the doubly counted orbit
pairs which have diagrams
\begin{equation}
  \begin{array}{c}
     \includegraphics[width=5cm]{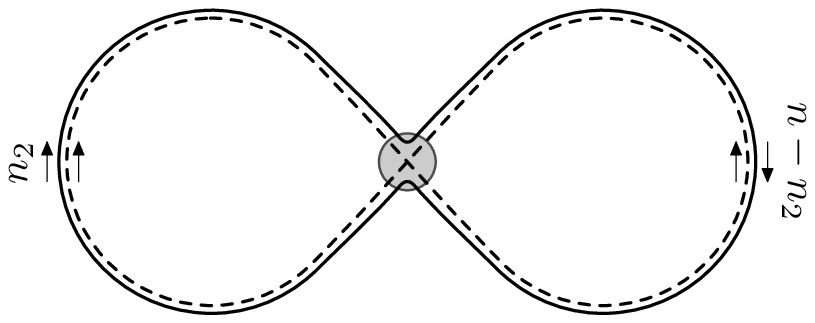}
  \end{array}=\frac{n^2}{B^2}=\tau^2\ .
\end{equation}
with $n_2=2,3,\dots,n-2$. These pairs have either already been included
in diagonal or transposition diagrams or 
they appear in two new diagrams with different $n_1$. 
As a consequence we have to subtract $n-3$ correction diagrams
with value $\frac{n^2}{B^2}$. Altogether the form-factor in the two-loop
expansion is
\begin{equation}
  \tilde{K}_n=2\tau+
  \tau^2(n-5 -(n-3))\rightarrow 2\tau -2\tau^2\ .
\end{equation}
This is equivalent to the short-time expansion of the
random-matrix prediction for the form factor in
time-reversal invariant systems (symmetry class $A$I) 
in the same order. 

Breaking time-reversal invariance (symmetry class $A$) 
effects all anti-parallel loops.
Instead of the classical evolution map $\mathcal{M}$ the dynamics
along the anti-parallel loops is governed by the complex matrix $\mathcal{R}$ \eqref{eq:timerev_evol}. A diagram which contains
an anti-parallel loop is suppressed exponentially and 
does not contribute in the limit $B\rightarrow \infty$. Thus only
the first diagonal diagram remains while all other
diagrams in the two-loop expansion are either cancelled by doubly
counted pairs or vanish due to anti-parallel loops.
The two-loop result $\tilde{K}_n=\frac{n}{B}=\tau$ is consistent
with the random-matrix prediction.  

It takes some effort to write down and calculate 
all three-loop diagrams
for the next order $\tau^3$
\cite{berkolaiko:2003a}. The building blocks of the loop expansion
are all known but it has been performed explicitly only for
a few orders. For Hamiltonian systems the loop expansion
in the semiclassical form factor
works analogously and has recently been
extended to all orders\cite{muller:2004,muller:2005}. 
One should be aware that
the loop expansion is valid only for $\tau \ll 1$.
Near the Heisenberg-time $\tau=\tau_H=1$ 
the probability that a bond is visited more than once is
unity and long loops have negligible importance.
Similar expansions which work near or beyond the Heisenberg time 
are not known.
The loop expansion also cannot estimate the error due
to all neglected diagrams. For a fixed $n$ the expansion in diagrams
is exact and finite, but in the limit $B\rightarrow \infty$, the number of
orbit pairs grows exponentially while only a small fraction
is included in the loop-expansion. 

\section{Graphs with spin and their spectral statistics}
\label{sec:spin_GSE}

So far we have only discussed two of the three Wigner-Dyson
symmetry classes on quantum graphs. The third symmetry class
($A$II) is relevant for fermionic systems
with time-reversal invariant dynamics and
broken spin-rotational symmetry. The time-reversed path of an electron
in such a system travels along the same trajectory in
opposite direction $\vec{p}\mapsto -\vec{p}$,
and has, additionally, opposite spin direction $\vec{s}\mapsto -\vec{s}$.
In quantum mechanics the time-reversal symmetry 
is described by an anti-unitary operator $\mathcal{T}$ 
which commutes with the Hamiltonian $\left[H,\mathcal{T}\right]=0$.
The two Wigner-Dyson symmetry classes of time-reversal
invariant systems are distinguished by
$\mathcal{T}^2=-\ONE$ in symmetry class $A$II
in contrast to $\mathcal{T}^2=\ONE$ in
symmetry class $A$I.

The spectra of a quantum system in symmetry class $A$II is doubly degenerate
due to Kramers' degeneracy. If $|\nu\rangle$ is an eigenstate
of the system $H|\nu\rangle = E |\nu \rangle$
the state $\mathcal{T}|\nu\rangle$ is an orthogonal eigenvector 
$\langle\nu|\mathcal{T}\nu\rangle$ to the same
energy 
$H \mathcal{T}|\nu\rangle= \mathcal{T}H|\nu\rangle=E\mathcal{T}|\nu\rangle$.
The different type of time-reversal symmetry
has impact on spectral statistics \cite{scharf:1988}.
The spectral fluctuations which are displayed
by 
complex quantum systems in symmetry class $A$II 
are described by the
Gaussian Symplectic Ensemble (GSE).
For more details we refer to the appendices \ref{app:symmetry_classes}
and \ref{app:random_matrix_theory} and to the literature \cite{Book:Haake}.

For quantum graphs this symmetry class has first been studied by 
Bolte and Harrison who 
considered the Dirac-operator on the graph with appropriate boundary
conditions at the vertices and discussed the spin contribution to
spectral statistics. 
Quantum graphs with a spin degree of freedom 
have also been discussed in the
context of localisation induced by spin-orbit coupling 
\cite{bercioux:2004}.
Here, we 
shall introduce spin degrees of freedom
which are only effected by the scattering at the vertices.
As in the previous section we shall discuss
only star graphs \cite{gnutzmann:2004a,gnutzmann:2004b}.
The spin degree of freedom is added to a graph 
by considering
wave functions $\psi_{b\sigma}(x_b)$
on the bond $b$ which have a spin up  $\sigma=1/2$
and a spin down $\sigma=- 1/2$ component. 
Spin rotational invariance is broken by 
allowing spin flips at the vertices.
As in the previous we will limit ourselves to
star graphs with spin and allow for spin flips only at the centre.

The construction of the star quantum evolution operator
in Section \ref{sec:stars}
and spectral theory is completely equivalent to
usual star graphs when one replaces the star graph with $B$ bonds
and a two-component wave function by a star graph with $2B$
bonds and a scalar wave function. 

Time-reversal symmetry of the star graph with spin requires
that the unitary $2B\times 2B$ 
star graph scattering matric $\scattering_\star$ has
the structure
\begin{equation}
 \scattering_\star=
 \begin{pmatrix}
   \mathcal{A} & \mathcal{B}\\
   \mathcal{C} & \mathcal{A}^T
 \end{pmatrix}
 \qquad \mathcal{B}=-\mathcal{B}^T\qquad
 \mathcal{C}=-\mathcal{C}^T
\end{equation}
where the explicit structure is in the spin index such that
the $B\times B$ matrix $\mathcal{A}$ ($\mathcal{A}^T$)
describes spin up (down) to spin up (down)
scattering from one bond to another while
$\mathcal{B}$, and $\mathcal{C}$ describe scattering processes
where the spin is changed.
That is the scattering amplitude $(b',\sigma') \rightarrow
(b,\sigma)$ equals the one of the time-reversed
process $(b,-\sigma) \rightarrow
(b,-\sigma)$ up to a minus sign which occurs when the 
spin component is flipped
\begin{equation}
  \scattering_{\star \; b\,\sigma, b'\,\sigma'}=
  (-1)^{\sigma-\sigma'}\scattering_{\star\; b'\,-\sigma',b\,-\sigma}.
\end{equation}
Since the number of spin flips along a periodic orbit $p$ is even,
its amplitude in the trace formula equals the amplitude of the time-reversed
periodic-orbit $A_p=A_{\hat{p}}$ where the time-reverse $\hat{p}$ of
an orbit $p$ with the code
 $p=\overline{(b_1,\sigma_1),(b_2,\sigma_2),\dots,(b_n,\sigma_n)}$
is now defined  by a reverse order of the bond indices $b_j$ and
flipped spins, 
$\hat{p}=\overline{(b_n,-\sigma_n),\dots,(b_2,-\sigma_2),(b_1,-\sigma_1)}$.

By convention every pair of degenerate eigenvalues 
in a spectrum that displays
Kramers' degeneracy is only counted once in the density of states.
This leads to an overall factor $1/2$ in the trace formula
and the mean level spacing between
different eigenvalues is $\overline{\Delta}=\frac{\pi}{B\overline{L}}$
as in a scalar graph with the same number of bonds.
The equivalence of the spectral form factor with the
discrete time form factor of the eigenphases in 
large graphs leads to the relation 
\begin{equation}
  \langle K(\tau=\frac{n}{B})\rangle_\tau = \langle \tilde{K}_n \rangle_n
\end{equation}
for time-reversal invariant spin star graphs, where
\begin{equation}
  \tilde{K}_n= 
  \frac{1}{4B}\langle |\mathrm{tr}\, \evol_{\star}(k)^n|^2 \rangle_k \ .
\end{equation}
We will discuss the form factor
in the
diagrammatic language introduced in the previous chapter
and assume that the classical evolution operator
$\mathcal{M}_{\star\, b\sigma, b'\sigma'}=
|\scattering_{\star\, b\sigma, b'\sigma'}|^2$ is strongly mixing
such that the decaying classical modes do not contribute
to any of the considered diagrams in the limit of large graphs.  

The loop expansion follows the same rules as in the previous
section. However one needs to add a sum over the $2n$ spin indices
along the two orbits. We start with the
two diagrams of the diagonal approximation. The
first diagram can be written in the form
\begin{equation}
  \begin{array}{c}
  \includegraphics[width=2cm]{./figures/diagonal_diagram1.eps}
  \end{array}
  = \frac{n}{4B}\mathrm{tr} 
  \begin{pmatrix}
    \mathcal{M} & \mathcal{J}\\
    \mathcal{L} & \mathcal{K}
  \end{pmatrix}^n
\end{equation}
where the matrix that appears on the right hand side
describes the propagation of two spins from one bond to another.
It contains the classical map $\mathcal{M}$ as a $2B\times 2B$ submatrix
which describes the propagation of two parallel spins that remain
parallel. Anti-parallel spins are propagated by the matrix
\begin{equation}
  \mathcal{K}_{b \sigma, b'\sigma'}=
  \scattering_{\star\; b\, \sigma, b'\,\sigma'}
  \scattering^*_{\star\; b \, -\sigma, b'\, -\sigma'}
\end{equation}
while anti-parallel spins and parallel spins are coupled by
\begin{equation}
  \mathcal{J}_{b \sigma, b'\sigma'}=
  \scattering_{\star\; b\, \sigma,B'\,\sigma'}
  \scattering^*_{\star\; b\, \sigma, b'\,-\sigma'}
  \;\; \text{and}
  \;\; 
  \mathcal{L}_{b \sigma, b'\sigma'}=
  \scattering_{\star\; b\, \sigma,B'\,\sigma'}
  \scattering^*_{\star\; b\, -\sigma, b'\,\sigma'}\ .
\end{equation}
Being defined in terms of a unitary matrix  $\mathcal{K}$, $\mathcal{L}$ and $\mathcal{J}$ 
all have eigenvalues inside the unit circle. Their contribution to
the trace is thus exponentially suppressed. Only the ergodic
mode of the classical map survives in the $n$-th trace when
$n$ is larger than all decay times. In the limit $B\rightarrow \infty$
the first diagonal diagram gives a contribution $\tau/4$ to the
form factor.
The second diagonal diagram gives the same contribution --
with the only difference that the ergodic modes corresponds to
antiparallel spins along the loop. Minus signs
occur in addition to the classical propagator whenever both spins flip
but since the number of spin flips is even along a periodic
orbit they all cancel. 
We thus have
\begin{equation}
  \langle \tilde{K}(\tau=\frac{n}{B})^\mathrm{diag} \rangle_\tau=
  \begin{array}{c}
    \includegraphics[width=1.5cm]{./figures/diagonal_diagram1.eps}
  \end{array}+
  \begin{array}{c}
    \includegraphics[width=1.5cm]{./figures/diagonal_diagram2.eps}
  \end{array}=
  \frac{\tau}{2}
\end{equation}
in accordance with the random-matrix result \eqref{eq:formfactor_short_time}.

Loops in other diagrams can be treated in a similar way. 
Each loop just gives a factor $1/2B$ and only transports
parallel spins for a loop that is traversed in the same direction by both
orbits and anti-parallel spins for loops that are traversed in opposite direction. For time-reversed loops
an additional factor $-1$ appears if the number of spin flips along the
loop is odd. The $\tau^2$ corrections to the diagonal
approximation are found again in the diagrams with two loops.
While the transposition diagrams cancel 
with the doubly counted orbits
in just the same way
as for scalar graphs one again gets a non-vanishing contribution from
the eight-shaped diagrams of the Sieber-Richter type
\begin{multline}  
  \begin{array}{c}
     \includegraphics[width=5cm]{./figures/diagram_tau2.eps}
  \end{array}= 
  \sum_{
  b_1,b_2,b_3,b_4,\sigma_1,\sigma_2,\sigma_3,\sigma_4}
  \frac{n^2 (-1)^{\sigma_2-\sigma_3}}{16 B^2} \times\\
  \scattering_{\star\, b_2\sigma_2,b_1\sigma_1}
  \scattering_{\star\, b_4\sigma_4,b_3\sigma_3}
  \scattering^*_{\star\, b_3-\sigma_3,b_1\sigma_1}
  \scattering^*_{\star\, b_4\sigma_4,b_2-\sigma_2}
  =-\frac{n^2}{8B^2} \ .
\end{multline}
The main difference to the scalar case is the different overall sign
(the other factors are just due to Kramers' degeneracy). The different sign
also appears in the diagram for the doubly counted orbits which has the
same value $-\frac{n^2}{8B^2}$. With $n-5$ eight-shaped diagrams
(not counting the transposition diagrams)
and $n-3$ diagrams accounting for double counting the two loop
correction is $\frac{n^2}{4B^2}=\frac{\tau^2}{4}$ in accordance
with the universal result \eqref{eq:formfactor_short_time}. 

\section[Andreev graphs and non-standard symmetry classes]{Andreev graphs\\ and non-standard symmetry classes}
\label{sec:pot_andreevstars}

Recently the threefold symmetry
classification of quantum systems introduced
by Wigner and Dyson has been extended to a ten-fold
classification
\cite{chiral1,chiral2,altlandzirnbauer,altlandzirnbauer2,zirnbauer:symmetry}.
This has been necessary  
for certain physical systems, such as a Dirac particle in
a random Gauge field, quasiparticles
in in a hybrid supercon\-ducting-normalconducting structure
or quasiparticles in a disordered superconductor. 
The new feature in these systems is a electron-hole
or particle-antiparticle symmetry.
The one-particle
excitations are described by a Hamiltonian 
(for instance, the Dirac operator or the Bogoliubov-de Gennes 
operator) that have
a positive spectrum corresponding to (quasi-) particle
excitations and a negative 
spectrum corresponding to antiparticle
excitations (while the excitation energies
of the full many-body field theory remain positive). 
That is
if $|\nu\rangle$ is an eigenstate with energy $E$, the
particle-antiparticle symmetry yields another eigenstate $|\nu'\rangle$
with energy $-E$.
There are two types of such particle-antiparticle symmetry operators.
They are either unitary or antiunitary.
A unitary symmetry operator $\mathcal{P}$
that anti-commutes with the Hamiltonian  
$\left[\mathcal{P},H\right]_+=0$ is a \define{chiral symmetry operator}
and an anti-unitary operator $\mathcal{C}$ which anti-commutes with the
Hamiltonian $\left[\mathcal{C},H\right]_+=0$ is a
\define{charge-conjugation symmetry}.
Both types of symmetry have impact on spectral statistics
near $E=0$ and lead to new quantum interference effects
which are not present in the Wigner-Dyson classes.
Sufficiently far away from the symmetry point
$E=0$ these systems are usually 
described well by the standard Wigner-Dyson classes.
Combining charge-conjugation and chiral symmetries with
time-reversal symmetries one can show
that there are seven non-standard symmetry classes beyond the
Wigner-Dyson classes \cite{zirnbauer:symmetry}.
A full account of these is given in appendix \ref{app:symmetry_classes}.
In each 
non-standard symmetry class 
random-matrix ensembles can be constructed which predict universal
statistical properties of their spectra in complex quantum systems.

In this section we will concentrate on two non-standard symmetry
classes, which have been called $C$ and $C$I (see appendix \ref{app:symmetry_classes}), and on chaotic systems with
spectral statistics which is dominated 
by the universal predictions of Gaussian random-matrix ensembles 
(see appendix \ref{app:random_matrix_theory}). The symmetry classes
$C$ and $C$I
can be realised by quasiparticles in a 
hybrid superconducting-normalconducting structure\cite{altlandzirnbauer,altlandzirnbauer2}. 
The wave function obeys
the Bogoliubov-de Gennes equation 
\begin{equation}
  \begin{pmatrix}
  \frac{\left(\frac{\hbar}{\I}\nabla + e \vec{A}(\vec{x})\right)^2}{2m}
  -E_F&
  \Delta_s(\vec{x})\\
  \Delta_s(\vec{x})^* & 
  -\frac{\left(\frac{\hbar}{\I}\nabla - e \vec{A}(\vec{x})\right)^2}{2m}
  +E_F
  \end{pmatrix}
  \begin{pmatrix}
  \Psi_e(\vec{x})\\
  \Psi_h(\vec{x})
  \end{pmatrix}
  =E
  \begin{pmatrix}
  \Psi_e(\vec{x})\\
  \Psi_h(\vec{x})
  \end{pmatrix}
\end{equation}
where $\Psi_e(\vec{x})$ and $\Psi_h(\vec{x})$ are the (quasi-)
electron and hole components of the wave function,
$ \pm \frac{\left(\frac{\hbar}{\I}\nabla \pm e \vec{A}(\vec{x})\right)^2}{2m}$
is the kinetic energy of the electron or hole in the presence
of a magnetic field $\vec{A}$, $E_F$ is the Fermi energy, and
$\Delta_s(\vec{x})$ is the pair potential of the superconducting
condensate which couples electron and hole components. The
pair potential vanishes in the normal conducting region.
There is an anti-unitary charge
conjugation symmetry $\mathcal{C}$
\begin{equation}
  \begin{pmatrix}
  \Psi_e(\vec{x})\\
  \Psi_h(\vec{x})
  \end{pmatrix}\mapsto
  \mathcal{C}
  \begin{pmatrix}
  \Psi_e(\vec{x})\\
  \Psi_h(\vec{x})
  \end{pmatrix}=
  \begin{pmatrix}
  \Psi_h(\vec{x})^*\\
  -\Psi_e(\vec{x})^*
  \end{pmatrix}
  \qquad \text{with} \qquad \mathcal{C}^2=-\ONE
\end{equation}
which transforms an eigenstate with energy $E$ to 
a different eigenstate with energy $-E$. 
The Bogoliubov-de Gennes equation is time-reversal invariant
with $\mathcal{T}^2=\ONE$
(symmetry class $C$I) if the magnetic field vanishes and the
pair potential is real (for broken time-reversal invariance the
symmetry class is $C$).

\begin{figure}[ht]
  \begin{center}
    \includegraphics[width=10cm]{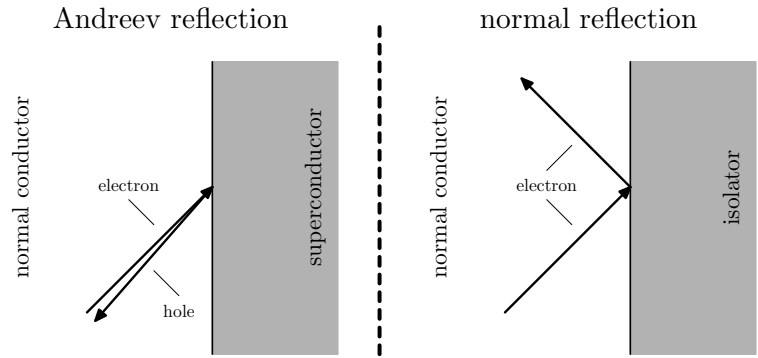}
  \end{center}
  \caption{Andreev reflection at a normalconducting-superconducting 
  interface and normal reflection at a potential wall.}
  \label{fig:Andreev}
\end{figure}

In a normalconducting region electrons and holes are not
coupled. In a hybrid
normalconducting-supercon\-ducting system
electrons and holes are coupled by
Andreev scattering (see figure \ref{fig:Andreev}) at
a normalconducting-super\-con\-duc\-ting interface \cite{andreev:scattering}. 
An electron (or hole)
with energy $|E|\ll |\Delta_s|\ll E_F$ in the 
normalconducting region ($\Delta_s=0$) hits the interface and is 
retroflected as a hole (electron) of the same energy in the direction
from which the incident quasiparticle came (as opposed to normal
specular reflection at a potential wall or normalconducting-isolating
interface). The physical process can be described 
as an electron which
enters the superconductor and recombines with an
electron-hole pair such that the two electrons build a new Cooper pair
and the hole is ejected out of the superconductor. Up to corrections
of order $E/\sqrt{E_F}$ the momentum is conserved in this process,
not only in magnitude but also in direction (again in contrast
to normal reflection at a potential wall). Note, that the velocity of
a hole is in opposite direction to its momentum due to the overall
minus sign in the kinetic energy (or `negative mass') of the hole.
  
If a non-magnetic chaotic system is coupled to a superconductor
Andreev reflection will destroy chaos near the Fermi level ($E=0$)
since almost every trajectory will hit the superconducting interface
and the incident particle travels back along the same trajectory.
Exponential divergence of nearby trajectories can only survive
on a short time scale between two Andreev reflections. 
Still signatures
of chaos in the normal system remain which are a topic of interesting
present research. For details we refer to the literature
\cite{beenakker:2004,melsen:1996,altland:review,taras,goorden:diss}.
Here we will only be interested in the regime where the
\emph{combined} electron-hole dynamics is chaotic\footnote{In the literature on Andreev billiards it has become a convention to call a system chaotic or integrable 
according to the dynamics of the normal system where electrons and holes are not coupled. We will always refer to the combined
electron-hole dynamics in presence of the superconductor.} -- this can be achieved
by a magnetic field which bends both electrons and holes in the
same direction such that hole trajectories do not travel back
along the electron trajectories. In general this breaks
time-reversal invariance (symmetry class $C$) but certain
reflection and point symmetries can restore (a non-conventional)
time-reversal invariance (symmetry class $C$I)\footnote{Another option to restore chaos is to introduce many point scatterers (disorder). Both symmetry classes can also be realised in a completely different
context like two coupled spins  where chaos
in the semiclassical limit (large spins) is not prevented by Andreev reflections.}. 

The charge conjugation symmetry in a chaotic system effects all
spectral correlation functions near $E=0$ and the deviations from
the Wigner-Dyson classes decrease on the scale of the
mean level spacing $\overline{\Delta}$. That is for $E\gg\overline{\Delta}$
chaotic systems in the symmetry classes $C$ and $C$I are
described by the GUE and GOE (symmetry classes $A$ and $A$I).
The effects can thus not be seen by any type of spectral average
in an individual system. Instead the spectral average 
has to be replaced by an average over some system parameter
like the Fermi energy, the magnetic field or the geometry of the
system. The deviations from Wigner-Dyson behaviour turn out to
be universal and are themselves described by Gaussian random matrix
ensembles which we will call the $C$-GE and $C$I-GE.

In contrast to the Wigner-Dyson symmetry classes there are 
universal interference effects
also in the mean density of states 
for the non-standard symmetry classes
(averaged over some system parameter). These will be in the
focus of the following discussion.

\begin{figure}[ht]
  \begin{center}
    \includegraphics[width=6cm]{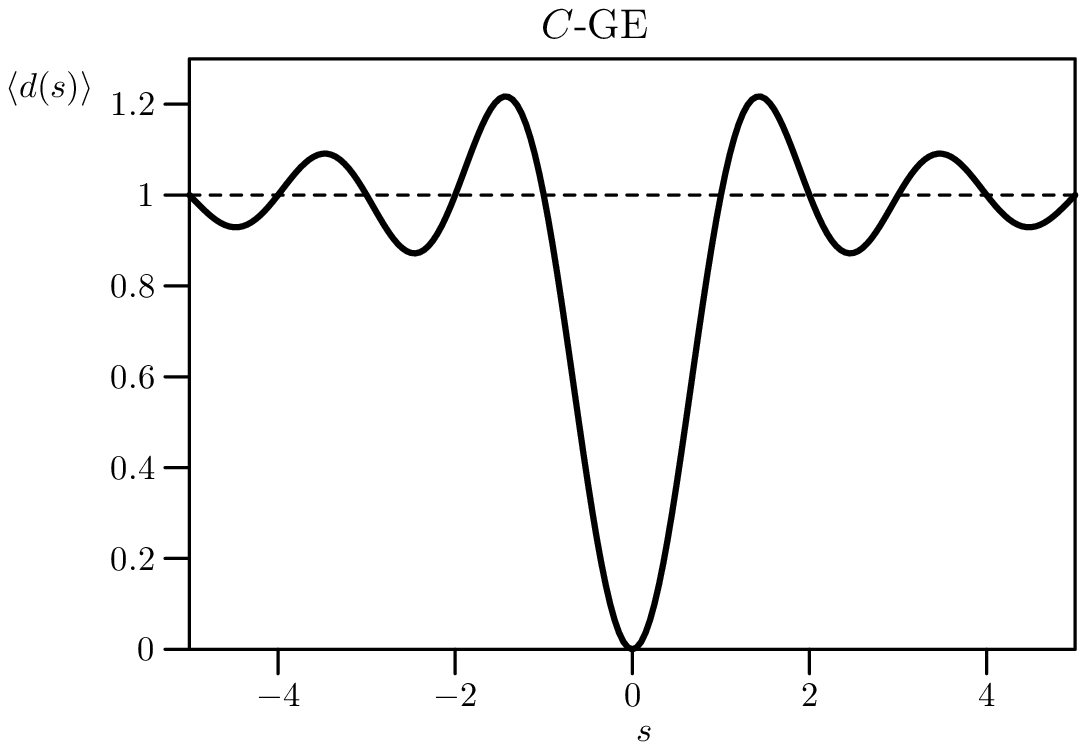}
    \includegraphics[width=6cm]{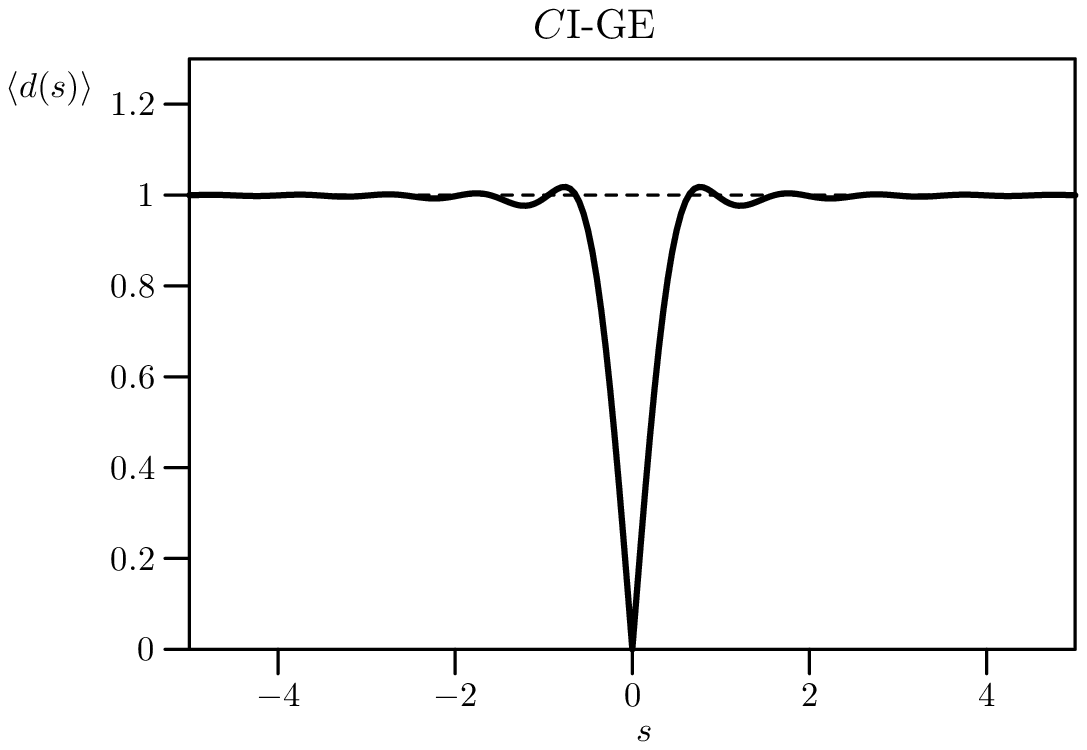}
  \end{center}
  \caption{The mean density of states in the random matrix ensembles 
  $C$-GE and $C$I-GE (full lines). The dashed lines give the flat
  density of states of a system without charge-conjugation symmetry
  and the same mean level spacing.}
  \label{fig:dos_andreev}
\end{figure}

\begin{figure}[ht]
  \begin{center}
    \includegraphics[width=6cm]{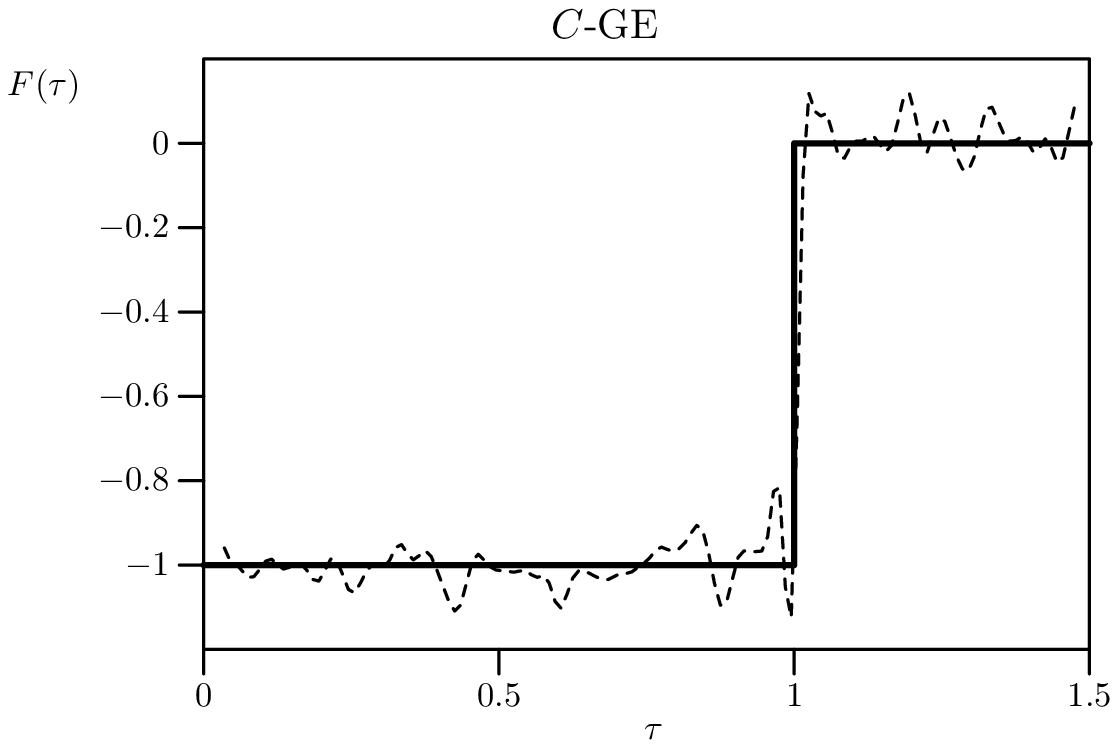}
    \includegraphics[width=6cm]{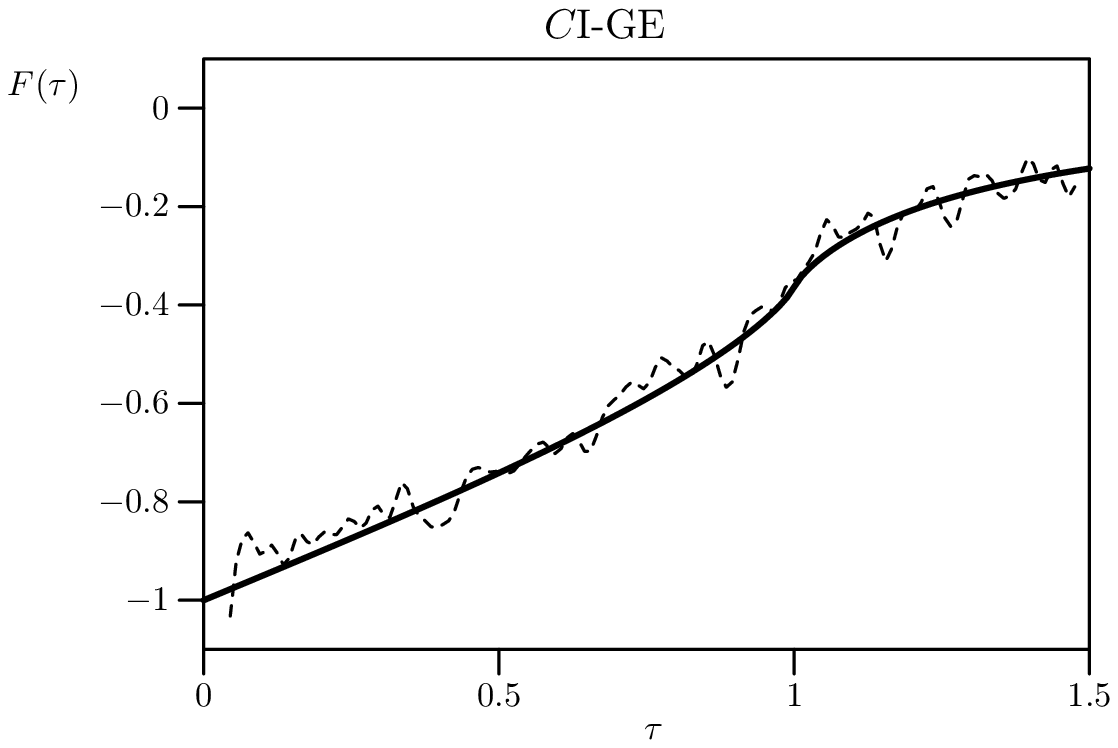}
  \end{center}
  \caption{Universal first-order form factor as predicted by
  the random matrix ensembles 
  $C$-GE and $C$I-GE (full lines). The dashed lines are an average over
  10000 realizations of Andreev star graphs with $B=50$ bonds.}
  \label{fig:formfactor_andreev}
\end{figure}

In figure \ref{fig:dos_andreev} the mean density of states for
the Gaussian random-matrix ensembles $C$-GE and $C$I-GE 
\cite{altlandzirnbauer,altlandzirnbauer2} is shown
(for explicit formulae see appendix \ref{app:random_matrix_theory})
as a function of the energy $s=E/\overline{\Delta}
E$ in units of the mean level spacing.
Quantum interference leads to a dip in the mean density of states
at $s=0$ where $\langle d(s) \rangle=0$, 
that is there are no states on the Fermi
level. For $s\gg 1$ (or $E\gg \overline{\Delta}$) the mean density of states approaches
the value $\langle d(s)\rangle \rightarrow 1$ 
(or $\langle d(E)\rangle \rightarrow 1/\overline{\Delta}$ 
which defines the mean level spacing here).
With
\begin{equation}
  \langle  d(s) \rangle = 1+ \INT{0}{\infty}{\tau} F(\tau) \cos 2\pi \tau s
\end{equation}
we introduce the Fourier transform of $\langle d(s)\rangle-1$
which we will call the \define{first-order form factor} (or just form factor
in this chapter) because its periodic-orbit treatment is to some
extent analogous to the (second-order) form factor in the Wigner-Dyson classes. We will show that the short-time
expansion 
\begin{equation}
\begin{split}
 F(\tau)^{\text{$C$}}=&-1\\
 F(\tau)^{\text{$C$I}}=&-1+\frac{\tau}{2}+\mathcal{O}(\tau^2)
\end{split}
\end{equation}
of this form factor 
for a quantum graph in the corresponding symmetry class 
is given by a loop expansion
analogously to the loop expansion of the second-order form factor
in the Wigner-Dyson case. 
In figure \ref{fig:formfactor_andreev}
the universal form factors are shown together
with a numerical average over Andreev star graphs of the corresponding
symmetry classes. The form factor is a strongly oscillating quantity.
To compare it to the universal result an additional time average
over a small interval $\Delta\tau\ll 1$ is needed to tame the oscillations. 

\define{Andreev graphs} \cite{gnutzmann:2004a,gnutzmann:2004b} are quantum graphs 
which are coupled to a superconductor such that
Andreev reflection occurs.
Such graphs can be constructed in the symmetry classes $C$ and $C$I
by assigning a two-component wave function $\Psi_{b\,\sigma}(x_b)$
to each bond where $\sigma=e,h$ corresponds to the electron
and hole components on the bond $b$ with additional restrictions
on the boundary conditions at the vertices. We will not give the most
general definition, instead we will look at star graphs. 
Each bond of the star graph is coupled to
a superconductor at the peripheral vertex where Andreev
reflection couples the electron and hole components. Along the bonds
and at the central vertex the graph is assumed to be normal
conducting (electrons are not coupled to holes). 

The quantum evolution map of an Andreev star graph 
with $B$ bonds in symmetry class $C$ (broken time-reversal invariance)
has the form
\begin{equation}
 \evol_{\star}(k)=
 \begin{pmatrix}
   \scattering_\star &0\\
   0 & \scattering_\star^*
 \end{pmatrix}
 \begin{pmatrix}
   0 & T_{eh}(k)\\
   T_{he}(k) &0
 \end{pmatrix}
\end{equation}
where the unitary $B\times B$ matrix $\scattering_\star$ is the
electron-electron scattering matrix of the central vertex.
Charge conjugation symmetry then requires that the hole-hole
scattering matrix is the complex conjugate $\scattering_\star^*$.
The diagonal $B\times B$ matrices $T_{eh}(k)$ ($T_{he}(k)$)
contain the phase factors acquired for a hole propagating
from the centre to a peripheral vertex where it is
scattered back as an electron which propagates back to the centre.
Altogether
\begin{equation}
 T_{eh\; bb'}(k)=-\I\delta_{bb'} \E^{\I \theta_b}\E^{\I 2 k L_b}
 \qquad
 T_{he\; bb'}(k)=-\I\delta_{bb'} \E^{-\I \theta_b}\E^{\I 2 k L_b},
\end{equation}
where $\E^{\I 2 k L_b}$ is the phase acquired during the propagating along the bond $b$ from the centre to the periphery and back\footnote{At first sight this does not
seem to be in accordance with the Bogoliubov-de Gennes equation.
Indeed an electron (a hole) propagating along a bond with energy $E$ would
aquire a phase $\E^{\I k_e L_b}$ ($\E^-\I k_h L_b$)
according to the Bogoliubov-de Gennes equation where electron (hole) momentum is $k_{e}=\sqrt{E_F+E}$ ($k_h=\sqrt{E_F-E}$)  -- the sign in the phase of the hole is different
since it propagates in opposite direction to its momentum.
Since we are interested in the limit $E_F\gg E$ one may 
expand the momenta $k_{e,h}=\sqrt{\E_F\pm E}=\sqrt{E_F}\pm \frac{E}{2\sqrt{E_F}}+\mathcal{O}(\frac{E^2}{E_F^{3/2}})$.
Adding the hole and electron phases $(k_e -k_h)L_b$ the
leading part cancels.
Keeping $k=\frac{E}{2\sqrt{E_F}}$ fixed in the limit $E_F \rightarrow
\infty$ only $2 k L_b$ remains.}  
and $-I\E^{\I\theta_b}$
($-I\E^{-\I\theta_b}$) is the phase acquired during the Andreev
reflection for a hole (an electron) hitting the superconductor --
$\E^{-\I \theta_b}$ is the phase of the superconductors pair potential.
An average over system parameters is easily obtained by 
taking the phases $\theta_b$ as independent random variables which are equidistributed on $0\le \theta_b <2\pi$.

For a time-reversal invariant Andreev star graph (symmetry class $C$I)
there are additional restrictions: the central electron-electron
scattering matrix (and thus also the hole-hole scattering matrix)
has to be symmetric $\scattering_\star=\scattering_\star^T$
and the pair potential has to be real, thus $\E^{\I\theta_b}=\pm 1$.
For time-reversal invariant Andreev graphs the
different signs of the pair potentials will be taken as independent random
variables in a system average. 
Note, that it is sufficient to break either the symmetry
of the central scattering matrix or the reality of the
pair potentials to break time-reversal invariance. Indeed, in the
numerical calculations performed in figure \ref{fig:formfactor_andreev}
we have taken $\scattering_\star=\scattering_{DFT}$ which is
symmetric and one finds very good agreement of the form factor with
the universal result.

We will assume that the $\scattering_{\star}$
is a full matrix and the magnitude of all
matrix elements is of order $1/\sqrt{B}$.
Since we do not perform any spectral
average there is no fundmental difference
between rationally dependent and incommensurable
bond lengths here. For simplicity
we choose all bond lengths
equal $L_b=L$. 
The mean level spacing is $\overline{\Delta}=\pi/2BL$.
Setting $k=\overline{\Delta}  s$ in the oscillating part of the density of states 
$ d^{\mathrm{osc}}(k)\D k = d^{\mathrm{osc}}(s) \D s $ we 
arrive (after a few minor manipulations) at 
\begin{equation}
  d^\mathrm{osc}(s)=
  \frac{2}{B}\sum_{n=1}^\infty (-1)^n \cos(2\pi s \frac{n}{B})
  \mathrm{tr}\left(\scattering_\star D \scattering_\star^* D^*\right)^n
\end{equation}
where $D_{{bb'}}=\delta_{bb'}\E^{\I\theta_b}$. One can immediately
read off the time averaged form factor
\begin{equation}
 \langle F(\tau=\frac{n}{B}) \rangle_\tau =\langle F_n \rangle_n
\end{equation}
where
\begin{equation}
 F_n= 2 \langle \mathrm{tr}\left(\scattering_\star D \scattering_\star^* 
 D^*\right)^n\rangle_\theta
\end{equation}
is a discrete variant. The trace appearing
in $F_n$ is equivalent to a sum over periodic orbits of
period $2n$ which are scattered
at the centre of the star alternately as electrons or holes.

Averaging over the phase factors $\E^{\I\theta_b}$ (for broken
time-reversal invariance) only those periodic orbits survive
in $F_n$ which visit each bond an even number of times -- one
half as electrons being Andreev reflected back to the centre
 as holes
the other half as holes being Andreev reflected as electrons.
The sum over the remaining orbits is still very complex and cannot
be performed in a closed form. It resembles the sum over pairs
of periodic orbits in the (second-order) spectral form factor
where only those pairs survive the average where both orbits 
visit the same bonds. Since there is only
one orbit there is no diagonal and off-diagonal
terms in $F_n$. However there are \define{self-dual}
periodic orbits which are invariant
under electron-hole interchange \cite{gnutzmann:2003}. That is, a self-dual 
periodic orbit visits the
same sequence of bonds twice with electrons and holes interchanged
at the second traversal. The self-dual approximation
takes into account only the coherent contribution of self-dual orbits.
Since self-dual periodic orbits only exist for odd $n$ one has
\begin{equation}
 F_n^\mathrm{self-dual}=
 \begin{cases}
 0 & \text{for odd $n$},\\
 -2\mathrm{tr} \mathcal{M}^n  & \text{for even $n$,}
 \end{cases} 
\end{equation}
where $\mathcal{M}_{bb'}=|\scattering_{\star \,bb'}|^2$ is the corresponding
classical map on the graph which does not distinguish between electrons and holes. We can use the classical sum rule
\eqref{eq:classicalsum} and replace
$\mathrm{tr} \mathcal{M}^n \rightarrow 1$
and an average over $n$ yields the
leading order of the universal term
\begin{equation}
 \langle F(\tau)^{\mathrm{self-dual}}\rangle_\tau=\langle 
 F_n^\mathrm{self-dual}\rangle_n=-1. 
\end{equation}
The overall minus sign is just a consequence of the $2n$ 
Andreev reflections each of which gives a factor $-\I$.
While the diagonal approximation to the second-order
form factor $\tilde{K}_n$ is linear in $\tau$
the self-dual approximation starts with a constant. The reason for
this difference is that the contribution of a pair of periodic
orbits in a spectral two-point correlator gets an extra factor $n$ for all cyclic permutations
of the second orbit with respect to the first one. 
In contrast we here deal with a single orbit where such a factor cannot
arise.

If time-reversal is broken the contribution of all remaining
periodic orbits have to vanish for universal spectral
statistics. In contrast time-reversal
invariant Andreev graphs in symmetry
class $C$I display corrections to the
self-dual approximation -- the first order being
linear in $\tau$. The average in a time-reversal invariant
Andreev graph is just over $\E^{\I\theta_b}=\pm 1$. The orbits
that survive this average outnumber the orbits that contributed
at broken time-reversal invariance. Every bond is still visited
an even number of times -- but there is no restriction that
half of them should come from the centre as an electron and the
other half as a hole. There are additional
orbits already on the level of the self-dual approximation.
However their contribution cancels as we will show now.
If an orbit first visits a sequence
of $n$ bonds and then traverses the same bonds backwards with electrons
and holes interchanged it is self-dual with respect to charge-conjugation
combined with time-reversal invariance. This is most easily seen by drawing
diagrams
\begin{equation}
  F_n^\mathrm{self-dual}=
  \begin{array}{c}
    \includegraphics[width=3cm]{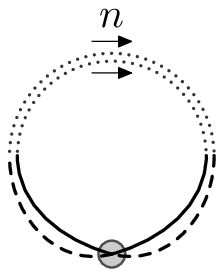}
  \end{array}+
  \begin{array}{c}
    \includegraphics[width=3cm]{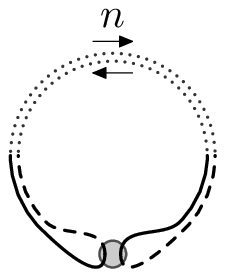}
  \end{array}
\end{equation}
where the first diagram contains the self-dual orbits that
we have already discussed for broken time-reversal symmetry
and the second diagram contains the additional
orbits which only survive in the time-reversal invariant case.
The first diagram gives a contribution $-1$ and the 
second diagram gives $(-)^n 2n$.
The additional factor $n$ stems from the reduced symmetry of the
diagram with respect to the first self-dual diagram. 
Averaging over a small time interval
the contribution of the second
diagram cancels
due to its alternating sign. 

Higher order corrections in both symmetry classes
can be obtained by drawing diagrams with more than one
loop \cite{gnutzmann:2004b}.
One should be aware that both types of loops parallel and anti-parallel
are present in the time-reversal invariant case but only
the parallel loops survive the ensemble average in the case of
broken time-reversal invariance. 
The linear correction to the self-dual approximation
is due to diagrams with two loops. We will not present a full
account of these diagrams but refer to the literature. It has been
shown that all two-loop diagrams in symmetry class $C$
give a vanishing contribution. In contrast there are two-loop diagrams
with a non-vanishing contribution in symmetry class $C$I -- these
have one parallel and one anti-parallel loop. Subtracting
doubly counted orbits they give the value $\tau/2$ which
is the linear order of the universal result.

We have shown that periodic-orbit theory in Andreev star graphs
can account for universal spectral statistics in non-standard symmetry classes. The remaining five non-standard symmetry classes can equally
be treated with star graphs but their treatment does not contain
any new ideas and we refer to the literature \cite{gnutzmann:2004b}.

\newpage
\thispagestyle{empty}
\chapter{The Supersymmetry approach to Quantum Graphs}
\label{chapter:susy}

In this chapter we will come back to universal spectral statistics in
the standard Wigner-Dyson symmetry classes and present a proof that the
statistics in large well-connected time-reversal invariant quantum graphs
follows the predictions of random-matrix theory
\cite{gnutzmann:2004c,gnutzmann:2005}.
This proof involves a field-theoretic description of 
spectral correlation functions in 
a specially adapted version of the 
supersymmetric non-linear $\sigma$-model. 
While the theory can be
presented for general simple graphs we restrict again to star graphs
which allows for some technical simplifications.
Similar methods have been used efficiently in disordered
systems \cite{Book:Efetov,mirlin,mirlin2,fyodorov,susy}
where an average over a disorder ensemble leads to
supersymmetric $\sigma$-models and universal spectral statistics
can be proven (for an ensemble of systems). 
The phases $\E^{\I k L_b}$ acquired during the propagation through
a bond are a source of disorder in an \emph{individual} quantum graph
with incommensurate bond lengths. This type of disorder eventually allows
for an \emph{exact mapping} to a variant of the nonlinear $\sigma$-model
as an additional powerful tool to analyse their spectral statistics. 
We will present the theory and the proof of universality for
the two-point correlator of the eigenphases of the quantum evolution map
of large star graphs. Generalisations to general graphs can be found in the
literature \cite{gnutzmann:2004c,gnutzmann:2005}.

\section{An exact supersymmetric model for spectral correlators}
\label{sec:susy_representation}

The starting point of the supersymmetry approach to the
two-point correlator of eigenphases of the quantum evolution
map is the expression
\begin{equation}
  \tilde{R}_2(s)=\frac{1}{8\pi^2} \left.\frac{\partial^2}{\partial j_+ \partial j_-}
  \right|_{j_\pm=0} \left\langle \xi(j_+,j_-;s) \right\rangle_k
\end{equation}
where 
\begin{equation}
  \xi(j_+,j_-;s)=
 \frac{
  \zetab(\theta_{+,\mathbf{F}};k)
  }{
  \zetab(\theta_{+,\mathbf{B}};k)
  }
  \frac{
  \zetab(\theta_{-,\mathbf{F}};k)^*}{
  \zetab(\theta_{-,\mathbf{B}};k)^*}
   \label{eq:generating_function}
\end{equation}
with
\begin{equation}
  \begin{split}
    \theta_{\pm,\mathbf{F}}=&
    \frac{2\pi}{B}\left( j_\pm  \pm \frac{s}{2}\right)\\
    \theta_{\pm,\mathbf{B}}=&
    \frac{2\pi}{B}\left(- j_\pm \pm \frac{s}{2} \right)\ .
  \end{split}
\end{equation}
$ \xi(j_+,j_-;s)$ 
is a generating function expressed as a quotient of secular functions.
A similar expression can be used for the spectral two-point correlator
$R_2(s)$ which we will not discuss here (for moderate bond length fluctuations we have shown in Chapter \ref{chapter:periodic_orbit}
that the two correlators are essentially equivalent for large graphs).

The generating function expressed
as a quotient of determinants is an ideal
starting point for supersymmetry approaches since it can easily be written
as a Gaussian integral over commuting and anti-commuting variables.
Powerful methods are available to perform averages on Gaussian
integrals which renders such an expression a desirable object.
We will not give an introduction to the supersymmetry method 
and the notions of \define{superdeterminant} $\mathrm{sdet} A$ of a \define{supermatrix}
$A$ or of its \define{supertrace} $\mathrm{str}\, A= \mathrm{tr} \,A_\mathbf{B}-\mathrm{tr}\,A_\mathbf{F}$ which
is available in many textbooks
and reviews 
\cite{Book:Efetov,Book:Haake,mirlin2,fyodorov}\footnote{One should 
be aware that there are different conventions for the 
definition of a supertrace
which might differ by an overall sign and consequently for the
superdeterminant which may be defined via $\mathrm{sdet}\, A= 
\E^{ \mathrm{str}\,\mathrm{ln} A}$. There are also different conventions
for the integration over anti-commuting numbers which differ by an overall constant.}, but we will give the main steps leading
to a convenient Gaussian expression for the generating function.

Defining the supervectors
\begin{equation}
  \psi=
  \begin{pmatrix}
    z_1\\
    \dots\\
    z_N\\
    \chi_1\\
    \dots\\
    \chi_N
  \end{pmatrix}
  \;\;
  \text{and}
  \;\;
  \tilde\psi=
  \begin{pmatrix}
    z_1^* & \dots & z_N^* & \tilde\chi_1 & \dots \tilde\chi_N
  \end{pmatrix},
\end{equation}
where $z_i$ are complex commuting  (\define{bosonic})
variables while $\chi_i$ and
$\tilde\chi_i$ are independent anti-commuting (\define{fermionic})
numbers, the quotient of
determinants of an $N \times N$ matrix $A_\mathbf{F}$ and a (positive definite)
$N \times N$ matrix $A_\mathbf{B}$ can be expressed as a Gaussian
integral
\begin{equation}
  \frac{\mathrm{det} \, A_\mathbf{F}}{
    \mathrm{det}\, A_\mathbf{B}}\equiv 
  \left(\mathrm{sdet}\, A\right)^{-1}=
  \int \D(\tilde\psi,\psi) e^{- \tilde\psi A \psi}.
\end{equation}
Here,
\begin{equation}
  A=
  \begin{pmatrix}
    A_\mathbf{B} & 0\\
    0 & A_\mathbf{F}
  \end{pmatrix}
\end{equation}
is a block-matrix in superspace (boson-fermion space) 
and the measure is given
by
\begin{equation}
  \D(\tilde\psi,\psi)= \pi^{-N}\prod_{i=1}^N 
  \D \mathrm{Re}(z_i) \D \mathrm{Im}(z_i) 
    \D \tilde\chi_i d\chi_i
\end{equation}
with $\int \D \chi_i \, \chi_i=1$ and $\int \D \chi_i=0$.  

Before applying this to \eqref{eq:generating_function}, it will be 
convenient
to double the matrix dimensions using
\begin{equation}
  \begin{split}
  \zetab(\theta;k)=&\mathrm{det}\!\! 
  \left( \ONE - \E^{-\I \theta} \evol_{\star}(k)
  \right)\\
  =&
  \mathrm{det}\!\!\left( \E^{-\I \theta} \scattering_\star\right)
  \mathrm{det}\!\!
  \begin{pmatrix}
    \mathbbm{1} & T_{\star}(k/2)\\
    T_{\star}(k/2) & \E^{\I\theta} \scattering_\star^\dagger
  \end{pmatrix}\ .
  \end{split} 
\end{equation}
At this point the doubling of dimension seems arbitrary --
it leads to simplifications at a later stage. 
Now we write the generating function as a Gaussian
superintegral
\begin{equation}
  \xi(j_+,j_-;s)=\int \D (\tilde\psi,\psi)\, \E^{-\I \frac{4\pi s}{B}}
  e^{-{\bm{S}}[\tilde\psi,\psi]}
\end{equation}
where
\begin{equation}
  \begin{split}
    \bm{S}[\tilde\psi,\psi]=& \tilde\psi_+
    \begin{pmatrix}
      \ONE & T_{\star}(k/2) \\
      T_{\star}(k/2) &z^*_+\scattering_\star^\dagger
    \end{pmatrix}
    \psi_+
    +\\
    & \tilde\psi_-
    \begin{pmatrix}
      \ONE & T_{\star}(k/2)^* \\
      T_{\star}(k/2)^* & z_- \scattering_\star
    \end{pmatrix}\psi_-.
  \end{split}
  \label{eq:action1}
\end{equation}
Here, $\psi=\{\psi_{a,s,x,b}\}$ is a $8B$-dimensional supervector
where, $a=\pm$ distinguishes between the retarded and the advanced
sector of the theory (components coupling to $\xi$ or $\xi^*$,
respectively).  The index $s=\mathbf{F},\mathbf{B}$ refers to
complex commuting and anti-commuting components (determinants in the
denominator and numerator, respectively), and $x=1,2$ to the internal
structure of the matrix kernel appearing in \eqref{eq:action1}.  The
$2\times 2$ matrices
\begin{equation}
  z_\pm \equiv
  \begin{pmatrix}
    \E^{-\I \theta_{\mathbf{B},\pm}} & 0\\
    0 & \E^{-\I \theta_{\mathbf{F},\pm}}
  \end{pmatrix}
  \label{eq:zpm}
\end{equation}
are diagonal matrices in superspace containing the appropriate 
phase factors in the boson-boson and fermion-fermion sector.

To account for the (optional) 
time-reversal invariance of the scattering matrix,
we introduce the doublets
\begin{equation}
  \begin{split}
    \Psi=& \frac{1}{\sqrt{2}}
    \begin{pmatrix}
      \psi\\
      \tilde\psi^T
    \end{pmatrix}
    \\
    \tilde\Psi=& \frac{1}{\sqrt{2}}
    \begin{pmatrix}
      \tilde\psi, & \psi^T \sigma_3^\mathbf{bf}
    \end{pmatrix},
  \end{split}
  \label{eq:doublets}
\end{equation}
where $\sigma_3^\mathbf{bf}\equiv \left(
\begin{smallmatrix}
  \mathbbm{1} & 0\\
  0 & -\mathbbm{1}
\end{smallmatrix}
\right)$ is the Pauli matrix in superspace. Notice that the lower
components of $\Psi$ emanate from the upper component by the time
reversal operation (transposition). 
For later reference, we note that the new fields
are related to each other through
\begin{equation}
  \Psi=\tau \tilde\Psi^T 
  \qquad 
  \tilde{\Psi}=\Psi^T \tau.
\end{equation}
The matrix $\tau$ is defined by
\begin{equation}
\label{tau_def}
  \begin{split}
    \tau=& E_\mathbf{B}\sigma_1^\mathbf{tr}
    -\I E_\mathbf{F}\sigma_2^\mathbf{tr},
  \end{split}
\end{equation}
where $\sigma_i^\mathbf{tr}$ are Pauli matrices in the newly
introduced `time-reversal' space and $E_\mathbf{B/F}$ are the
projectors on the bosonic and fermionic sectors, respectively. 
All we will need
to know to proceed is that $\tau$ obeys the conditions
\begin{equation}
\label{tau_prop}
  \tau^T=\tau^{-1}
  \qquad  \text{and}
  \qquad 
  \tau^2= \sigZ^\mathbf{bf}.
\end{equation}
The appearance of the matrix $\tau$ in \eqref{eq:doublets}
suggests to introduce the generalised matrix
transposition
\begin{equation}
  A^\tau \equiv \tau A^T \tau^{-1}.
\end{equation}
Using \eqref{tau_prop} and
$(A^T)^T=\sigZ^\mathbf{bf} A \sigZ^\mathbf{bf}$
for a supermatrix
\cite{Book:Efetov}, one finds that
the generalised transposition is an involution,
\begin{equation}
  (A^\tau)^\tau=A.
\end{equation}
For later reference we also note that
\begin{equation}
  \tilde{\Psi}_+ 
  A 
  \Psi_- 
  =
  \Psi_-^T 
  \sigZ^\mathbf{bf} 
  A^T 
  \tilde{\Psi}^T_+
  =
  \tilde{\Psi}_- 
  A^\tau 
  \Psi_+.
\end{equation}
With all these definitions, the action \eqref{eq:action1} now takes
the form
\begin{equation} 
  \begin{split}
    \bm{S}[\tilde\Psi,\Psi]=& \tilde\Psi_+
    \begin{pmatrix}
      \ONE & T_{\star}(k/2)\\
      T_{\star}(k/2) &z_+^*\mathcal{S}^\dagger
    \end{pmatrix}
    \Psi_+
    +\\
    & \tilde\Psi_-
    \begin{pmatrix}
      \ONE & T_{\star}(k/2)^* \\
      T_{\star}(k/2)^* & z_- \mathcal{S}
    \end{pmatrix}\Psi_-.
  \end{split}
  \label{eq:action2}
\end{equation}
where the matrix structure is again in the auxiliary index $x$ and we
have introduced the matrix
\begin{equation}
  \mathcal{S}=
  \begin{pmatrix}
    \scattering_\star & 0\\
    0 & \scattering_\star^T
  \end{pmatrix}
\end{equation} 
where the matrix structure is in time-reversal space.

We are now in a position to subject the generating functional to the
spectral average.
We do this by invoking the method discussed
in Section \ref{sec:phase_ergodicity}, whereby the average
over $k$ is replaced exactly by phase space averaging 
on the $B$-torus.
The only $k$-dependence is in $T_{\star}(k/2) $ 
which is replaced by
replace $T_{\star}(\{\phi_b\})=
\mathrm{diag}(\E^{\I\phi_b})$.
The averaging is written explicitly as
\begin{equation}
  \langle \xi(j_+,j_-;s)\rangle_\phi= \E^{-\I\frac{4\pi s}{B}}
  \int \D (\tilde\psi,\psi)
  \E^{-\bm{S}_0} \prod_{b=1}^B \int \frac{\D \phi_b}{2\pi}
  \E^{-\bm{S}_{1,b}}.
  \label{eq:action_phaseaverage}
\end{equation}
Here,
\begin{equation}
  \begin{split}
    \bm{S}_0=&\tilde\Psi_{+,1}\Psi_{+,1}+ \tilde\Psi_{-,1}\Psi_{-,1}
    +\\
    &\tilde\Psi_{+,2}z_+^*\mathcal{S}^\dagger\Psi_{+,2}
    +\tilde\Psi_{-,2}z_-\mathcal{S}\Psi_{-,2}
  \end{split}
\end{equation}
is the phase-independent part of the action and
\begin{equation}
  \bm{S}_{1,b}=2\tilde\Psi_{+,1,b} \E^{\I \phi_b}\Psi_{+,2,b}
    +2\tilde\Psi_{-,2,b}\E^{-\I\phi_b}\Psi_{-,1,b}.
\end{equation}
So far, we have not achieved much other than representing the spectral
determinants by a complicated Gaussian integral, averaged over phase
degrees of freedom. The most important step in our analysis will now
be to subject the generating function to an integral transform known
as the colour-flavour transformation \cite{zirnbauer:1996,zirnbauer_Brisbane:1998,zirnbauer_Cambridge:1999}. The
colour-flavour transformation amounts to a replacement of the
phase-integral by an integral over a new degree of freedom, $Z$.
Much better than the original degrees of freedom, the $Z$-field will
be suited to describe the long time behaviour of the system,
which is equivalent to the low energy sector $s\ll 1$ of the
field theory.

In a variant adopted to the present context (a single `colour' and $F$
`flavours') the colour-flavour transformation assumes the form
\begin{equation}
  \int \frac{\D \phi}{2\pi} \E^{ \eta_{+}^T \E^{\I \phi} \nu_{+} +
    \nu_{-}^T \E^{-\I\phi}\eta_{-} }
  =
  \int \D (\tilde{Z},Z)\  \mathrm{sdet}\left(\ONE-Z\tilde{Z}\right) \E^{
    \eta_{+}^T Z\eta_{-}+ \nu_{-}^T \tilde{Z}\nu_{+} },
  \label{eq:CF_trafo}
\end{equation}
where $\eta_\pm$ and $\nu_\pm$ are arbitrary $2F$ dimensional
supervectors and $Z$, $\tilde{Z}$ are $2F$-dimensional supermatrices.
The boson-boson and fermion-fermion block of these supermatrices are
related by $\tilde{Z}_{\mathbf{BB}}=Z_\mathbf{BB}^\dagger$, and
$\tilde{Z}_{\mathbf{FF}}=-Z_\mathbf{FF}^\dagger$, while the entries of
the fermion-boson and boson-fermion blocks are independent
anti-commuting integration variables. The integration
$\D (\tilde{Z},Z)$ runs over all independent matrix elements of $Z$ and
$\tilde{Z}$ such that all eigenvalues of
$Z_\mathbf{BB}Z_\mathbf{BB}^\dagger$ are less than unity and the
measure is normalised such that
\begin{equation}
  \int \D (\tilde{Z},Z)\, \mathrm{sdet}(\ONE-Z\tilde{Z})=1.
\end{equation}

We apply the colour-flavour 
transformation $B$ times -- once for each
phase $\phi_b$. As a result, we obtain a $B$-fold integral over
supermatrices $Z_b$. There is one flavour corresponding to the 
time-reversal index $t=1,2$. We combine all matrices $Z_b$
($\tilde{Z}_b$) into a single block-diagonal $4B$-dimensional
supermatrix $Z$ ($\tilde{Z}$) such that
\begin{equation}
  Z_{bts,b't's'}=\delta_{b,b'} Z_{b\, ts,t's'}.
\end{equation} 
The averaged generating function now has the form
\begin{equation}
  \begin{split}
    \langle \xi(j_+,j_-;s) \rangle=& \E^{\I \frac{4\pi s}{B}}
    \int \D (\tilde\psi,\psi) \int \D (\tilde{Z},Z)\\
    & \mathrm{sdet}(\ONE-\tilde{Z}Z)\,
    e^{-\bm{S}(\tilde{\Psi},\Psi,\tilde{Z},Z)}
  \end{split}
\end{equation}
where
\begin{equation}
  \begin{split}
    \bm{S}(\tilde{\Psi},\Psi,\tilde{Z},Z)=& \tilde{\Psi}_1
    \begin{pmatrix}
      \ONE & Z\\
      Z^\tau & \ONE
    \end{pmatrix}
    \Psi_1+\\
    & \tilde\Psi_2
    \begin{pmatrix}
      z_+^*\mathcal{S}^\dagger & \tilde{Z}^\tau\\
      \tilde{Z} & z_- \mathcal{S}
    \end{pmatrix}
    \Psi_2,
  \end{split}
\end{equation}
and we used $2\tilde\Psi_{1}Z\Psi_1=\tilde\Psi_{1}Z\Psi_1+
\tilde\Psi_{1}Z^\tau\Psi_1$,
$2\tilde\Psi_{2}\tilde{Z}\Psi_2=\tilde\Psi_{2}\tilde{Z} \Psi_2+
\tilde\Psi_{2}\tilde{Z}^\tau\Psi_2$. Here, the indices $1$, $2$ refer
to the auxiliary index $x$, and the matrix structure is in
advanced-retarded space.  Integrating the Gaussian fields $\tilde\Psi$
and $\Psi$ we arrive at the (exact) representation
\begin{equation} 
  \langle \xi(j_+,j_-;s) \rangle=
  \int \D (\tilde{Z},Z)\ \E^{-\bm{S}(\tilde{Z},Z)}
  \label{eq:field}
\end{equation}
where the action is given by
\begin{equation}
  \begin{split}
    \bm{S}(\tilde{Z},Z)=& -\mathrm{str}\,\mathrm{log}\, 
    (\ONE-\tilde{Z}Z)
    +\frac{1}{2} \mathrm{str}\,\mathrm{log}\,(\ONE - Z^\tau Z) \\&
    +\frac{1}{2} \mathrm{str}\,\mathrm{log}\,(\ONE - \mathcal{S}z_+
    \tilde{Z}^\tau z_-^* \mathcal{S}^\dagger \tilde{Z}),
  \end{split}
  \label{eq:exact_action}
\end{equation}
where the prefactor $\E^{-\I\frac{4\pi s}{B}}$ has cancelled.

What have we gained with expression \eqref{eq:exact_action}
apart from an exact reformulation of the two-point correlator
in terms of a complicated supersymmetric field theory? 
This question seems quite urgent since instead of
an integral over $B$ phase factors $\E^{\I\phi_b}$ we now have to
deal with an integral over $B$ pairs $Z_b$, $\tilde{Z}_b$ of supermatrices -- each 
of size $4\times 4$. 
The main difference of expression \eqref{eq:exact_action} with
respect to the definition \eqref{eq:generating_function} is the
direct coupling of the retarded and advanced sectors of the
theory. In the defining expression the retarded sector contributes
via (products of) periodic orbits of type $\mathrm{tr} 
\left( \bondprop_\star(\phi) \scattering_\star\right)^n$
while the advanced sector contributes further factors
involving $\scattering_\star^*$ instead. Such periodic orbits
gather quasi-random phases. In contrast, after the colour-flavour transformation we obtain a theory which may be expressed
as a sum over periodic orbits 
$\mathrm{str} \left(
\tilde{Z}^\tau \mathcal{S}\tilde{Z} \mathcal{S}^\dagger\right)^n$
of a quite different type where forward and backward scattering
alternate. The phase of such an orbit can be expected to
be less random. We will show in the next section how a saddle-point
analysis can be used to extract the universal contribution
to spectral statistics together with sufficient conditions
for a dominance of universal spectral correlation over
small deviations. It is not clear if the
supersymmetric method may also be useful for graphs which 
deviate strongly from universality like the Neumann star graph.

\section[The mean-field approximation and universality]{The mean-field approximation\\ and universality}
\label{sec:proof_universality}

The expressions \eqref{eq:field} and \eqref{eq:exact_action} 
for the generating function of the two-point correlation function
of the eigenphases of the quantum map of a quantum graph
are an exact identity.
While the integral over the modes $Z,\tilde{Z}$ cannot be performed
analytically in a closed form this expression is an ideal
starting point for saddle-point analysis for large graphs
$B\rightarrow \infty$. We will only be 
interested in the correlator for $s\ll B$
which allows us to expand the sources
$z_\pm$ defined in \eqref{eq:zpm} as
\begin{equation}
  z_\pm=\ONE - \I \frac{2\pi}{B}
  \left(\sigZ^\mathbf{bf}j_\pm \mp \frac{s}{2}\right)\ .
\end{equation}
Higher orders will vanish in the limit $B\rightarrow \infty$.
The resulting action can be written as a sum
\begin{equation}
    \bm{S}(Z,\tilde{Z})=\bm{S}_0(Z,\tilde{Z})+\bm{S}_s(Z,\tilde{Z})s
    +\bm{S}_+(Z,\tilde{Z})j_++\bm{S}_-(Z,\tilde{Z})j_-
\end{equation}
where
\begin{equation}
  \begin{split}
  \bm{S}_0(Z,\tilde{Z})=&
  -\mathrm{str}\,\mathrm{log}\left(\ONE-\tilde{Z}Z\right)+\\
  &
  +\frac{1}{2}\mathrm{str}\,\mathrm{log}\left(\ONE-Z^\tau Z\right)
  +\frac{1}{2}\mathrm{str}\,\mathrm{log}\left(\ONE-\mathcal{S}
  \tilde{Z}^\tau 
  \mathcal{S}^\dagger \tilde{Z}\right)\\
  \bm{S}_s(Z,\tilde{Z})=&
  -\I \frac{\pi}{B}\mathrm{str}\,
  \frac{
  \mathcal{S}\tilde{Z}^\tau 
  \mathcal{S}^\dagger\tilde{Z}}{
  \ONE-\mathcal{S}\tilde{Z}^\tau \mathcal{S}^\dagger \tilde{Z}}\\
  \bm{S}_+(Z,\tilde{Z})=&\I\frac{\pi}{B}\mathrm{str}\,
  \frac{\sigZ^\mathbf{bf} \mathcal{S} \tilde{Z}^\tau
  \mathcal{S}^\dagger \tilde{Z}
  }{\ONE-\mathcal{S}\tilde{Z}^\tau\mathcal{S}^\dagger \tilde{Z}}\\
  \bm{S}_-(Z,\tilde{Z})=&-\I\frac{\pi}{B}\mathrm{str}\,
  \frac{\mathcal{S} \tilde{Z}^\tau \sigZ^\mathbf{bf}
  \mathcal{S}^\dagger \tilde{Z}
  }{\ONE-\mathcal{S}\tilde{Z}^\tau\mathcal{S}^\dagger \tilde{Z}}.
  \end{split}
  \label{eq:linear_action}
\end{equation}

The saddle-point
manifold (which is also called the zero-mode for reasons to become 
clear below) is equivalent to a mean-field description in which all
system-dependent features (encoded in the matrix $\mathcal{S}$)
drop out.
We will show that a reduction of the action \eqref{eq:linear_action}
to the
saddle-point manifold  
yields an exact expression
for universal spectral two-point correlators. Later, in Section
\ref{sec:conditions} we will
analyse the validity of this approximation
and give sufficient conditions under which
deviations from the mean-field can be neglected in the limit
$B\rightarrow \infty$. We will assume time-reversal invariance $\scattering_\star=\scattering_\star^T$ 
(and thus $\mathcal{S}=\scattering_\star \otimes \ONE^\mathbf{tr}\equiv \scattering_\star$) throughout -- the
effect of breaking time-reversal invariance will be discussed
in Section \ref{sec:GOE_GUE}.

Let us start with the saddle-point 
equations for the action \eqref{eq:linear_action}.
Note that the source terms $j_\pm$ are the only terms
which break the supersymmetry -- in other words $\xi(j_+=0,j_-=0;s)=1$
which can be seen directly from the defining expression \eqref{eq:generating_function}. The two-point correlator
as a derivative of the generating function with respect to the
sources is a measure of the response of the supersymmetric
action at $j_\pm=0$ to breaking this symmetry.
A saddle-point analysis is justified
because the largest response
can be expected where the supersymmetric action is small.
This will be shown explicitly when we analyse the validity
of this approximation.

The first saddle point equation at $j_\pm=0$ and $s=0$ takes the form
\begin{equation}
  \frac{\delta \bm{S}_0}{\delta Z}=0=
  \mathrm{str}\, \frac{\tilde{Z}}{\ONE-\tilde{Z}Z}
  -\mathrm{str}\, \frac{Z^\tau}{\ONE-Z^\tau Z}
  \label{eq:first_saddle_point_equation}
\end{equation}
and has the solution
\begin{equation}
  \tilde{Z}=Z^\tau.
\end{equation}
The second saddle-point equation reads
\begin{equation}
\frac{\delta \bm{S}_0}{\delta \tilde{Z}}=0=
    \mathrm{str}\, \frac{Z}{\ONE-Z Z^\tau}
    -\mathrm{str}\, 
    \frac{\mathcal{S} Z \mathcal{S}^\dagger}{
    \ONE-\mathcal{S} Z \mathcal{S}^\dagger Z^\tau} 
\end{equation}
where we have used explicitly the solution $Z=\tilde{Z}^\tau$ 
of the first saddle-point equation and the
time-reversal invariance of the system
via $\mathcal{S}^\tau = \mathcal{S}$.
It is solved for field configurations $Z$ that commute with
the scattering matrix $\left[\mathcal{S}, Z\right]=0$.
This  
implies equidistribution of the field $Z$ over the bond index $b$.
The saddle-point manifold is thus given by the mean-field
configurations
\begin{equation}
  Z_{0\, b\,ts,t's'}=
  Y_{ts,t's'} \qquad \text{and} \qquad
  \tilde{Z}_{0\, b\, ts, t's'}= \tilde{Y}_{ts,t's'},
\end{equation}
where $Y=\tilde{Y}^\tau$ is a $4 \times 4$ supermatrix.
The commuting parts of $Y$  obey 
$Y^\tau_\mathbf{BB}=\tilde{Y}_{\mathbf{BB}}=Y^*_{\mathbf{BB}}$ and 
$Y^\tau_\mathbf{FF}=\tilde{Y}_{\mathbf{FF}} =-Y^*_{\mathbf{FF}}$ 
while the non-commuting entries of $Y$ are all independent integration
variables. The fermion-fermion part is integrated over
$\mathbbm{R}^4\simeq \mathbbm{C}^2$ while boson-boson part is
restricted to the compact region where all eigenvalues of
$Y_\mathbf{BB}^\dagger Y_\mathbf {BB}$ are less than unity.

Reducing the action \eqref{eq:linear_action} to the zero-mode the
first contribution vanishes exactly $\bm{S}_0(Z_0,\tilde{Z}_0)=0$
while the remaining term becomes\footnote{Note, that for $s\neq 0$ or $j_\pm\neq 0$
the second saddle-point equation is changed which leaves only two
saddle points in the reduced action \eqref{eq:action_GOE}. These can
serve as starting points for perturbative treatments of the correlator.
The loop expansion in the diagrammatic periodic-orbit treatment of the form-factor
in Chapter \ref{chapter:periodic_orbit} is essentially equivalent to an expansion around
the saddle-point $Z=\tilde{Z}=0$. Keeping the full saddle-point manifold
of $\bm{S}_0$ we are able to go beyond such perturbative expansions.}
\begin{equation}
  \bm{S}^\mathrm{GOE}
  (\tilde{Y},Y)=
  -\I \pi s\ \mathrm{str}\,
  \frac{Y \tilde{Y}}{
  \ONE-Y \tilde{Y}}+
  \I \pi j_+ \ \mathrm{str}\,
  \frac{\sigZ^\mathbf{bf} Y \tilde{Y}
  }{\ONE-Y\tilde{Y}}
  -\I \pi j_- \ \mathrm{str}\,
  \frac{\sigZ^\mathbf{bf} \tilde{Y}Y
  }{\ONE-\tilde{Y} Y}.
  \label{eq:action_GOE}
\end{equation}
Restricting the integration to the saddle-point manifold, we obtain
\begin{equation}
  \langle \xi(j_+,j_- ; s)\rangle\simeq   
  \xi^\mathrm{GOE}(j_+,j_- ; s)\rangle
  \equiv \int \D (\tilde{Y},Y)
  e^{-\bm{S}^\mathrm{GOE}(Y,\tilde{Y})},
\end{equation}
where the denotation $\xi^\mathrm{GOE}$ indicates that the matrix
integral over $Y$ obtains but an exact representation of the GOE
correlation function. 
The right hand side can be represented in more widely recognisable
form. Let us define the $8 \times 8$
supermatrix
\begin{equation}
  \begin{split}
    Q=&
    \begin{pmatrix}
      \ONE & Y\\
      \tilde{Y} & \ONE
    \end{pmatrix}
    \Sigma_z
    \begin{pmatrix}
      \ONE & Y\\
      \tilde{Y} & \ONE
    \end{pmatrix}^{-1}\\
    =&
    \begin{pmatrix}
      \ONE+2 Y\tilde{Y}/(\ONE-Y\tilde{Y}) &
      -2 Y/(\ONE-\tilde{Y}Y)\\
      2 \tilde{Y}/(\ONE-Y\tilde{Y})& -\ONE -2\tilde{Y}Y/(\ONE-\tilde{Y}Y)
    \end{pmatrix},
  \end{split}
\end{equation}
where $\Sigma_z=\left(
  \begin{smallmatrix} \mathbbm{1}& 0\\
    0 & -\mathbbm{1} \end{smallmatrix}\right)$.  It is then a
straightforward matter to show that the action $\bm{S}(\tilde{Y},Y)$
takes the form of Efetov's action \cite{Book:Efetov} 
for the GOE
correlation function
\begin{equation}
  \label{eq:Q_int}
  \xi^\mathrm{GOE}(j_+,j_-;s)=\int \D Q\, e^{\I \bm{S}(Q)}
\end{equation}
where the measure is given by $\D Q\equiv \D (\tilde Y,Y)$,
\begin{equation}
  \bm{S}(Q)=\frac{\pi}{2} \mathrm{str}\,
  (Q-\Sigma_z) \hat{\epsilon}
\end{equation}
and $\hat{\epsilon}=-\left(
\begin{smallmatrix}
  j_+\sigZ^\mathbf{bf}+\frac{s}{2} & 0 \\
  0 & j_-\sigZ^\mathbf{bf}-\frac{s}{2}
\end{smallmatrix}\right)$. 
For a discussion of the integral \eqref{eq:Q_int}, and the ways
random matrix predictions are obtained by integration over $Q$, we refer to the textbook \cite{Book:Efetov}.

\section{Validity of the mean-field approximation and sufficient conditions for universality}
\label{sec:conditions}

So far, we have shown that the reduction of the exact
supersymmetric field integral for the generating function $\xi$
to an integral over the saddle-point manifold (that is over mean-field configurations)
results in the GOE spectral correlations. However, we
have not yet shown under which conditions 
this reduction is actually
legitimate. Let us turn to this subject now.

In a full saddle-point analysis one writes the fields
as a sum $Z = Z_0+ \delta Z $ where $Z_0$ parameterises
the saddle-point manifold and $\delta Z $ describes the 
fluctuations around the saddle-point manifold. The reduction
of the exact field integral to the integral over the
saddle-point manifold is obtained by an expansion of
the action $\bm{S}(Z,\tilde{Z})$ in the fluctuations $\delta Z$
to second order and a subsequent Gaussian
integral over these modes. In general this will lead to
deviations from the mean-field result presented above.
We will investigate under which conditions these deviations
are small and vanish in the limit $B\rightarrow \infty$
of large graphs. 

As we will show later it is sufficient for this
purpose to consider
the expansion of the full action \eqref{eq:linear_action} to second
order in the fields $Z$,
\begin{equation}
  \begin{split}
    \bm{S}^{(2)}(Z,\tilde{Z})=&
    \bm{S}_0^{(2)}(Z,\tilde{Z})
    - \frac{\I \pi s}{B}
    \mathrm{str} 
    \left(\mathcal{S}\tilde{Z}^\tau
    \mathcal{S}^\dagger \tilde{Z}\right)
    +\\
    &\frac{\I \pi j_+}{B}
    \mathrm{str}\left(\sigZ^\mathbf{bf}
    \mathcal{S}\tilde{Z}^\tau
    \mathcal{S}^\dagger \tilde{Z}\right)
    -\frac{\I \pi j_-}{B}
    \mathrm{str}\left(
    \mathcal{S}\tilde{Z}^\tau \sigZ^\mathbf{bf}
    \mathcal{S}^\dagger \tilde{Z}\right)
    ,
  \end{split}
\end{equation}
where
\begin{equation}
  \begin{split}
    \bm{S}_0^{(2)}(Z,\tilde{Z})=&
    \mathrm{str}\left(
      \tilde{Z}Z-\frac{1}{2}Z^\tau Z -\frac{1}{2}
      \mathcal{S}\tilde{Z}^\tau\mathcal{S}^\dagger 
      \tilde{Z}
    \right)\\
    =&\frac{1}{2}\mathrm{str}\left[(Z-\tilde{Z}^\tau)(\tilde{Z}-Z^\tau)
    +\tilde{Z}^\tau \tilde{Z}
    -\mathcal{S}\tilde{Z}^\tau\mathcal{S}^\dagger\tilde{Z}\right].
  \end{split}
  \label{eq:quadratic_action}
\end{equation}
Physically, the quadratic action describes the joint propagation of a
retarded and an advanced Feynman amplitude along the same path in
configuration space. It thus carries information similar to that
obtained from the diagonal approximation 
in the periodic orbit approach. More precisely it is equivalent
to neglecting deviations from $\tilde{K}_n^{\mathrm{diag}}= 2n \mathrm{tr}\,
\mathcal{M}^n /B$ which come from repetitions and self-retracing orbits.
The second
order expansion is justified if the fluctuations of the fields $Z$ are
massively damped (in the sense that the matrix elements of $Z$
effectively contributing to the integral are much smaller than unity.)
Under these conditions, the integration over matrix elements of $Z$
may be extended to infinity and we obtain a genuine Gaussian integral.
In fact, one is forced to extend the integration
to infinity in order to preserve $\xi(j_\pm;s)=1$ at every level
of approximation.

The eigenvalues $m_\ell$ of the quadratic form appearing in $\bm{S}^{(2)}_0$ determine the damping $m_\ell$ -- or the \define{mass}, in a field
theoretical jargon -- inhibiting fluctuations of an eigenmode.
As indicated by its name, the zero-mode $Z_0$ carries zero
mass. With the observation that the quadratic form involves the
classical map $\mathcal{M}_{\star\, bb'}=|\scattering_{\star\, bb'}|^2$
of the graph via 
\begin{equation}
  \mathrm{str}\, \mathcal{S}\tilde{Z}^\tau
  \mathcal{S}^\dagger \tilde{Z}\equiv
  \mathrm{str}\sum_{bb'} \scattering_{\star\, bb'}\tilde{Z}^\tau_{b'}
  \scattering^*_{\star\, bb'}\tilde{Z}_b=
  \mathrm{str}\sum_{bb'} \tilde{Z}_b M_{\star\, bb'} \tilde{Z}_{b'}
\end{equation}
the action is easily diagonalised by an orthogonal
transformation $Z_b=\sum_{b'}O_{b\ell}Z'_{\ell}$ 
(we will drop the primes of $Z'_\ell$ in the sequel)
and can then be written as a sum over separate contributions
\begin{equation}
  \bm{S}^{(2)}(Z,\tilde{Z})=\sum_{\ell=1}^B
  \bm{S}^{(2)}_\ell(Z_\ell,\tilde{Z}_\ell)
\end{equation}
where $\bm{S}^{(2)}_{\ell}$ is the contribution 
\begin{equation}
  \begin{split}
    \bm{S}^{(2)}_\ell(Z_\ell,\tilde{Z}_\ell)=&
    \frac{1}{2}\mathrm{str}\left[(Z_\ell-\tilde{Z}^\tau_\ell)(
    \tilde{Z}_\ell-{Z}^\tau_\ell)
    +(1-\nu_\ell)\tilde{Z}^\tau_\ell \tilde{Z}_\ell\right]-\\
    &
    \nu_\ell\frac{\pi \I}{B}\mathrm{str} 
    \left[
    s\tilde{Z}_\ell^\tau\tilde{Z}_\ell
    -\left(j_+\sigZ^\mathbf{bf}\tilde{Z}^\tau_\ell
    -j_- \tilde{Z}_\ell^\tau \sigZ^\mathbf{bf}\right)\tilde{Z}_\ell
    \right]
  \end{split}
  \label{eq:action_mode}
\end{equation}
of the eigenmode of the classical map $\mathcal{M}$ with eigenvalue 
$\nu_\ell$. 
Let us now focus on the integral over one mode $Z_\ell$,
$\tilde{Z}_\ell$.
We can split the action of a single mode into contributions from
a pair fields $Z_\ell$, $\tilde{Z}_\ell$ that
satisfies the first saddle-point equation $Z=\tilde{Z}^\tau$ 
and contributions violating it. 
One can show that only the configurations $Z=\tilde{Z}^\tau$
contribute non-trivially while
the integral over configurations violating this
condition just give a factor unity\footnote{
  For the integral over anti-commuting numbers this can easily seen 
  from the action \eqref{eq:action_mode} where the shifts 
  $Z_{\ell, 
  \mathbf{BF/FB}}\mapsto 
  Z_{\ell\, \mathbf{BF}}-\tilde{Z}^\tau_{\ell\,\mathbf{BF/FB}}$ 
  decouple
  the anticommuting parts of $Z_\ell$ from the anti-commuting part of 
  $\tilde{Z}$. For the commuting entries one can write down the integral
  over all real and imaginary parts of the entries
  $Z_{\ell\, \mathbf{BB/FF}}$ and $\tilde{Z}_{\ell\, \mathbf{BB/FF}}$ --
  shifting the entries and changing the contours of integration
  leads to a similar decoupling such that the action 
  at $s=j_\pm=0$ is
  of the form 
  $\bm{S}_\ell= \frac{1}{2}\mathrm{str}\left[
  +(1-\nu_\ell) \tilde{Z}_{\ell\, \mathrm{even}}Z_{\ell\, 
  \mathrm{even}})\right]$
  where $Z_{\ell\, \mathrm{even}}=\tilde{Z}^\tau_{\ell\, \mathrm{even}}$
  and $Z_{\ell\, \mathrm{odd}}=-\tilde{Z}^\tau_{\ell\, \mathrm{odd}}$ and
  the terms at $s\neq0$ and $j_\pm\neq 0$ only 
  involve $Z_{\ell\, \mathrm{even}}$.}. 
We are left with half the number
of integration variables and an action 
$\bm{S}_\ell^{(2)}=\frac{1-\nu_\ell}{2}
\mathrm{str}\, Z\tilde{Z}=\frac{m_\ell}{2}\mathrm{str}\, Z \tilde{Z}$ 
for vanishing sources $j_\pm=s=0$. In this 
expression we see that the eigenvalues $\nu_\ell$ of the classical map
are related to the masses $m_\ell=1-\nu_\ell$ of the supersymmetric field  theory. Performing the remaining integral we will see that the mass $m_\ell$
sets the scale for the response to the sources $j_\pm$ which
break the supersymmetry of the model. That is, a large mass $m_\ell$
corresponds to a strongly damped contribution to the two-point correlator.
The remaining Gaussian integral over one mode gives
\begin{equation}
  \begin{split}
  I_\ell=&\int \D (Z_\ell,\tilde{Z}_\ell) 
  \E^{-\bm{S}_\ell^{(2)}(Z_\ell,\tilde{Z}_\ell)}\\
  =&\frac{[1+i\frac{\pi(m_\ell-1)(s+j_\Sigma)}{m_\ell B}]^2
  [1+i\frac{\pi(m_\ell-1)(s-j_\Sigma)}{m_\ell B}]^2
  }{[1+i\frac{\pi(m_\ell-1)(s+j_\Delta)}{m_\ell B}]^2
  [1+i\frac{\pi(m_\ell-1)(s-j_\Delta)}{m_\ell B}]^2}
  \end{split}
\end{equation}
where $j_\Delta=j_+-j_-$ and $j_\Sigma=j_++j_-$.
The generating function in this approximation is just the 
product
\begin{equation}
  \xi^{(2)}(j_+,j_-;s)=\prod_{\ell=1}^B I_\ell
\end{equation}
over the contributions from each mode.
Differentiating
with respect to the sources $j_\pm$ we finally obtain the quadratic approximation to
the correlation function,
\begin{equation}
  \begin{split}
    \tilde{R}^{(2)}_2(s)=&\sum_{\ell=1}^B \frac{1}{8\pi^2}\mathrm{Re}\,
    \frac{\partial^2}{\partial j_+ \partial j_-}
    I_\ell \big|_{j_\pm=0}\\
    =&\sum_{\ell=1}^B \frac{(m_\ell-1)^2(m_\ell^2 B^2 - 
    \pi^2 (m_\ell-1)^2 s^2)}{ (m_\ell^2 B^2 +
      \pi^2 (m_\ell-1)^2 s^2)^2}.
  \end{split}
\end{equation}
The contribution of the zero mode ($m_1=0$) is given by $-\frac{1}{\pi^2
  s^2}$ and coincides with the diagonal approximation to the GOE
correlation function. Later on we shall see that in the case of
broken time reversal invariance, one half of the matrix elements of
$Z_0$ become massive implying that the contribution of the zero mode
reduces to the GUE expression $-\frac{1}{2\pi^2 s^2}$.

In the limit $B\to \infty$, the $s$-dependence of the contribution of
massive modes to the correlation function is negligible for our
purpose, i.e.  individual modes contribute maximally as
$\sim(m_\ell-1)^2/2m_\ell^2 B^2\sim (m_\ell B)^{-2}$.  Only modes
of mass $m_\ell\sim B^{-\alpha}$, where $\alpha$ is a non--vanishing
positive exponent, can survive the limit $B\to \infty$.  The contribution
of an individual mode is negligible if the exponent $0\le \alpha <1$.
There are at most $\mathcal{O}(B)$ nearly massless
modes, and we are led to require that $ B^{2\alpha-1}$ must vanish in
the limit of large graphs $B\rightarrow \infty$, or that $0\le \alpha
<1/2$. 

This is a slightly stronger condition than the one discussed
within the context of the periodic-orbit analysis of Section
\ref{sec:diagonal_approximation}, where we have seen
that for \emph{fixed} $\tau$ the form factor
is universal if $0\le\alpha<1$. The two results are 
fully consistent as the condition $0\le \alpha <1/2$ 
implies the stronger statement that
deviations to the two-point correlator
vanish uniformly in $s$.

In the
intermediate region $1/2 \le \alpha <1$ -- permissible by Tanner's
criterion -- non-universal corrections vanish only if the number $r$
of classical modes with a small mass remains constant (or does not
grow too fast) such that $B^2 \Delta_g^2/r \rightarrow \infty$. If,
however, the number of low energy modes is extensive, $r\sim B$, the
stricter condition $0\le \alpha<1/2$ has to be imposed. 
An example of a graph with
$\mathcal{O}(B)$ almost massless modes is the Neumann star
graph, for which all the classical modes (apart from the zero-mode
$m_1=0$) have a mass
$m\sim 1/B$. The Neumann star thus strongly violates
the condition for universal GOE statistics -- as a single
mode with mass $m\sim 1/B$ is already sufficient to give non-vanishing
corrections.

Above we have shown that in the limit $B\to \infty$ only the zero mode
effectively contributes to the correlation function (provided, of
course, the master condition $\Delta_g \sim B^{-\alpha}$ is met.)
While the zero mode integral must be performed rigorously, all other
modes are strongly overdamped and may be treated in a quadratic
approximation. This is the a posteriori justification for the
quadratic approximation on which our analysis of the mass spectrum was
based.

\section{Breaking time-reversal invariance}
\label{sec:GOE_GUE}

The analysis above applies to time reversal invariant graphs. In this
section we discuss what happens if time reversal invariance gets
gradually broken.  
We assume full universality, i.e. $B \Delta_g^2 \gg 1$ such that only
the zero-mode contributes to $R_2(s)$. Our aim is to derive a
condition for the crossover between GOE-statistics in
the time-reversal invariant case
and GUE--statistics for fully broken time-reversal
 invariance.

The substructure of the $Z$--fields with $Z=\tilde{Z}^\tau$
in time reversal space is given by
\begin{equation}
    Z_b=
    \begin{pmatrix}
      Z_{\mathbf{D}\,b}& Z_\mathbf{C\, b}\\
      \tilde{Z}_{\mathbf{C}\,b}^T\sigma_3^\mathbf{BF} &
      \tilde{Z}_{\mathbf{D}\,b}^T
    \end{pmatrix},\qquad
    \tilde{Z}_b=
    \begin{pmatrix}
      \tilde{Z}_{\mathbf{D}\,b} & \sigma_3^\mathbf{BF}
      Z_{\mathbf{C}\,b}^T\\
      \tilde{Z}_{\mathbf{C}\,b} & Z_{\mathbf{D}\,b}^T
    \end{pmatrix},
  \label{eq:zmode_goe}
\end{equation}
where $Z_{\mathbf{D/C}\, b}$ and $\tilde Z_{\mathbf{D/C}\,b}$ are $2\times 2$
supermatrices subject to the constraint $\tilde{Z}_{\mathbf{D/C}\,\mathbf{BB}}
=Z^*_{\mathbf{C/D}\, \mathbf{BB}}$ and $\tilde{Z}_{\mathbf{C/D}\,\mathbf{FF}} =-Z^*_{\mathbf{C/D}\, \mathbf{FF}}$,
while the non-commuting entries of these matrices are independent
integration variables. The subscripts $\mathbf{D} (\mathbf{C})$ allude
to the fact that in disordered fermion systems, the modes
$Z_\mathbf{D}$ ($Z_\mathbf{C}$) generate the so-called diffusion
(cooperon) excitations. Physically, the former (latter) describe the
interference of two states as they propagate along the same path (the
same path yet in opposite direction) in configuration space. Cooperon
modes are susceptible to time reversal invariant breaking
perturbations.

Substituting this representation into the quadratic action, we obtain
\begin{equation}
   \bm{S}^{(2)}=\mathrm{str}\left(
   \tilde{Z}_\mathbf{D}(\ONE-\mathcal{M}_{\star})Z_\mathbf{D}
+\tilde{Z}_\mathbf{C}(\ONE-\mathcal{R}_{\star})Z_\mathbf{C}\right)
  \label{eq:second_order_action_CD}
\end{equation}
for the action of the zero mode at $j_\pm=s=0$. Here,
$\mathcal{M}_{\star\,bb'}=\{|\scattering_{\star\, b,b'}|^2\}$ 
is the classical evolution map and
$\mathcal{R}_{\star\, bb'}= \scattering_{\star\, b,b'}
\scattering_{\star\,b'b}^*$. For broken time-reversal invariance
$\mathcal{M}_{\star}\neq \mathcal{R}_{\star}$ and the symmetry of the action in time
reversal invariance space gets lost.

Noting that $B=\sum_{b,b'} |\scattering_{\star\, bb'}|^2$, we conclude that the
Cooperon zero mode $Z_{\mathbf{C},b}=Y_{\mathbf{C}}$ 
acquires a mass term $\sim B
m_\mathbf{C}\,\mathrm{str}(Y_\mathbf{C}\tilde Y_\mathbf{C})$, where
the coefficient
\begin{equation}
  \begin{split}
    m_\mathbf{C}&=\frac{1}{B} \left|\sum_{bb'}
      \scattering_{\star\,b b'}
      (\scattering^*_{\star \, b b'}-
      \scattering^*_{\star\,  b' b})\right|\\
    & =\frac{1}{B} \left|\mathrm{tr}\,
      \scattering_\star^\dagger(\scattering-\scattering_\star^T
      )\right|
  \end{split}
\end{equation} 
measures the degree of the breaking of the symmetry $S=S^T$.
We have encountered the Cooperon mass before
in Section \ref{sec:diagonal_approximation} as
the inverse time $n_T=1/m_{\mathbf{C}}$ on which a pair of time reversed
orbits decays in the diagonal approximation. 
The Cooperon mode may be neglected once $B m_\mathbf{C} \rightarrow
\infty$ as $B\rightarrow \infty$.

\section{On higher order correlation functions and the gap condition}
\label{sec:higher_orders}

The supersymmetry method described above can be generalised
in a straight forward manner to higher order correlation functions with
a generating function that contains an appropriate number
of spectral determinants $\zetab$ in its enumerator and denominator.
A reduction to a saddle-point manifold 
gives the universal correlations known from random-matrix theory.
It all works the same way as for the two-point correlator and
the condition that this reduction becomes exact in the limit of
large graphs also remains unchanged.
For a finite graph this prooves universal spectral correlators
upto small deviations in correlators of order $n$ where $n\ll B$.

\section*{Acknowledgments}

We are grateful to Alexander Altland, Gregory Berkolaiko, Peter Braun, 
Fritz Haake, Jonathan Keating, Tsampikos Kottos,  
Marek Ku\'s, Felix von Oppen,
Holger Schanz, Michail Solomyak, and Karol \.{Z}yczkowski for discussions and help
through the years. Both authors would like to thank the School of
Mathematics at Bristol university for the hospitality extended
during the period this manuscript was prepared. Support is
acknowledged from the Minerva foundation, the Einstein (Minerva)
Center and The Minerva Center for non-linear physics (Weizmann
Institute), the Israel Science Foundation, the German-Israeli
Foundation, the EPSRC grant GR/T06872/01 and the Institute of
Advance Studies, Bristol University.

\newpage
\thispagestyle{empty}

\appendix

\chapter{The symmetry classes of quantum systems}
\label{app:symmetry_classes}

Based on earlier ideas of Wigner \cite{wigner:1958}, Dyson introduced a three-fold
classification of quantum systems according to their behaviour under 
time-reversal, spin and rotational invariance \cite{dyson:1962,dyson:1962a,dyson:1962b}.  
This symmetry classification turned out to be very useful, for
instance in semiclassical, disordered and random-matrix
approaches to complex quantum systems. In this appendix
we give a short summary of this three-fold symmetry classification
and a recent extension to a ten-fold classification which is related 
with the ten classes of Riemannian symmetric spaces.
We only present a description of the ten symmetry classes
and refer to the literature \cite{zirnbauer:symmetry} for a proof that this classification is complete.
Table \ref{table:symmetry_classes} summarises this appendix.

\begin{table}[b]
  \begin{center}
    \begin{tabular}{c|ccc|c}
      symmetry class &
      $\mathcal{T}$& $\mathcal{P}$ & $\mathcal{C}$ & symmetric space
      \\
      \hline
      \hline
      &&&&\\[-0.3cm]
      $A$&0&0&0& $U(N)$\\
      $A$I&+1&0&0& $U(N)/O(N)$\\
      $A$II&-1&0&0&$U(2N)/Sp(N)$\\
      $A$III&0&+1&0&$U(p+q)/U(p)\times U(q)$\\
      $BD$I&+1&+1&+1&$SO(p+q)/SO(p)\times SO(q)$\\
      $C$II&-1&+1&-1&$Sp(p+q)/Sp(p)\times Sp(q)$\\
      $C$&0&0&-1&$Sp(N)$\\
      $C$I&+1&-1&-1&$Sp(N)/U(N)$\\
      $BD$ ($D$) &0&0&+1&$SO(N)$\\
      $D$III &-1&-1&+1&$SO(2N)/U(N)$
    \end{tabular}
  \end{center}
  
  \caption{   \label{table:symmetry_classes}
    The ten symmetry classes of quantum systems. If a symmetry class
    obeys time-reversal symmetry or a spectral mirror symmetry
    the entry $\pm 1$ in the corresponding column indicates
    if the symmetry operator squares to $\pm \ONE$. 
    The entry $0$ indicates that the corresponding symmetry is broken.
    The last column gives the corresponding Riemannian symmetric space 
    (of compact type).
  }
\end{table}

In the literature the symmetry classes are often denoted by the
Gaussian random-matrix ensemble of hermitian matrices respecting the
corresponding symmetry restrictions. Such a notation is
misleading and often quite confusing.
Here, we stick to the notation introduced by Zirnbauer \cite{zirnbauer:symmetry}.

According to a theorem by Wigner a symmetry of a physical
system is represented in quantum mechanics 
either by a unitary operator or by an anti-unitary operator 
that commutes (or anti-commutes) with the Hamilton operator $H$.  
Unitary symmetry
operators which commute with $H$ are associated with constants of the motion. In the eigenbasis
of such a unitary symmetry operator the Hamilton operator is represented by a block diagonal matrix -- each block connected to one eigenvalue of the symmetry operator which serves as a quantum number of the state. 
In the following we will always assume that any unitary symmetry has been used to reduce the Hilbert space.

Anti-unitary symmetry operators which commute with $H$ 
do not lead to a constant of motion, yet they effect the form
of the representation of the Hamilton operator in a suitable basis.
More importantly, they have an effect on statistical properties
of the spectrum and of the wavefunctions
in complex quantum systems.
The same is true for (unitary and anti-unitary) symmetry operators
that anti-commute with $H$.

Though every symmetry of a physical system is represented by a unitary or anti-unitary operator in quantum mechanics
the reverse is not true. An arbitrary unitary or anti-unitary operator $\mathcal{U}$ in general \emph{does not} qualify as a symmetry of the system.
E.g. for any Hamilton operator $H$ one may uniquely define a anti-unitary operator $\mathcal{A}$ 
by its action $\mathcal{A}\sum_n a_n |n\rangle = \sum_n a_n^* |n\rangle$
in the eigenbasis $H|n\rangle= E_n |n\rangle$ of the Hamilton operator.
Trivially, $\mathcal{A}$ commutes with the Hamilton operator 
$[\mathcal{A},H]=0$ without being connected to any symmetry of the system.
In contrast a symmetry operator can be defined representation independent
which includes that it can be defined independent of the value one formally
assigns to $\hbar$. This immediately rules out the trivial operator $\mathcal{A}$ since the eigenbasis will generally depend on the value of $\hbar$. 

\section{Time-reversal invariance}
\label{sec:time-reversal}

Consider a physical system represented by the Hamilton operator $H$.
The system is \define{time-reversal invariant} if  
an anti-unitary symmetry operator $\mathcal{T}$, the 
\define{(generalised) time-reversal operator} exists
which commutes with the Hamilton operator
\begin{equation}
  \left[\mathcal{H},\mathcal{T}\right]=0.
  \label{eq:timerevinv}
\end{equation}
It can be shown \cite{Book:Haake,gnutzmann:2004a} 
that such an operator obeys either
\begin{equation}
  \mathcal{T}^2= \ONE \qquad \text{or} \qquad \mathcal{T}^2= -\ONE.
  \label{eq:time_rev_phase}
\end{equation} 
and $\mathcal{T}$ reverses the direction of time in the 
Schr\"odinger equation.

It follows that there are three symmetry classes, the \define{Wigner-Dyson classes} connected to time-reversal invariance:

\begin{center}
  \begin{tabular}{ll}
  &\\
  Class $A$:& 
  \begin{minipage}{8cm}
    Systems with broken time--reversal invariance.
  \end{minipage}
  \\
  \\
  Class $A$I:&
  \begin{minipage}{8cm} 
    Time-reversal invariant systems  
    and $\mathcal{T}^2=\ONE$.
  \end{minipage}\\
  \\
  Class $A$II:& 
  \begin{minipage}{8cm}
    Time-reversal invariant systems 
    and $\mathcal{T}^2=-\ONE$.
  \end{minipage}\\
  \\
  \end{tabular}
\end{center}

A particle in a potential described by the
Hamilton operator $H=\frac{\vec{p}^2}{2m}+ V(x)$ is time-reversal invariant.
The corresponding time-reversal operator 
$\mathcal{T}_{\text{$A$I}}$
is defined to be the anti-unitary operator obeying
$\mathcal{T}_{\text{$A$I}}\, \vec{x}\,
\mathcal{T}_{\text{$A$I}}^{-1}=\vec{x}$ and
$\mathcal{T}_{\text{$A$I}}\, \vec{p}\, \mathcal{T}_{\text{$A$I}}^{-1}=-\vec{p}$. 
In coordinate representation $\mathcal{T}_{\text{$A$I}}$ 
is represented as complex conjugation of the wave function 
$\langle \vec{x}| \mathcal{T}_{\text{$A$I}}|\psi\rangle=\langle \vec{x}|\psi\rangle^*$. Obviously, $\mathcal{T}_{\text{$A$I}}^2=\ONE$
and such systems belong to class $A$I.

For particles with spin $s=1/2$
and broken spin rotation invariance one may define the time-reversal
operator $\mathcal{T}_{\text{$A$II}}$, 
which acts as $\mathcal{T}_{\text{$A$I}}$
on $\vec{x}$ and $\vec{p}$ and additionally obeys $\mathcal{T}_{\text{$A$II}}
\, \vec{s}\, \mathcal{T}_{\text{$A$II}}^{-1}=-\vec{s}$.
For this operator $\mathcal{T}_{\text{$A$II}}^2=-\ONE$ and the system
is in class $A$II if $[H,\mathcal{T}_{\text{$A$II}}]=0$. 

In both cases time-reversal is broken by the presence of a magnetic field.
Thus a particle in a magnetic field generally belongs to class $A$.
Note however, that we made a special choice of the time-reversal operator
in our examples.  

\section{Kramers' degeneracy}
\label{sec:kramers_degeneracy}

The energy spectrum of any physical
system which is time-reversal invariant with
$\mathcal{T}^2=-\ONE$ is (at least) two-fold degenerate.
If $|n\rangle$ (normalised such that $\langle n| n\rangle=1$)
solves the stationary
Schr\"odinger equation $H|n\rangle =E_n |n\rangle$ so does
$\mathcal{T}|n\rangle$. Since $\langle \mathcal{T} n|\mathcal{T} n\rangle
= \langle n| n\rangle^*=1$ the state $\mathcal{T}|n\rangle$ does not vanish.
At the same time $|n\rangle$ and $\mathcal{T}|n\rangle$ are orthogonal
$\langle n | \mathcal{T} n\rangle=\langle \mathcal{T}n |
\mathcal{T}^2 n\rangle^*=-\langle n | \mathcal{T} n\rangle=0$ such
that $E_n$ is two-fold degenerate.

\section[Chiral symmetries and charge conjugation symmetries]{Chiral symmetries\\ and charge conjugation symmetries}
\label{sec:novel_symmetries}

We will now consider certain unitary 
symmetry operators $\mathcal{P}$ and anti-unitary operators $\mathcal{C}$
which anti-commute with the Hamilton operator,
\begin{equation}
  [H,\mathcal{P}]_+=0 \qquad \text{or}
  \qquad  [H,\mathcal{C}]_+=0\ .
  \label{eq:anticommuting_symmetries}
\end{equation}
One may have to shift the Hamilton operator by a constant to reveal
such a kind of symmetry which is tacitly assumed in the following.
Equation \eqref{eq:anticommuting_symmetries} 
leads to an energy spectrum which is symmetric. If $E$ is in the spectrum,
so is $-E$. 
Since the Hamilton operator of a proper quantum mechanical system
should be bounded from below such symmetries 
either occur in systems with finite-dimensional Hilbert space
(e.g. a finite number of coupled spins) or
in the framework of one-particle
equations of a field theory. Examples are the Dirac and Bogoliubov-de-Gennes equations 
which are formally equivalent to the Schr\"odinger equation, that is
they can be written in the form 
$\I \hbar \frac{\partial}{\partial t} |\Psi(t)\rangle= H |\Psi(t)\rangle$. 
The negative part of spectrum of these
operators is related to positive energy excitations of the corresponding
field theory. Since we will not consider the complete field theory
here we will call the one-particle operator $H$ the Hamilton operator 
of the system. 

A \define{chiral symmetry operator} $\mathcal{P}$ is unitary
and obeys $\mathcal{P}^2=\ONE$. 
A physical system with a chiral
symmetry can have an additional time-reversal invariance.
In that case the time-reversal operator $\mathcal{T}$ and the chiral
symmetry operator $\mathcal{P}$ are required to commute
$[\mathcal{T},\mathcal{P}]=0$. Thus there are three
\define{chiral symmetry classes}:

\begin{center}
  \begin{tabular}{ll}
  &\\
  Class $A$III:& 
  \begin{minipage}{8cm}
    Systems with a chiral symmetry and broken time--reversal   
    invariance.
  \end{minipage}
  \\
  \\
  Class $BD$I:&
  \begin{minipage}{8cm} 
    Time-reversal invariant systems with a chiral symmetry 
    and $\mathcal{T}^2=\ONE$.
  \end{minipage}\\
  \\
  Class $C$II:& 
  \begin{minipage}{8cm}
    Time-reversal invariant systems with a chiral symmetry
    and $\mathcal{T}^2=-\ONE$.
  \end{minipage}\\
  \\
  \end{tabular}
\end{center}

The chiral symmetry classes have first been considered in Quantum 
Chromodynamics as
Dirac particles in a random gauge potential \cite{chiral1,chiral2}. 
They also occur
for Bogoliubov-de-Gennes quasiparticle excitations in the certain types of disordered superconductors \cite{altland}. 

Finally a \define{charge conjugation symmetry} is described by
an \emph{anti-unitary} operator $\mathcal{C}$ that \emph{anti-commutes}
with the system Hamilton operator and which obeys either $\mathcal{C}^2=\ONE$
or $\mathcal{C}^2=-\ONE$. Accounting for the possibility of additional
time-reversal invariance there are six symmetry classes which have
a charge conjugation symmetry all of which can be found in certain types of
disordered superconductors. Two of them have been presented as
the time-reversal invariant chiral symmetry classes. Indeed,
if $\mathcal{P}$ is a chiral symmetry operator and $\mathcal{T}$
with $[\mathcal{P},\mathcal{T}]=0$ a time-reversal operator
$\mathcal{C}=\mathcal{P}\mathcal{T}$ is a charge conjugation
operator with $[\mathcal{C},\mathcal{T}]=0$ and $\mathcal{C}^2=\mathcal{T}^2$.
The remaining four symmetry classes have not been given any satisfactory
and consistent name. We will call them \define{charge conjugation symmetry classes} though the symmetry operator $\mathcal{C}$ may not be related to any
physical charge conjugation (see below). 
In any case the distinction 
between the three chiral symmetry classes and the four charge conjugation 
classes has historical origins and should not be taken as a mathematical
or physical relevant distinction. The four charge conjugation symmetry classes are:

\begin{center}
  \begin{tabular}{ll}
  &\\
  Class $C$:& 
  \begin{minipage}{8cm}
    Systems with broken time-reversal invariance and
    a charge conjugation symmetry obeying $\mathcal{C}^2=-\ONE$.
  \end{minipage}
  \\
  \\
  Class $C$I:&
  \begin{minipage}{8cm} 
    Time-reversal invariant systems with a charge conjugation symmetry 
    obeying both,  $\mathcal{T}^2=\ONE$ and $\mathcal{C}^2=-\ONE$.
  \end{minipage}\\
  \\
  Class $BD$:& 
  \begin{minipage}{8cm}
    Systems with broken time--reversal invariance and
    a charge conjugation symmetry obeying $\mathcal{C}^2=\ONE$.
    This symmetry class is denoted as class $D$ if 
    the Hilbert space has even dimension.
  \end{minipage}\\
  \\
  Class $D$III:& 
  \begin{minipage}{8cm}
    Time-reversal invariant systems with a charge conjugation symmetry 
    obeying both,  $\mathcal{T}^2=-\ONE$ and $\mathcal{C}^2=\ONE$.
  \end{minipage}\\
  \\
  \end{tabular}
\end{center}
They have first been discussed in connection to hybrid
normalconducting-superconducting structures and
disordered superconductors \cite{altlandzirnbauer,altlandzirnbauer2}.

Note, that the notation for these symmetries
stems from the systems for which
they have first been discussed in detail
where they were connected to the charge conjugation of a quasiparticle or to
the chirality of a Dirac particle. In general these symmetries need not
have this physical interpretation: any of the seven symmetry classes
can be realised by a Hamilton operator for two coupled
spins $\vec{S}_{1,2}$ with spin quantum numbers $s_{1,2}$ where 
the (generalised) chiral symmetry operator or (generalised) charge conjugation operator are not related to the chirality or electric charge of a particle.

\section{The symmetry classes for quantum graphs}
\label{sec:graphs_symmetry_classes}

The symmetries of the Hamilton operator $H$ of a physical system can be
translated to symmetries for the quantum evolution map
\cite{gnutzmann:2004b}
$\evol(k)=\bondprop(k) \scattering(k)$. It will be more
convenient to define an equivalent quantum evolution map by
\begin{equation}
  \evol_{1/2}(k)=\bondprop(k/2) \scattering(k) 
  \bondprop(k/2)
\end{equation}
and discuss this symmetrised variant.
Here, the right factor $\bondprop(k/2)$ propagates the wave function 
from the centre of a directed bond to
the next vertex where it is scattered by $\scattering(k)$
to the next directed bond. Finally, the left factor
$\bondprop(k/2)$ propagates the wave function along the next directed
bond to its center. 
Some symmetry classes are more naturally discussed for
graphs with a multi-component wavefunction 
(spin or particle-hole components). 

A time-reversal operator $\mathcal{T}$ 
acts on the evolution map as
\begin{equation}
 \evol_{1/2}(k) \mapsto \mathcal{T} \big[ \evol_{1/2}(k) \big] = 
 \hat{\mathcal{T}}
 \evol_{1/2}(k)^* \hat{\mathcal{T}}^\dagger
 \label{eq:evol_time_reversal}
\end{equation}
where $\hat{\mathcal{T}}$ is a unitary matrix with the additional property
$\hat{\mathcal{T}}^* \hat{\mathcal{T}}=\pm\ONE$. Time-reversal invariance
of a graph means that the evolution map is equal
to its time-reversed under the action of $\mathcal{T}$ 
\begin{equation}
  \evol (k)=\mathcal{T}
  \big[\evol(k)^\dagger\big]=
  \hat{\mathcal{T}} \evol(k)^T\, \hat{\mathcal{T}}^{-1}.
\end{equation}
For symmetry class $A$I one requires
$\hat{\mathcal{T}}^*\hat{\mathcal{T}}=\ONE$ whereas
symmetry class $A$II requires 
$\hat{\mathcal{T}}^*\hat{\mathcal{T}}=-\ONE$.
The natural choice for the time-reversal symmetry operator
in symmetry class $A$I is
given by
\begin{equation}
  \hat{\mathcal{T}}=\sigX^\DIR
\end{equation}
which acts like the corresponding Pauli-matrix on the direction indices
$\omega$ while bond indices $b$ remain uneffected. 
Equivalently, the evolution map of a time-reversal invariant graph in class $A$I has the property
\begin{equation}
  \evol(k)_{1/2\, \alpha \beta}=\evol(k)_{1/2\, \hat{\beta} \hat{\alpha}}\ .
\end{equation}

Time-reversal invariant graphs with $\mathcal{T}^2=-\ONE$ (class $A$II)  
can be realised as graphs with spin $s=1/2$ and
\begin{equation}
  \hat{\mathcal{T}}=\sigX^\DIR \otimes \sigY^\SPIN\ .
\end{equation}
Thus time-reversal involves a change in direction and in spin.

The chiral and charge conjugation symmetry classes on 
graphs can all be realised
by either putting the Bogoliubov-de-Gennes or Dirac equations
on the graph. Thus they always involve electron and hole components
of a wavefunction (one needs an additional spin $\sigma=\pm 1/2$ in some classes).
We will not give a complete construction of the symmetries
for all seven classes, for the classes $C$ and $C$I an example for
a star graph is given in Section \ref{sec:pot_andreevstars}.

A chiral symmetry operator $\mathcal{P}$ acts on the
evolution map as 
\begin{equation}
  \evol_{1/2}(E) \mapsto\mathcal{P}\big[ \evol_{1/2}(E) \big]= 
  \hat{\mathcal{P}} 
  \evol_{1/2}(-E)\hat{\mathcal{P}}^{-1}
\end{equation}
where $\hat{\mathcal{P}}$ is a unitary matrix with $\hat{\mathcal{P}}^2=\ONE$ 
and $E$ is the
energy.
A quantum graph has a chiral symmetry if
\begin{equation}
  \evol_{1/2}(E)=\mathcal{P}\big[ \evol_{1/2}(E)^\dagger \big]=
   \hat{\mathcal{P}} \evol(-E)^\dagger \hat{\mathcal{P}}^{-1}
  \label{eq:chiral_symmetry}
\end{equation}
where taking $\evol(-E)^\dagger$ reflects the anti-commuting
character of the Hamilton operator.
It is easy to see from equation \eqref{eq:chiral_symmetry}
that the corresponding energy spectrum is symmetric.
If $\evol_{1/2}(E)$ has an eigenvalue unity and there is a chiral
symmetry $\evol_{1/2}(-E)$ also has an eigenvalue unity.

Finally, a charge conjugation symmetry operator $\mathcal{C}$
acts on the evolution map as
\begin{equation}
  \evol_{1/2}(E)\mapsto\mathcal{C}\big[ \evol_{1/2}(E) \big]= 
  \hat{\mathcal{C}} \evol_{1/2}(-E)^*\hat{\mathcal{C}}^{-1}
\end{equation}
where $\hat{\mathcal{C}}$ is a unitary matrix with 
$\hat{\mathcal{C}}^2=\pm \ONE$.
A quantum graph with a charge conjugation symmetry obeys
\begin{equation}
  \evol_{1/2}(E)=\mathcal{C}\big[ \evol_{1/2}(E)^\dagger \big]=
   \hat{\mathcal{C}} \evol_{1/2}(-E)^T\, \hat{\mathcal{C}}^{-1}
  \label{eq:charge_conj_symmetry}.
\end{equation}

Note, that we have not used the momentum $k$ but the energy $E$ as the
argument of the evolution map. For the chiral and charge conjugation symmetry classes
one should be aware that the momentum is in general not a constant
of motion and thus the energy has to be used. E.g. in the Bogoliubov-de-Gennes
equation an electron and a hole moving in the positive $x$-direction at energy $E$ are
described by the wavefunctions
\begin{equation}
  \psi^\mathrm{electron}(x)\sim 
  \begin{pmatrix}
    1\\
    0
  \end{pmatrix}  
  \E^{\I \sqrt{\mu+E}\,x}
  \qquad 
  \text{and}
  \qquad
  \psi^\mathrm{hole}(x)\sim 
  \begin{pmatrix}
    0\\
    1
  \end{pmatrix}  
  \E^{-\I \sqrt{\mu-E}\,x}
\end{equation}
where $\mu$ is the Fermi energy. If $\mu\gg E$ one may expand
$k_\pm=\sqrt{\mu \pm E}\approx\sqrt{\mu} \pm \frac{E}{2 \sqrt{\mu}}
=k_F \pm \delta k$. In the limit $\frac{\mu}{E}\rightarrow \infty$
one may thus replace $E \mapsto \delta k$
and the symmetry relations of the evolution map hold for $\delta k$.  \newpage
\thispagestyle{empty}
\chapter{Some relevant results of random-matrix theory}
\label{app:random_matrix_theory}

Random matrices were first applied to physical systems by Wigner
and Dyson \cite{wigner:1958,dyson:1962,dyson:1962a,dyson:1962b}
as a model for complex nuclei. Since, random-matrix theory 
has had an enormous impact
on various areas in physics and is today one of the main tools to describe
statistical properties of complex quantum systems.
There are many books and reviews on random-matrix theory\cite{guhr:1998,Book:mehta,beenakker:review,brouwer:diss} and its applications
in physics. In this appendix we summarise some of relevant results 
in the context of this review. 
We will neither discuss the various methods to calculate
properties of random matrices nor give any proof of the presented results,
and we will only summarise results
on the Gaussian ensembles of Hermitian matrices.
  
\section{Universality and universality classes}
\label{sec:universality}

The success of random-matrix theory relies on the fact that many
statistical properties of the spectrum and of the wavefunctions for a large class of
complex quantum systems 
are universal (system independent) yet not trivial. 
When properly scaled they only depend on some
general properties of the system such as time-reversal
invariance. In appendix \ref{app:symmetry_classes} we have
described the ten symmetry classes of quantum systems
which was completely general. 
Here, we relate these symmetries to the statistical properties
of complex quantum systems described by a random Hamilton operator. 

A \define{universality class} is a subset of a symmetry class which shares
the same statistical properties -- or at least some 
universal correlation functions upto small system dependent deviations. 
In general there
are many universality classes within one symmetry class.
Each of them can
be described (and defined) by some ensemble of random matrices
in the limit of large matrices.
 Usually there will be a lot of different ensembles that share
the same universal statistical properties \cite{hackenbroich}.

We will focus on the so called \define{ergodic
universality classes} that can be described by Gaussian ensembles 
of Hermitian matrices in each of the ten symmetry classes
which describe strongly chaotic systems or disordered systems
in the delocalised regime.
For each of the three Wigner-Dyson symmetry classes ($A$, $A$I, and $A$II)
and for two
of the charge conjugation symmetry classes ($C$ and $C$I) 
there is a unique ergodic universality class. 
In contrast for the other symmetry classes there is an additional
parameter. For instance, the symmetry class
$BD$ can be realised in an even or in an odd dimensional Hilbert space
and this will effect the spectral statistics.
In the odd-dimensional case the symmetry fixes one eigenvalue at $E=0$
which changes the statistical properties of the spectrum near $E=0$
while in the even-dimensional case (also denoted as class $D$) the charge
conjugation symmetry does not fix a vanishing eigenvalue.
Similarly, a chiral symmetry operator may fix some integer number $\nu=0,1,2\dots$ of eigenvalues $E=0$. The number $\nu$ is known
as the topological quantum number and is important for applications
of random-matrix theory to quantum chromodynamics. We will only present
the results for those ergodic universality classes where no eigenvalue
$E=0$ is fixed by the symmetry operators $\mathcal{P}$ or $\mathcal{C}$
(thus $\nu=0$ for the chiral classes).  

\section[The Gaussian ensembles of random-matrix theory]{The Gaussian ensembles\\ of random-matrix theory}
\label{sec:gaussian_ensembles}

A Gaussian random-matrix ensemble \cite{Book:mehta} consists of
$N \times N$ Hermitian matrices $H=H^\dagger$ with a Gaussian distribution
\begin{equation}
  P(H)\, \D\!H= \E^{-\lambda\, \mathrm{tr}\, H^2} \D\!H\ .
  \label{eq:rmt_distribution}
\end{equation}
This form is the same for all ergodic universality
classes. However the symmetries of the Hamilton operator restrict the
number of independent matrix elements. The flat measure
$\D H= \mathcal{N}^{-1}\D H_{11} \D\mathrm{Re}(H_{12})\D\mathrm{Im}(H_{12})\dots$
runs over all \emph{independent} real and imaginary parts of the
Hermitian matrix $H$ with the additional symmetry requirements
of the symmetry class. The real parameter $\lambda>0$ rescales the
spectrum of the system and $\mathcal{N}$ is a normalisation constant. 

If there are no further symmetry restrictions on $H=H^\dagger$
the ensemble is called the
\define{Gaussian unitary ensemble }(GUE). It describes 
ergodic systems with broken time-reversal invariance
in class $A$. 
Time-reversal invariance with $\mathcal{T}^2=\ONE$ (class $A$I)
restricts the matrices $H=H^\dagger$ to be 
real (and hence symmetric) $H=H^*=H^T$.
The corresponding ensemble is called the 
\define{Gaussian orthogonal ensemble }(GOE).

Time-reversal invariant systems in symmetry class $A$II 
are realised only in even dimensional Hilbert spaces
due to Kramers' degeneracy.
The random Hamiltonian is a $2N\times 2 N$ matrix
restricted by the symmetry condition
$
\left(
  \begin{smallmatrix}
    0 & \ONE_N\\
    -\ONE_N &0
  \end{smallmatrix}
\right)
H=H^T
\left(
  \begin{smallmatrix}
    0 & \ONE_N\\
    -\ONE_N &0
  \end{smallmatrix}
\right)$
and the corresponding ensemble
is called the \define{Gaussian symplectic ensemble }(GSE).
Note, that we have made a special choice for the time-reversal
operators $\mathcal{T}$ for the GOE and GSE.

The Gaussian ensembles for the chiral and charge conjugation
symmetry classes can easily be constructed from the three
Wigner-Dyson ensembles GUE, GOE, and GSE by reducing the number of independent
matrix elements further as required. In the chiral case the
ensembles have been called \define{chiral Gaussian unitary ensemble }(chGUE)
for class $A$III, the \define{chiral Gaussian orthogonal ensemble }(chGOE)
for class $BD$I, and the \define{chiral Gaussian symplectic ensemble }(chGSE)
for class $C$II. The Gaussian ensembles for the charge conjugation
symmetry classes do not have any established name. We will refer to them
as {${C}$-GE}, {${C}$I-GE}, {${D}$-GE}, and {${D}$III-GE}.

\section{Spectral statistics for Gaussian random matrices}
\label{sec:rmt_spectral_statistics}

The mean density of states for a random-matrix ensemble is defined
as
\begin{equation}
  \langle d(s) \rangle = \frac{1}{g} \langle \mathrm{tr}\,
  \delta(s-H)\rangle
  = \frac{1}{g} \int \D H \, P(H)\, \mathrm{tr}\, \delta(s-H)
\end{equation}
where $g=2$ in symmetry classes with Kramers' degeneracy (every eigenvalue is counted only once) while $g=1$ in other ensembles. For large matrices
in the Wigner-Dyson ensembles the mean density of states is given 
by Wigner's semicircle \cite{wigner:1958,Book:Haake,Book:mehta} law
\begin{equation}
  \langle d(s) \rangle_{\mathrm{GUE,GOE,GSE}}=
  \sqrt{1-\frac{\pi^2 s^2}{4 N^2}}= 1 
  +\mathcal{O}\left(\frac{s^2}{N^2}\right)
  \label{eq:semicircle_law} 
\end{equation}
where we have 
scaled the spectrum such that $\langle d\rangle=1$ in the centre
of the semi-circle by an appropriate choice of the scaling parameter 
$\lambda$ in the definition \eqref{eq:rmt_distribution}. Here $N$
is the size of the matrix ($N \times N$ for GUE and GOE, $2N\times 2N$ for
GSE). The result \eqref{eq:semicircle_law} is exact in the
limit $s,N \rightarrow \infty$ where $s/N$ is kept constant. 
The relevant limit for our purposes is $N\rightarrow \infty$ keeping
$s$ constant. In that limit the density of states is flat and
the mean level spacing, $\Delta s=\langle d \rangle^{-1}=1$.

In the presence of chiral or charge conjugation symmetries, there
are deviations from Wigner's semicircle law near $s=0$ due to
the symmetric spectrum. On the scale of the mean level spacing
the mean densities of states for these ensembles are given by
\cite{altlandzirnbauer2,forrester,verbaarschot,nagao,ivanov}
\begin{equation}
  \begin{split} 
    \langle d (s) \rangle_\text{chGUE}=&
    \frac{\pi^2 |s|}{2}
    \left(
      J_0^2(\pi s) +
      J_1^2(\pi s)
    \right)
    \\
    \langle d (s)\rangle_\text{chGOE}=&
    \langle d (s) \rangle_\text{chGUE}
    +
    \frac{\pi}{2}
    J_0(\pi s)
    \left( 
      1- \INT{0}{\pi|s|}{\xi} J_0(\xi)
    \right)
    \\
    \langle d(s)\rangle_\text{chGSE}=& 
    \langle d (2s) \rangle_\text{chGUE}
    -\frac{\pi}{2}
    J_0(2\pi s)\,\INT{0}{2\pi|s|}{\xi}
    \, J_0(\xi)
    \\ 
    \langle d (s)\rangle_\text{$C$-GE}=&
    1-\frac{\sin 2 \pi s}{2 \pi s}\\
    \langle d(s)\rangle_\text{$C$I-GE}=&
    \langle d (s) \rangle_\text{chGUE}
    -
    \frac{\pi}{2}
    J_0(\pi s)J_1(\pi|s|)
    \\
    \langle d(s)\rangle_\text{$D$-GE}=&
    1+\frac{\sin 2 \pi s}{2 \pi s}
    \\
    \langle d(s)\rangle_\text{$D$III-GE}=&
    \langle d(2s)\rangle_\text{$C$I-GE}
    +
    \frac{\pi}{2}J_1(2\pi|s|)\ .
  \end{split}
\end{equation}
For all seven ensembles the deviation from the flat density of states
is pronounced at $s=0$ and decays for $s\gg 1$. 
The deviations near $s=0$
are universal interference effects that can be
seen in complex quantum systems of
the corresponding symmetry class. For quantum graphs 
this is discussed 
in Section \ref{sec:pot_andreevstars}.

The \define{spectral two-point correlation function}
for an ensemble of matrices is defined as
\begin{equation}
  \begin{split}
    R_2(s;s_0)=&
    \langle 
    d(s_0+\frac{s}{2})
    d(s_0-\frac{s}{2}) \rangle-1
    \\
    =&
    \frac{1}{g^2}\langle \mathrm{tr}\,
    \delta(s_0+\frac{s}{2}-H)\,\mathrm{tr}\,
    \delta(s_0-\frac{s}{2}-H) \rangle -1\ .
  \end{split}
\end{equation}
For the Wigner-Dyson ensembles they do not depend on 
the central energy $s_0$ (in the limit $N\rightarrow \infty$;
$s,s_0=const$, as before) and are given by \cite{Book:mehta,Book:Haake}
\begin{equation}
  \begin{split}
  R_2(s)_\text{GUE}=&
  \delta(s)-
  \frac{\sin^2 \pi s }{\pi^2 s^2}
  \\
  R_2(s)_\text{GOE}=&
  R_2(s)_\text{GUE} 
  +
  \frac{
    \left(
      \pi |s| \cos \pi s - 
      \sin \pi |s|
    \right)
    \left(
      2\, \mathrm{Si}(\pi |s|)-\pi
    \right)
    }{
    2 \pi^2 s^2
  }
  \\
  R_2(s)_\text{GSE}=&
  R_2(2s)_\text{GUE}
  +
  \frac{
    2 \pi |s| \cos 2 \pi s - \sin 2 \pi
    |s|}{4 \pi^2 s^2} 
  \,\mathrm{Si}(2\pi|s|)\ .
  \end{split}
\end{equation}
The Fourier transform
\begin{equation}
  K(\tau)=\INT{-\infty}{\infty}{s} R_2(s)\E^{-2\pi \I 
  s \tau}
\end{equation}
is known as the \define{spectral form factor}. 
For the Wigner-Dyson ensembles the form factors are given by
\begin{equation}
  \begin{split}
    K(\tau)_\mathrm{GUE}=&
    \begin{cases}
      |\tau| & \text{for $|\tau|\le 1$,}\\
      1 & \text{for $|\tau| \ge 1$,}
    \end{cases}\\
    K(\tau)_\mathrm{GOE}=&
    \begin{cases}
      |\tau|\left(2- \mathrm{ln}\left(2 |\tau| +1\right)\right) &
      \text{for $|\tau|\le 1$,}\\
      2-|\tau| \mathrm{ln} \frac{2|\tau|+1}{2|\tau|-1}
      & \text{for $|\tau| \ge 1$,}
    \end{cases}
    \\
    K(\tau)_\mathrm{GSE}=&
    \begin{cases}
      \frac{|\tau|}{4}\left(2-\mathrm{ln}
      \left|1- \left|\tau\right| \right|
      \right)&
      \text{for $|\tau|\le 2$,}\\
      1 & \text{for $|\tau| \ge 2$.}
    \end{cases}
  \end{split}
\end{equation}

Since we do not discuss second-order correlation functions for
systems with chiral or charge conjugation symmetry we will not
state further results. Note however, that all spectral correlation
function go over to the correlation functions of the Wigner-Dyson ensembles
for energies $s \gg 1$

Another frequently used statistics that depends on correlation-functions
of any order is the \define{level spacing distribution}.
Ordering the spectrum such that $s_n \le s_{n+1}$
the level spacings the differences $s_i=s_{i+1}-s_i$
between two subsequent eigenvalues. Their distribution 
for the Wigner-Dyson ensembles can be approximated very well by the
Wigner surmises 
\begin{equation}
  \begin{split}
  P(s)_\text{GUE}=&
  \frac{32 s^2}{\pi^2} \,\E^{-\frac{4}{\pi}s^2}\\
  P(s)_\text{GOE}=&
  \frac{\pi s}{2}\, \E^{-\frac{\pi}{4}s^2}\\
  P(s)_\text{GSE}=&
  \frac{2^{18} s^4}{3^6 \pi^3}\,  \E^{-\frac{64}{9 \pi}s^2}\ 
  \end{split}
  \label{eq:Wigner_surmises}
\end{equation}
which are exact for $2 \times 2$ (GUE, GOE) or $4 \times 4$ (GSE)
matrices.  
 \newpage
\thispagestyle{empty}

\end{document}